\PassOptionsToPackage{justification=justified}{caption}
\documentclass[superscriptaddress,reprint,english,aps,floatfix,longbibliography,10pt,twocolumn,pra,nofootinbib,notitlepage,fleqn]{revtex4-1}
\usepackage{graphicx}
\usepackage[usenames,dvipsnames]{color}
\definecolor{purple}{RGB}{128,0,128}
\definecolor{ultramarine}{RGB}{63, 0, 255}
\definecolor{medblue}{RGB}{0, 0, 100}
\definecolor{panblue}{RGB}{0,24,150}
\definecolor{carmine}{RGB}{150, 0, 24}
\definecolor{gray}{RGB}{150, 150, 150}
\usepackage[pdftex,pdfversion=1.7,pdfencoding=auto,pdfnewwindow=true,pdfusetitle=true,bookmarks=true,bookmarksnumbered=true,bookmarksopen=true,pdfpagemode=UseThumbs,bookmarksopenlevel=1,pdfpagelabels=false,breaklinks=true]{hyperref}
\hypersetup{colorlinks,
linkcolor=carmine,
citecolor=medblue,
urlcolor=panblue,
anchorcolor=OliveGreen}
\usepackage[section]{placeins}
\usepackage{amsmath}
\usepackage{amsthm}
\usepackage{amssymb}
\usepackage{mathtools}
\usepackage{lmodern}
\usepackage{dsfont}
\usepackage{bbold}
\usepackage{bm}
\usepackage{paralist}
\usepackage{comment}
\usepackage{breqn}
\usepackage{float}
\usepackage[justification=justified,labelfont=small]{subcaption}
\usepackage{ragged2e}
\DeclareCaptionJustification{justified}{\justifying}
\usepackage{enumerate}
\usepackage[capitalise]{cleveref}
\Crefname{eqs}{Eqs.}{Eqs.}
\creflabelformat{eqs}{(#2#1#3)}
\crefrangelabelformat{equation}{(#3#1#4-#5#2#6)}
\Crefmultiformat{equation}{Eqs.~(#2#1#3}{,#2#1#3)}{,#2#1#3)}{,#2#1#3)}
\crefrangelabelformat{eqs}{(#3#1#4-#5#2#6)}
\Crefmultiformat{eqs}{Eqs.~(#2#1#3}{,#2#1#3)}{,#2#1#3)}{,#2#1#3)}
\usepackage{dutchcal}
\newtheorem*{theorem*}{Theorem}

\newtheorem*{lemma*}{Lemma}

\newcommand{\bit}{\begin{itemize}}
\newcommand{\eit}{\end{itemize}\par\noindent}
\newcommand{\ben}{\begin{enumerate}}
\newcommand{\een}{\end{enumerate}\par\noindent}
\newcommand{\beq}{\begin{equation}}
\newcommand{\eeq}{\end{equation}\par\noindent}
\newcommand{\beqa}{\begin{eqnarray}}
\newcommand{\eeqa}{\end{eqnarray}\par\noindent}
\newcommand{\beqn}{\begin{eqnarray}}
\newcommand{\eeqn}{\end{eqnarray}\par\noindent}

\newcommand{\avg}[1]{\langle #1 \rangle}
\newcommand{\pr}[1]{\operatorname{pr}\{#1\}}

\def\M{\mathcal{M}}

\def\S{\mathcal{S}}

\newcommand{\tr}[1]{\operatorname{tr}\left\{ #1 \right\}}

\newcommand{\obs}[1]{\operatorname{O}_{#1}}
\newcommand{\ens}[1]{\operatorname{\mathcal{E}}_{#1}}

\newcommand{\classicalOutputSrc}[1]{s_{#1}}
\newcommand{\classicalOutputSrcEqv}[1]{\mathcal{s}_{#1}}

\newcommand{\classicalOutputMsmt}[1]{m_{#1}}
\newcommand{\classicalOutputMsmtEqv}[1]{\mathcal{m}_{#1}}

\newcommand{\src}[1]{\operatorname{S}_{#1}}

\newcommand{\msmt}[1]{\operatorname{M}_{#1}}

\newcommand{\eqvClassSrc}[1]{\S_{#1}}

\newcommand{\eqvClassMsmt}[1]{\M_{#1}}

\newcommand{\opAvg}[1]{\langle #1 \rangle_{\msmt{},\src{}}}
\newcommand{\onticAvg}[1]{\langle #1 \rangle_{\lambda}}

\def\1{\mathbb{1}}

\usepackage{yfonts}
\usepackage[T1]{fontenc}
\usepackage{mathrsfs}

\makeatletter
\def\p@subsection{\thesection.}
\def\p@subsubsection{\thesection.\thesubsection.}
\makeatother

\begin{document}

%\title{Testing noncontextuality without assuming determinism: operational inequalities from the Peres-Mermin magic square}
%\title{Tests of noncontextuality based on the Peres-Mermin square: From algebraic proofs of the Kochen-Specker theorem to operational noncontextuality inequalities}
\title{Deriving robust noncontextuality inequalities from algebraic proofs of the Kochen-Specker theorem: the Peres-Mermin square}

\author{Anirudh Krishna}
\email{anirudh.krishna@usherbrooke.ca}
\affiliation{Perimeter Institute for Theoretical Physics\(,\) Waterloo\(,\) Ontario\(,\) Canada\(,\) N2L 2Y5}
\affiliation{Universit{\'e} de Sherbrooke\(,\) Sherbrooke\(,\) Qu{\'e}bec\(,\) Canada\(,\) J1K 2R1}

\author{Robert W. Spekkens}
\email{rspekkens@perimeterinstitute.ca}
\affiliation{Perimeter Institute for Theoretical Physics\(,\) Waterloo\(,\) Ontario\(,\) Canada\(,\) N2L 2Y5}

\author{Elie Wolfe}
\email{ewolfe@perimeterinstitute.ca}
\affiliation{Perimeter Institute for Theoretical Physics\(,\) Waterloo\(,\) Ontario\(,\) Canada\(,\) N2L 2Y5}

%\affiliation{Perimeter Institute for Theoretical Physics, Waterloo, Ontario, Canada, N2L 2Y5}

\date{\today}                                           % Activate to display a given date or no date

\begin{abstract}
When a measurement is compatible with each of two other measurements that are incompatible with one another, these define distinct {\em contexts} for the given measurement. The Kochen-Specker theorem rules out models of quantum theory that satisfy a particular assumption of context-independence: that sharp measurements are assigned outcomes both deterministically and independently of their context. This notion of noncontextuality is not suited to a direct experimental test because realistic measurements always have some degree of unsharpness due to noise.  However, a generalized notion of noncontextuality has been proposed that is applicable to {\em any} experimental procedure, including unsharp measurements, but also preparations as well, and for which a quantum no-go result still holds.  According to this notion, the model need only specify a {\em probability distribution} over the outcomes of a measurement in a context-independent way, rather than specifying a particular outcome.   It also implies novel constraints of context-independence for the representation of preparations.  In this article, we describe a general technique for translating proofs of the Kochen-Specker theorem into inequality constraints on realistic experimental statistics, the violation of which witnesses the impossibility of a noncontextual model.  We focus on algebraic state-independent proofs, using the Peres-Mermin square as our illustrative example. Our technique yields the necessary and sufficient conditions for a particular set of correlations (between the preparations and the measurements) to admit a noncontextual model. The inequalities thus derived are demonstrably robust to noise.  We specify how experimental data must be processed in order to achieve a test of these inequalities. We also provide a criticism of prior proposals for experimental tests of noncontextuality based on the Peres-Mermin square.
\end{abstract}

\maketitle

{\hypersetup{linkcolor=black}
% or \hypersetup{linkcolor=black}, if the colorlinks=true option of hyperref is used
\tableofcontents
}

%------------------------------------------------------------------------------------------------
\section{Introduction}
%------------------------------------------------------------------------------------------------

Ontological models of quantum theory are an attempt to explain the statistical
predictions of quantum theory.  They take every system to be associated with
a space of possible physical states, termed {\em ontic states}, every quantum
state to be represented by a statistical distribution over these ontic states, and
every measurement to be represented by a conditional probability distribution for the outcome given the ontic state~\cite{HarriganSpekkens}.  Hidden variable models are examples of ontological models, but so is the physicist's orthodox conception of quantum
theory, wherein the ontic states are simply the pure quantum states, not supplemented
by any additional variables.\footnote{The latter is termed the $\psi$-complete ontological model in Ref.~\cite{HarriganSpekkens}.}

%The standard Kochen-Specker notion of noncontextuality
The principle of noncontextuality is an assumption
about ontological models that seeks to capture a  notion of classicality.
It started its life as an assumption about {\em outcome-deterministic ontological models} of quantum
theory, that is, ontological models wherein the outcome of every measurement was
fixed deterministically by the ontic state (in contrast to the orthodox conception).
This assumption was famously demonstrated to be in contradiction with the predictions of quantum
theory by Kochen and Specker \cite{KS} and Bell \cite{Bell}.  The Kochen-Specker theorem is one of the strongest constraints on the intepretation of quantum theory. Furthermore, failing to admit of a noncontextual model
appears to be a resource.  For instance, in the context of the state
injection model for quantum computation~\cite{BravyiKitaev, Knill2005}, the failure of noncontextuality has been shown in some cases to be necessary for achieving
universal quantum computation \cite{HowardEtAl,bermejo2016contextuality}.
%Furthermore,  the fact that quantum theory fails to admit of a noncontextual model has been shown to be a resource for computation: in the context of the state injection model for quantum computation~\cite{BravyiKitaev, Knill2005}, violating the principle of noncontextuality is a necessary and sufficient condition for achieving universal quantum computation \cite{HowardEtAl}. [add Raussendorf reference?]

In Ref.~\cite{Spekkens05}, a generalized notion of noncontextuality was proposed.
%defined, one that applied to measurement procedures but also to preparation procedures.
For measurements, it constitutes a relaxation of what was assumed by Kochen-Specker and by Bell.
Specifically, it allows the assignment of measurement outcomes by ontic states to be {\em indeterministic}.
%did not presume that measurement outcomes were assigned deterministically by the ontic states.
In this way, it redefined the notion of noncontextuality for measurements in a way that excised the notion of determinism.  This is desirable from a foundational
perspective as it allows one to separate the issue of noncontextuality from that of
determinism (recall that Bell's notion of local causality does not presume that the
outcomes of measurements are fixed deterministically).  Additionally, it can be shown
that the assumption of outcome determinism is unwarranted for any unsharp measurement (i.e., a measurement for which one cannot find a basis of preparations relative to which it is perfectly predictable), and every measurement appearing in a real experiment is of this sort~\cite{Spekkens14}.
As such, this generalization is important if one hopes to turn the proven theoretical advantages for computation into practical advantages, because in practice,
%for the feasibility of  experimental tests.  It is also important if
%if the failure of noncontextuaity is ever to yield an advantage for computation not just in theory but in practice, where
sharpness is an idealization that is never strictly satisfied.
%for  if the failure of noncontextuality is to yield an advantage for computation not just in theory but in practice, then .  al realistic experiments, where measurements are necessarily unsharp, one must move beyond the assumption of outcome determinism, and adopting the revised notion of noncontextuality proposed in Ref.~\cite{Spekkens05} is a means of doing so.

% the orthodox conception of an ontological model, wherein the ontic states are the pure states and measurements respond indeterministically, is found to represent measurements noncontextually.

Although the revised notion of noncontextuality yields a {\em weaker} constraint on the representation of measurements in the ontological
 model than did the traditional notion\footnote{A consequence of this relaxation is that, by its lights, the $\psi$-complete ontological model is found to represent measurements noncontextually.  However,  it fails to represent {\em preparations} noncontextually, as noted in Ref.~\cite{Spekkens05}. },
 %.  In another sense, however, it imposed a stronger constraint than the traditional notion because
 it naturally applies not only to measurements but to preparations as well, and thereby implies novel
 constraints on how quantum states can be represented by distributions over ontic states in the model.\footnote{It also implies novel constraints on how {\em quantum channels} can be represented by conditional probability distributions from the space of ontic states to itself.}  It was argued in Ref.~\cite{Spekkens05} that whatever motivations can be given for assuming noncontextuality for one type of procedure, such as a measurement, this same motivation can be given for assuming it of any other type of procedure, such as a preparation.  Consequently, the only {\em natural} assumption to consider in this approach is that the revised notion of noncontextuality applies to {\em all} procedures.  This assumption is termed  {\em universal noncontextuality} or simply {\em noncontextuality}.  We will henceforth refer to the traditional notion of noncontextuality as {\em KS-noncontextuality} (for Kochen-Specker) to avoid any confusion.

 In Ref.~\cite{Spekkens05}, it was shown that quantum theory does not admit of a universally noncontextual ontological model.  It was also demonstrated that if one replaces the assumption of KS-noncontextuality for measurements with the assumption of universal noncontextuality for all procedures,
 %in transitioning from assuming KS-noncontextuality for measurements to assuming the generalized notion of noncontextuality for all procedures,
  the relaxation of the constraints on the representation of measurements is compensated by the strengthening of the constraints on the representation of preparations in such a way that any proof that quantum theory fails to admit of a KS-noncontextual model
  %(of which there are many)
   can be translated into a proof that it fails to admit of a universally noncontextual model.

Much of the research on noncontextuality to date has centered on the question of whether {\em quantum theory} admits of a noncontextual model.
%Our discussion so far has  noncontextual models {\rm for quantum theory}.
A more general question, which has been the impetus for much recent work,  is whether one
can devise a {\em direct experimental test} of the assumption of a noncontextual ontological model, one that is independent of the validity of quantum theory.  Just as
a Bell inequality is a constraint on experimental statistics that follows directly from
the assumption of a locally causal ontological model, without any reference to the quantum formalism, what one
wants of a test of noncontextuality is a constraint on experimental statistics that
follows directly from the assumption of a noncontextual ontological model, without any reference to the quantum
formalism.
Such constraints will here be termed {\em noncontextuality inequalities}.
%Such a test would allow one to determine whether nature herself satisfies the principle of noncontextuality.
%If one finds a conflict between experimental statistics and noncontextuality,
If experimental statistics are found to violate these inequalities, then one can conclude that not just quantum theory
but {\em any} operational theory that can do justice to the experimental statistics---and therefore nature itself---must fail to admit of such a model, thereby constraining the form of all future physical theories.

The generalized notion of noncontextuality proposed in Ref.~\cite{Spekkens05} was
defined in such a way as to be applicable to any operational theory, not just quantum theory, such that if an experiment yields data supporting an operational theory distinct from quantum theory, the question of whether it admits of a noncontextual model is still meaningful.
The definition asserts that an ontological model of an operational theory is noncontextual if two
experimental procedures that are statistically indistinguishable at the operational
level are statistically indistinguishable at the ontological level.  They key point is that the notion of statistical indistinguishability at the operational level can be assessed in any operational theory\footnote{The fact that the notion of statistical indistinguishability is applicable not just pairs of measurements, but to pairs of preparations and transformations respectively as well, is what allows the generalized notion of noncontextuality to be applicable to these other procedures.}.

It has been shown that violations of noncontextuality inequalities defined in terms of this notion can imply advantages for information processing which are independent of the validity of quantum theory. For example, they imply an advantage for the cryptographic task of parity-oblivious random access codes~\cite{SpekkensParityOblivious,SikoraParityOblivious,AmbainisParityOblivious}.
% It may even be possible to use such violations in the place of Bell-inequality violations to achieve device-indepdent protocols.
%\color{red} [Repeat some of the QIP abstract for the \citet{MazurekEtAl} paper] \color{black}
Such inequalities also hold promise for making the results on quantum computational advantages discussed above robust to noise and for expressing the origin of the advantage in a manner that is independent of the validity of quantum theory.
% into a demonstration that computational advantages follow directly from the failure of noncontextuality, independent of the validity of quantum theory.
\color{black}

Several recent works have considered the question of how to derive noncontextuality inequalities and how to subject them to experiment test~\cite{MazurekEtAl,KunjwalSpekkens,Pusey}.  The present work is concerned with a special case of this problem, namely, how to derive noncontextuality inequalities {\em starting from any given proof of the Kochen-Specker theorem}, that is, from a proof of the failure of KS-noncontextuality in quantum theory.
As noted above, Ref.~\cite{Spekkens05} showed how, in general, to convert a proof of the failure of KS-noncontextuality in quantum theory into a proof of the failure of universal noncontextuality in quantum theory, so the outstanding problem is how to convert a proof of the failure of universal noncontextuality in quantum theory into an operational noncontextuality inequality.
%If this problem could be solved in full generality, one could convert any proof of the failure of KS-noncontextuality in quantum theory---and there is now a wide diversity of such proofs---into an operational noncontextuality inequality.

Note that any test of noncontextuality that is devised from a particular no-go theorem requires an experimentalist to target a particular set of preparations and a particular set of measurements, each with specified
%compatibility and operational equivalence
relations holding among their members (we will say more about the nature of these relations in due course).
A more general version of the problem, however, is to figure out how to infer from {\em any} experimental data---that is, from an experiment that was not designed to target particular preparations or measurements or any particular relations among them---whether or not it admits of a noncontextual model.  Because a test of noncontextuality is a test of classicality, having the capability to test the assumption of noncontextuality on {\em any} experimental data is clearly of greater utility than merely knowing how to implement a dedicated experiment for testing the hypothesis of noncontextuality.
%There has been some progress on this problem.
%The only prior work that addressed the question of how to derive {\em all} of the noncontextuality inequalities in a given experimental scenario was that of Pusey~ \cite{Pusey}.
Pusey~\cite{Pusey} identified the conditions that are both necessary and sufficient for the existence of a noncontextual model for experimental data derived from the simplest experimental scenario in which such conditions are expected to be nontrivial.  Unfortunately, this simplest scenario does not arise within operational quantum theory\footnote{Recall that a set of measurements is said to be {\em tomographically complete} for a system if the
statistics for any measurement on the system can be computed from the statistics of
the measurements in this set.  Pusey's simplest scenario is one wherein
a tomographically complete set of measurements consists of just two binary-outcome
measurements.  This scenario does not arise in operational quantum theory, because the
simplest quantum system, a qubit, requires {\em three} binary-outcome measurements
for tomographic completeness.}.
Extending Pusey's analysis to more general scenarios is an important open problem.

Nonetheless, there are also advantages to building sets of noncontextuality inequalities from specific proofs of the Kochen-Specker theorem, because such proofs have nontrivial structural properties.  Different proofs---and there is now a great diversity of these---capture what is surprising about the failure
of noncontextuality in different ways, and these intuitions are likely to be helpful in identifying the applications thereof.

We here focus on deriving noncontextuality inequalities from {\em state-independent} proofs of the Kochen-Specker theorem.

Ref.~\cite{KunjwalSpekkens} has already demonstrated how one can derive one such inequality from any state-independent {\em geometric} proof of the Kochen-Specker theorem, that is, any proof expressed in terms of an uncolourable set of rays.  Here, we extend this work in two important ways:  (1) we provide a technique for finding {\em all} of the noncontextuality inequalities that apply to a certain set of correlations starting from any state-independent proof of the Kochen-Specker theorem, and (2) we show how to do so for proofs that are expressed algebraically rather than geometrically.
 We expand on each of these points presently, in reverse order.

%For the case of uncolourability proofs of the Kochen-Specker theorem, the way to do so was recently described in Kunjwal and Spekkens~\cite{Kunjwal Spekkens}.
%This article contributes to this research programme in several ways, which we now outline.

The distinction between geometric and algebraic proofs of the failure of KS-noncontextuality in quantum theory is not fundamental  because one can convert any
algebraic proof into a geometric form and vice-versa.
Nonetheless, each proof style has its advantages.  The first known proofs were geometric uncolorability proofs.  Algebraic proofs arose later, but in many respects they have a logic that is easier to grasp. Indeed, the paradigm example of a proof of the Kochen-Specker theorem is now arguably the algebraic version of the Peres-Mermin square proof~\cite{Peres, Mermin}, which will be the example we focus on here.

Furthermore, the algebraic structure suggests generalizations of these proofs that might not be obvious from the geometric perspective~\cite{WaegellPMGeneralizationToN,WaegellPMGeneralizationTo3}.
Although one could derive a noncontextuality inequality for the Peres-Mermin square by first expressing the latter as a geometric proof (as in Ref.~\cite{Peres}) and then applying the technique described in Ref.~\cite{KunjwalSpekkens}, it is more useful to have
 a technique for deriving noncontextuality inequalities that is native to the algebraic approach.  We here provide such a technique.

In order to turn  a proof of the failure of universal noncontextuality in quantum theory into a noncontextuality inequality, one must {\em operationalize} the description of the experiment provided in the no-go theorem, purging it of any reference of the quantum formalism, and one must {\em robustify} the constraints on experimental data that are derived from noncontextuality, which means that these constraints must provide quantitative bounds that can be violated in principle even
if the experimental operations are noisy.  This progression was achieved in Ref.~\cite{KunjwalSpekkens}, but the resulting inequality provided an upper bound on just a single operational quantity (an average, over certain preparation-measurement pairs, of the degree of correlation between them).  The technique described in the present article goes much further towards providing a means of deriving {\em all} of the noncontextuality inequalities that hold for a given set of preparations and measurements.
% with specified relations holding among them.
Although we focus on a subset of the correlations between preparations and measurements that arise in the construction,
for this restricted set of experimental data, satisfaction of the inequalities that we derive is both necessary and sufficient for the existence of a noncontextual model.

Finally, we note a difference in the way experiments are described in this article relative to previous treatments of inequalities for universal noncontextuality~\cite{SpekkensParityOblivious,MazurekEtAl,KunjwalSpekkens}.  We here use the notion of a {\em source}, that is, a process which samples a classical variable from a distribution, chooses which preparation procedure to implement on the system based on the value sampled and outputs both the system and the variable.  This choice ensures that our derived noncontexuality inequalities are easier to compare with Bell inequalities.

The remainder of the paper is structured as follows.

In \cref{sec:Background}, we provide an overview of operational theories (\ref{sec:Operational}) and ontological models (\ref{sec:Ontological}).  In particular, we discuss the concepts of operational equivalence and of compatibility (applied to measurements and sources) and illustrate the concepts with quantum examples.
We provide formal definitions of measurement noncontextuality and preparation noncontextuality, in particular, a characterization of these assumptions in terms of expectation values for the outcomes of measurements and sources given the ontic state.

In \cref{NoGo}, we review the well-known proof of the failure of KS-noncontextuality in quantum theory based on the Peres-Mermin square (\ref{sec:NoGoKS}), and we show how to translate this no-go theorem into one that demonstrates the failure of universal noncontextuality in quantum theory (\ref{NoGoUniversal}).

\cref{sec:III} is the heart of the article, describing our technique for turning quantum no-go theorems into operational noncontextuality inequalities.  In the first subsection (\ref{sec:operationalPM}), we operationalize the description of the quantum measurements and sources that appear in the Peres-Mermin-inspired proof of the failure of universal noncontextuality, thereby obtaining a notion of a Peres-Mermin experimental scenario that is purged of any reference to quantum theory.  This provides a template for how to achieve this operationalization for any such construction.
The following five subsections (\ref{sec:ontic}-\ref{sec:NCInequalities})
%\ref{sec:verticesCpolytope})
describe how to derive noncontextuality inequalities from such an operational construction, using Peres-Mermin as the illustrative example.
 %In \cref{sec:NCInequalities}, we present the noncontextuality inequalities relevant for Peres-Mermin.
We also show how the ideal quantum realization of the measurements and sources in the Peres-Mermin scenario violate these inequalities (\ref{sec:QViolationInequalities}), and we demonstrate the robustness of these inequalities to noise (\ref{sec:Robustness}), by showing how they can be violated by partially depolarized versions of the ideal quantum realizations of the measurements and sources.

In \cref{sec:ExperimentalTest}, we clarify what must be done experimentally in order to test the noncontextuality inequalities we have derived, and in \cref{sec:Conclusions} we provide our concluding remarks.

Appendix~\ref{sec:repconversion} discusses the problem of computationally converting between the vertex and halfspace representations of a polytope., Appendix~\ref{symmetries} discusses the symmetries of our noncontextuality inequalities under deterministic processings of the experimental procedures, and  Appendix~\ref{TrivialIneqs} demonstrates that a certain class of inequalities on experimental statistics are trivial.  Finally, Appendix~\ref{subsec:prevProposals} reviews a previous proposal for how to implement an experimental test of noncontextuality based on the Peres-Mermin square, and argues against its adequacy.
%contrasts it with the proposal of this article.

\section{Preliminaries}\label{sec:Background}

\subsection{Operational concepts}\label{sec:Operational}

\subsubsection{Operational theories}

The primitive elements of an operational theory are preparations and measurements, each
specified as lists of instructions to be performed in the laboratory.

A {\em source} is a device that implements one of a set of preparation procedures on a system, sampled from some probability distribution, and has a classical outcome that heralds which preparation has in fact been implemented. (The use of the term ``source'' to refer to such a device is conventional in both classical and quantum Shannon theory, where it is the standard way of modelling the input to a communication channel~\cite{preskillnotes}.)
We will denote a source by $\src{}$ and the variable describing its classical outcome by $\classicalOutputSrc{}$.
% and the system prepared by $\quantumSystem$.
%system, may be a classical variable, or a quantum system, or a system described by a GPT, labelled $\quantumSystem$.

A measurement, denoted $\msmt{}$, accepts as input a system
%, denoted $\quantumSystem$,
and returns a classical outcome, denoted by the variable $\classicalOutputMsmt{}$.

An operational theory provides an algorithm for computing the probability distribution for the outcome of any measurement acting on any preparation, and consequently it allows the computation of the joint probability distribution over the outcome of any measurement $\msmt{}$ and the outcome of any source $S$, $\pr{\classicalOutputMsmt{},\classicalOutputSrc{}|\msmt{},\src{}}$.
%The operational framework provides a way to compute the probability
%$\pr{\classicalOutputSrc{},\classicalOutputMsmt{}|\msmt{},\src{}}$.
%Let $\pr{\classicalOutputMsmt{1},\classicalOutputSrc{}|\msmt{1},\src{}}$ be termed the {\em joint distribution} on the measurement-source pair $(M,S)$.
We refer to this as simply the joint distribution on the measurement-source pair $(M,S)$.

\subsubsection{Operational equivalence}

Consider two measurement procedures, $\msmt{1}$ and $\msmt{2}$, whose outcomes are random
variables, denoted $\classicalOutputMsmt{1}$ and $\classicalOutputMsmt{2}$ respectively.
$\msmt{1}$ and $\msmt{2}$ are said to be operationally equivalent if they define the same joint distribution for all possible sources:
\begin{align}
% \forall S, \forall \alpha: \pr{\classicalOutputSrc{},\classicalOutputMsmt{1}{\rm =}\alpha|\msmt{1},\src{}} =
%  \pr{\classicalOutputSrc{},\classicalOutputMsmt{2}{\rm =}\alpha|\msmt{2},\src{}}.
\forall S: \pr{\classicalOutputMsmt{1},\classicalOutputSrc{}|\msmt{1},\src{}} =
  \pr{\classicalOutputMsmt{2},\classicalOutputSrc{}|\msmt{2},\src{}}.
\end{align}
%for all sources $\src{}$.
%This shall be denoted $\msmt{1} \simeq \msmt{2}$.
Letting $\eqvClassMsmt{}$ denote an operational equivalence class of measurements, we can express the operational equivalence of two measurements $\msmt{1}$ and $\msmt{2}$ by specifying that they belong to the same class,
%Such an operational equivalence relation can also be expressed by specifying that the two measurements belong to the same operational equivalence class $\eqvClassMsmt{}$,
\begin{align}
\msmt{1},\msmt{2} \in \eqvClassMsmt{}.
\end{align}
%The particular element of an equivalence class that is implemented is called the {\em measurement context}.

%The standard example from quantum theory of a pair of measurements that are physically distinct but operationally equivalent.

Similarly, two sources, $\src{1}$ and $\src{2}$, whose outcomes are random variables $\classicalOutputSrc{1}$ and $\classicalOutputSrc{2}$ respectively, are said to be operationally
equivalent if they define the same joint distibution for all possible measurements:
\begin{align}
 \forall \msmt{}: \pr{\classicalOutputMsmt{},\classicalOutputSrc{1}|\msmt{},\src{1}} =
  \pr{\classicalOutputMsmt{},\classicalOutputSrc{2}|\msmt{},\src{2}}.
% \forall \msmt{}, \forall \alpha : \pr{\classicalOutputSrc{1}{\rm =}\alpha,\classicalOutputMsmt{}|\msmt{},\src{1}} =
%  \pr{\classicalOutputSrc{2}{\rm =}\alpha,\classicalOutputMsmt{}|\msmt{},\src{2}}.
\end{align}
%for all measurement procedures $\msmt{}$.
%This shall be denoted $\src{1} \simeq \src{2}$.
Letting $\eqvClassSrc{}$ denote an operational equivalence class of sources, we can express the operational equivalence of two sources,  $\src{1}$ and $\src{2}$, by stipulating that they are in the same class,
\begin{align}
\src{1},\src{2} \in  \eqvClassSrc{} .
\end{align}

%This defines an equivalence class $\eqvClassSrc{}$ between sources
%\begin{align}
%  \eqvClassSrc{} = \{\src{1},\src{2}\}
%\end{align}

Because the joint distribution $\pr{\classicalOutputSrc{},\classicalOutputMsmt{}|\msmt{},\src{}}$ is computable from the operational theory, so too is the correlation between the
%The operational theory allows us to compute the correlations between the
outcome $\classicalOutputSrc{}$ of the source $\src{}$ and the outcome
$\classicalOutputMsmt{}$ of the measurement $\msmt{}$,
\begin{align}\label{eq:correlationsXY}
  \opAvg{\classicalOutputSrc{} \classicalOutputMsmt{}} =
  \sum_{\classicalOutputSrc{},\classicalOutputMsmt{}} \classicalOutputSrc{}
  \classicalOutputMsmt{}
  \pr{\classicalOutputSrc{},\classicalOutputMsmt{}|\msmt{},\src{}}~.
\end{align}
Such correlations will be the quantities appearing in our noncontextuality inequalities.
%This will be a quantity of interest later.

\subsubsection{Compatibility}\label{Compatibility}

In this section, we briefly review the notion of compatibility. The interested reader
is pointed to \cite{LiangWisemanSpekkens} for a detailed overview.

Informally, two or more devices
are said to be compatible if their output can be obtained by classical post-processing of the
output of a single device.  More precisely, one can define the notion of compatibility in terms of a notion of simulatability.

Consider two measurements, $\msmt{}$ and $\msmt{}'$, which accept an input system and output random variables $\classicalOutputMsmt{}$ and
$\classicalOutputMsmt{}'$ respectively. We say that $\msmt{}$ can {\em simulate}
$\msmt{}'$ if there exists a conditional distribution
$\pr{\classicalOutputMsmt{}'|\classicalOutputMsmt{}}$ such that
\begin{align}
\hspace{-\mathindent}\forall \src{} :  \pr{\classicalOutputMsmt{}'\!,\classicalOutputSrc{} |\msmt{}'\!,\src{}} =
  \sum_{\classicalOutputMsmt{}} \pr{\classicalOutputMsmt{}' |\classicalOutputMsmt{}}
  \pr{\classicalOutputMsmt{},\classicalOutputSrc{} |\msmt{},\src{}}\!.
\end{align}
Two measurements $\msmt{1}$ and $\msmt{2}$ are said to be {\em compatible} if both of them can
be simulated by some third measurement $\msmt{}$, that is, if there exists $\pr{\classicalOutputMsmt{1}|\classicalOutputMsmt{}}$ and $\pr{\classicalOutputMsmt{2}|\classicalOutputMsmt{}}$ such that
\begin{align}\label{compatibilitymsmts}\begin{split}
\hspace{-\mathindent}\forall \src{} :  \pr{\classicalOutputMsmt{1},\classicalOutputSrc{} |\msmt{1},\src{}} \!=
  \!\sum_{\classicalOutputMsmt{}} \pr{\classicalOutputMsmt{1} |\classicalOutputMsmt{}} \!
  \pr{\classicalOutputMsmt{},\classicalOutputSrc{} |\msmt{},\src{}},\\\hspace{-\mathindent}
  \forall \src{} :  \pr{\classicalOutputMsmt{2},\classicalOutputSrc{} |\msmt{2},\src{}} \!=
  \!\sum_{\classicalOutputMsmt{}} \pr{\classicalOutputMsmt{2} |\classicalOutputMsmt{}} \!
  \pr{\classicalOutputMsmt{},\classicalOutputSrc{} |\msmt{},\src{}}.
\end{split}\end{align}

Similar definitions hold for sources.  We say that a source $\src{}$ with classical outcome $\classicalOutputSrc{}$ {\em simulates} source
$\src{}'$ with classical outcome $\classicalOutputSrc{}'$ if there exists
a conditional distribution
$\pr{\classicalOutputSrc{}'|\classicalOutputSrc{}}$ such that
\begin{align}\label{compatibilitysrcs}
\hspace{-\mathindent}\forall \msmt{}:  \pr{\classicalOutputMsmt{},\classicalOutputSrc{}' |\msmt{},\src{}'}\! = \!
  \sum_{\classicalOutputSrc{}} \pr{\classicalOutputSrc{}'|\classicalOutputSrc{}} \!
  \pr{\classicalOutputMsmt{},\classicalOutputSrc{} |\msmt{},\src{}}.
\end{align}

When it comes to defining a notion of compatibility of sources, there is a nuance relative to the case of measurements.  The definition we adopt will apply
%concerning the notion of compatibility of sources that we make use of in this article: it applies
{\em only} to those pairs of sources, $\src{1}$ and $\src{2}$, that are operationally equivalent when one marginalizes over their outcomes, that is, it presumes that $\src{1}$ and $\src{2}$ are such that
\begin{align}\label{sameavgstate}
\hspace{-\mathindent} \forall \msmt{}: \sum_{\classicalOutputSrc{1}} \pr{\classicalOutputMsmt{},\classicalOutputSrc{1}|\msmt{},\src{1}} =
 \sum_{\classicalOutputSrc{2}}  \pr{\classicalOutputMsmt{},\classicalOutputSrc{2}|\msmt{},\src{2}}.
% \forall \msmt{}, \forall \alpha : \pr{\classicalOutputSrc{1}{\rm =}\alpha,\classicalOutputMsmt{}|\msmt{},\src{1}} =
%  \pr{\classicalOutputSrc{2}{\rm =}\alpha,\classicalOutputMsmt{}|\msmt{},\src{2}}.
\end{align}
Every set of sources we consider in this article will have this property.
Two such sources, $\src{1}$ and $\src{2}$, are said to be {\em compatible} if there exists
a third source $\src{}$ that simulates them both, that is, if there exists $\pr{\classicalOutputSrc{1}|\classicalOutputSrc{}}$ and $\pr{\classicalOutputSrc{2}|\classicalOutputSrc{}}$ such that
\begin{align}\label{compatibilitysources}\begin{split}
\hspace{-\mathindent}\forall \msmt{}:  \pr{\classicalOutputMsmt{},\classicalOutputSrc{1} |\msmt{},\src{1}} \!=\!
  \sum_{\classicalOutputSrc{}} \pr{\classicalOutputSrc{1}|\classicalOutputSrc{}} \!
  \pr{\classicalOutputMsmt{},\classicalOutputSrc{} |\msmt{},\src{}},
  \\\hspace{-\mathindent}
  \forall \msmt{}:  \pr{\classicalOutputMsmt{},\classicalOutputSrc{2} |\msmt{},\src{2}} \!=\!
  \sum_{\classicalOutputSrc{}} \pr{\classicalOutputSrc{2}|\classicalOutputSrc{}} \!
  \pr{\classicalOutputMsmt{},\classicalOutputSrc{} |\msmt{},\src{}}.
\end{split}\end{align}

It is instructive to consider what these notions of compatibility correspond to in quantum theory.

Recall that every measurement in quantum theory is represented by a positive operator-valued measure (POVM) whose elements are labelled by the outcome $m$, that is, by $\{ E_{m}\}_m$ where $\forall m : E_m \ge 0$ and $\sum_m E_m = \1$.

Every source in quantum theory is represented by an ensemble of subnormalized quantum states
 %that are labelled by the outcome $s$, that is, by
  $\{ p_s \rho_s \}_s$ where $\forall s : \rho_s \ge 0, {\rm Tr}(\rho_s) =1$ and $p_s$ is a probability distribution over $s$.
In other words, the ensemble $\{ p_s \rho_s \}_s$ defines the source that samples $s$ from the probability distribution $p_s$, then prepares the quantum system in the normalized state $\rho_s$ and outputs $s$ as its outcome.

If a measurement $\msmt{}$ is represented by the POVM $\{ E_{m}\}_m$ and a source $S$ is represented by the ensemble $\{ p_s \rho_s \}_s$, then the probability of the source generating outcome $s$ and the measurement generating outcome $m$ is given by the Born rule as $\pr{\classicalOutputMsmt{} ,\classicalOutputSrc{}| \msmt{},\src{}} = {\rm Tr} (E_{m} p_s \rho_{s} )$.

Substituting this Born rule expression into Eq.~\eqref{compatibilitymsmts}, we infer
%It follows
 that if two measurements, associated to POVMs $\{ E_{m_1} \}_{m_1}$ and $\{ E_{m_2} \}_{m_2}$, are compatible, then there exists a third POVM $\{ E_{m}\}_m$ and conditional distributions $\pr{\classicalOutputMsmt{1} |\classicalOutputMsmt{}} $ and $\pr{\classicalOutputMsmt{2} |\classicalOutputMsmt{}} $ such that
%\begin{align}
%\forall \rho : {\rm Tr} (\rho E_{m_1} )  =
%  \sum_{\classicalOutputMsmt{}} \pr{\classicalOutputMsmt{1} |\classicalOutputMsmt{}} {\rm Tr} (\rho E_{m} ),\nonumber\\
%\forall \rho :   {\rm Tr} (\rho E_{m_2} )  =
%  \sum_{\classicalOutputMsmt{}} \pr{\classicalOutputMsmt{2} |\classicalOutputMsmt{}}  {\rm Tr} (\rho E_{m} ).
%\end{align}
%which in turn implies that there must exists a third POVM $\{ E_{m}\}$ and conditional distributions $\pr{\classicalOutputMsmt{1} |\classicalOutputMsmt{}} $ and $\pr{\classicalOutputMsmt{2} |\classicalOutputMsmt{}} $ such that
\begin{align}\begin{split}\label{defncompmmts}
E_{m_1}  =
  \sum_{\classicalOutputMsmt{}} \pr{\classicalOutputMsmt{1} |\classicalOutputMsmt{}}  E_{m},\nonumber\\
 E_{m_2}  =
  \sum_{\classicalOutputMsmt{}} \pr{\classicalOutputMsmt{2} |\classicalOutputMsmt{}}  E_{m}.
\end{split}\end{align}

For two quantum sources, associated to ensembles $\{ p_{s_1} \rho_{s_1} \}_{s_1} $ and $\{ p_{s_2} \rho_{s_2} \}_{s_2}$, the property that ensures the applicability of our definition of compatibility, Eq.~\eqref{sameavgstate}, is that the two ensembles define the same average state\footnote{The specialization of our notion of compatibility to the quantum case allows us to clarify our motivation for restricting the scope of the notion to pairs of sources satisfying Eq.~\eqref{sameavgstate}: if we did not restrict the notion in this manner, then two sources could be incompatible simply by virtue of averaging to different states, while the component states in the two sources were all diagonal in the same basis. For the purposes of evaluating the possibility of a noncontextual model, one prefers to have a notion of compatibility wherein sources being incompatible guarantees that they are not jointly diagonalizable.}, $\sum_{s_1} p_{s_1} \rho_{s_1} = \sum_{s_2} p_{s_2} \rho_{s_2}$.
Substituting the Born rule expression into Eq.~\eqref{compatibilitysrcs}, we infer that if the two quantum sources are compatible,
%In a similar fashion, substituting the Born rule expression into Eq.~\eqref{compatibilitysrcs}, we infer that if two quantum sources, associated to ensembles $\{ p_{s_1} \rho_{s_1} \}_{s_1} $ and $\{ p_{s_2} \rho_{s_2} \}_{s_2} $, are compatible,
then there exists a third ensemble ${\{ p_{s} \rho_{s} \}_{s}}$ and conditional distributions ${\pr{\classicalOutputSrc{1}|\classicalOutputSrc{}}}$ and ${\pr{\classicalOutputSrc{2}|\classicalOutputSrc{}}}$ such that
\begin{align}\begin{split}
p_{s_1} \rho_{s_1}  =
  \sum_{\classicalOutputMsmt{}} \pr{\classicalOutputSrc{1} |\classicalOutputSrc{}}  p_s \rho_{s},\\
 p_{s_2} \rho_{s_2}   =
  \sum_{\classicalOutputMsmt{}} \pr{\classicalOutputSrc{2} |\classicalOutputSrc{}}  p_s \rho_{s}.
\end{split}\end{align}

Note that the notion of compatibility for quantum measurements that we have articulated above~\cite{busch1997operational}  concerns only their retrodictive aspect and makes no reference to how the quantum state of a system evolves as a result of the measurement.  In other words, it is sufficient to know which POVM is associated to the measurement, while the {\em instrument} that is associated to it, i.e., the set of update maps for each outcome, is irrelevant (indeed, it is not even required that there {\em be} an update map---the quantum system could be destroyed in the measurement process).   This contrasts with the notion of compatibility that is the focus of many other works seeking to devise experimental tests of noncontextuality~\cite{Kirchmair}, where two measurements on a system are deemed compatible if implementing them in one temporal order gives the same statistics as implementing them in the opposite temporal order.
%, namely, whether the temporal order in which a pair of measurements are implemented can be reversed without effect.
%The latter notion concerns measurements implemented in temporal succesion and whether the order in which they are implmented can be reversed without effect.
%, is one of joint simulatability, not commutativity. It is therefore makes no reference to the choice of update map for the measurement. The latter is irrelevant to compatibility.
%Note also that
The joint-simulatability notion of compatibility articulated above
%, unlike the commutativity notion,
pertains not just to sharp measurements (represented by projector-valued measures) but to all unsharp measurements as well (represented by positive operator-valued measures). In particular, it allows nontrivial compatibility relations among unsharp measurements that are associated to POVMs wherein the different elements of the POVM do not commute with one another, whereas such POVMs need not even compatible with themselves according to the temporal-reordering notion of compatibility.  The wide scope of applicability of the joint-simulatability notion of compatibility makes it particulary well-equipped to contend with experimental noise and imperfections in tests of noncontextuality.
% It is this applicability to noisy measurements that makes it indispensable for deriving noncontextuality inequalities that are robust to noise.

%Finally, note that this notion of compatibility serves to define a notion of operational equivalence that will be the relevant ones for our construction.  We can again illustrate this most easily by focusing on the case of quantum theory.

A given POVM defines an operational equivalence class of measurements insofar as it can be implemented in many different ways.  This is because any given POVM is generally compatible with many other POVMs which are not compatible with one another, and for each such compatible set of POVMs, there is a different experimental procedure, and hence a different concrete realization of the given POVM.  The compatible set of which the POVM is considered a part is therefore an example of a measurement context.  Note that if one particularizes the definition of compatibility to projector-valued measures, then the condition for compatibility becomes commutativity of the associated observables, and we recover the standard notion of a measurement context of an observable as the commuting set of observables of which it is considered a part.  The important point, however, is that in addition to recovering the standard notion of measurement context for sharp (i.e., projective) measurements, one has a notion of measurement context also for unsharp measurements.

In a similar fashion, a given ensemble of quantum states defines an operational equivalence class of sources because it too can be implemented in many different ways, depending on which compatible set of ensembles it is considered to be a member of.   The compatible set of ensembles of which it is a member constitutes the {\em source context}.
% Importantly, these examples hold for nonprojective POVMs, not just projective measurements.  It is conventional to refer to a quantum measurement as {\em sharp} when the POVM representing it consists entirely of projectors, so we can summarize our conclusion as: the notion of compatibility applies to both sharp and unsharp quantum measurements.
It will later be useful to distinguish between sharp and unsharp quantum sources, where sharp sources are those consisting entirely of states that are normalizd projectors.  Clearly, the notion of compatibility of sources that we have introduced applies equally well to sharp and unsharp quantum sources, just as the notion for measurements applies to the sharp and unsharp cases alike.

\subsection{Ontological concepts}\label{sec:Ontological}

\subsubsection{Ontological models}

As proposed in Ref.~\cite{Spekkens05}, generalized noncontextuality is a constraint on an ontological model of an operational theory.
An ontological model is an attempt to reproduce the predictions of the operational theory by
imagining that the correlations between the outcome of the source and that of the measurement are explained by the physical system that acts as a causal mediary between them.
%modelling the physical states of the system that is being subject to the experiment.
All of the physical attributes of the system at any given point in time is termed the {\em ontic state} of the system at that time. We shall
denote this by $\lambda$, and the space of all possible ontic states of the system will be denoted by $\Lambda$.

Consider the most general way of representing a measurement procedure $\msmt{}$ in an ontological model. The output $\classicalOutputMsmt{}$ might not be completely determined by the ontic state $\lambda$ of the system. Instead, specifying  $\msmt{}$ might only specify the conditional probability of obtaining output $\classicalOutputMsmt{}$ if the system is in the ontic state $\lambda$.  This could arise because of objective indeterminism or because the
outcome of the measurement depends not only on the input system but also on degrees of freedom of the
measurement apparatus.   We denote this conditional probability by $\xi\left(\classicalOutputMsmt{}|\lambda,\msmt{}\right)$ and refer to it as the {\em response function} associated to  $\msmt{}$.

Similarly, the most general way of representing a preparation procedure in an ontological model is to allow that the preparation does not uniquely fix the ontic state of the system, but rather that the ontic state might only be sampled probabilistically from a distribution that is specified by the preparation.  This implies that the most general way of representing a source $\src{}$ in an ontological model is as a joint distribution over its outome $\classicalOutputSrc{}$ and the ontic state $\lambda$ that it outputs,
%The source might not necessarily uniquely pin down the ontic state of
%the system. Instead, it might only specify the probability
$\mu(\lambda,\classicalOutputSrc{}|\src{})$.
% of the system being in an ontic state $\lambda$
%and obtaining output $\classicalOutputSrc{}$.

The purpose of an ontological model for an operational theory is to reproduce the statistics of that theory.  This occurs if the ontological model is such that
%For the ontological model to be consistent, we require that
for all sources $\src{}$ and measurements $\msmt{}$,
\begin{align}\label{eq:consistencyConditionSources}
  \pr{\classicalOutputMsmt{}, \classicalOutputSrc{}|\msmt{},\src{}} =
  \sum_{\lambda \in \Lambda}
  \xi\left( \classicalOutputMsmt{}|\lambda,\msmt{} \right)
  \mu(\lambda,\classicalOutputSrc{}|\src{})~.
\end{align}
%In other words, we require that it reproduce the statistics of the operational theory.

It is useful to express the connection between the ontological model and the operational theory in terms of expectation values as well.

The expectation value of outcome $m$ of measurement $\msmt{}$ for the ontic state
$\lambda$ is
\begin{align}\label{eq:onticExpectationX}
 \langle m \rangle_{\lambda,\msmt{}}
  \equiv
  \sum_{m} m  \;\xi\left(m|\lambda,\msmt{} \right).
\end{align}
The expectation value of outcome $\classicalOutputSrc{}$ of measurement $\src{}$ for the ontic state $\lambda$ is a {\em retrodictive} expectation value and so is a bit more subtle to express.  Consider a source $\src{}$ with classical outcome $\classicalOutputSrc{}$ that is associated with the joint distribution $\mu(\lambda,\classicalOutputSrc{}|\src{})$.  What probability ought one to assign to the outcome variable $\classicalOutputSrc{}$ having taken a particular value if one knows that the ontic state emitted by the source was $\lambda$?  The answer is given by a simple Bayesian inversion:
\begin{align}\label{Binversion}
%  \mu(\classicalOutputSrc{}|\lambda,\src{}) =
%  \frac{\mu(\classicalOutputSrc{},\lambda|\src{})}{\sum_{\classicalOutputSrc{}'} \mu(\classicalOutputSrc{}',\lambda|\src{})}.
  \mu(\classicalOutputSrc{}|\lambda,\src{}) =
  \frac{\mu(\classicalOutputSrc{},\lambda|\src{})}{\mu(\lambda|\src{})}
\shortintertext{where}\label{marginalization}
\mu(\lambda|\src{}) \equiv  \sum_{\classicalOutputSrc{}'} \mu(\classicalOutputSrc{}',\lambda|\src{}).
\end{align}
We can then use this conditional probability to define an expectation value for an outcome $\classicalOutputSrc{}$ of a source $\src{}$ given knowledge of the ontic state $\lambda$, as:
\begin{align}\label{eq:onticExpectationY}
 \langle \classicalOutputSrc{} \rangle_{\lambda,S} \equiv
  \sum_{\classicalOutputSrc{}} \classicalOutputSrc{}
  \mu(\classicalOutputSrc{}|\lambda,\src{})~,
\end{align}

It follows that the correlation between $\classicalOutputMsmt{}$ and $\classicalOutputSrc{}$ can be expressed as
\begin{align}\begin{split}\label{correlationasexpectationvalues}
  \opAvg{\classicalOutputMsmt{}\classicalOutputSrc{} }
  &=
\sum_{m,s} ms \pr{\classicalOutputMsmt{}, \classicalOutputSrc{}|\msmt{},\src{}} \\
    &=    \sum_{\lambda \in \Lambda}  \langle \classicalOutputMsmt{} \rangle_{\lambda,M}  \langle \classicalOutputSrc{} \rangle_{\lambda,S}\; \mu(\lambda |\src{}),
\end{split}\end{align}
where we have used Eqs.~(\ref{eq:consistencyConditionSources},~\ref{eq:onticExpectationX},~\ref{Binversion},~and~\ref{eq:onticExpectationY}).  We will use the latter expression when deriving our quantum no-go theorems and noncontextuality inequalities.

\subsubsection{Measurement noncontextuality}\label{MNC}

The assumption of measurement noncontextuality stipulates that if two measurements
$\msmt{1}$ and $\msmt{2}$ are operationally equivalent, then the response functions associated to these measurements in the ontological model are equal.  Equivalently, we can express measurement noncontextuality as the assumption that the response function associated to a measurement depends only on its operational equivalence class and not on the measurement context (which explains the appropriateness of the term ``noncontextual'').  Denoting the operational equivalence class of $\msmt{1}$ and $\msmt{2}$ by $\eqvClassMsmt{}$ (with outcome denoted by $\classicalOutputMsmt{}$), and denoting the response function for $\msmt{1}$ and $\msmt{2}$ and $\eqvClassMsmt{}$ by $ \xi\left(\classicalOutputMsmt{1} |\lambda, \msmt{1} \right)$, $\xi\left(\classicalOutputMsmt{2} |\lambda, \msmt{2} \right)$ and $\xi\left(\classicalOutputMsmt{}|\lambda, \msmt{} \right)$ respectively, the assumption of measurement noncontextuality can be formalized as follows:
%if the operational equivalence the corresponding response functions must also be equal
\begin{align}\label{eq:MNC}
%  \text{If } \msmt{1},\msmt{2} \in \eqvClassMsmt{},
%  %\forall \src{}:
% % \pr{\classicalOutputSrc{},\classicalOutputMsmt{}|\msmt{1},\src{}} &=
%  %\pr{\classicalOutputSrc{},\classicalOutputMsmt{}|\msmt{2},\src{}}
%   \nonumber\\
%%   \text{Then } \forall \lambda  \in \Lambda, \forall \alpha: \xi\left(\classicalOutputMsmt{1}=\alpha |\lambda, \msmt{1} \right) &=  \xi\left(\classicalOutputMsmt{2}=\alpha|\lambda, \msmt{2} \right)\nonumber\\  &\equiv \xi\left( \classicalOutputMsmtEqv{}=\alpha | \lambda, \eqvClassMsmt{} \right).
%   \text{Then } \forall \lambda  \in \Lambda:
%  \xi\left(\classicalOutputMsmt{1} |\lambda, \msmt{1} \right) &=
%  \xi\left(\classicalOutputMsmt{2}|\lambda, \msmt{2} \right)\nonumber\\
%  &\equiv \xi\left( \classicalOutputMsmtEqv{} | \lambda, \eqvClassMsmt{} \right).
%	&\nonumber\text{If }\msmt{1},\msmt{2} \in \eqvClassMsmt{}\text{, then}:
%	\\&\nonumber\forall\;\; {m\in \operatorname{ValuesRange}(\eqvClassMsmt{}),\; \lambda\in \Lambda}:
\begin{split}&\text{If }
  \msmt{1},\msmt{2} \in \eqvClassMsmt{}\text{, then}
\;\;\forall\; {\lambda\in \Lambda}:
\\&\quad \xi\left(\classicalOutputMsmt{1}|\lambda, \msmt{1} \right) =
  \xi\left(\classicalOutputMsmt{2}|\lambda, \msmt{2} \right)\equiv \xi\left( \classicalOutputMsmtEqv{} |\lambda, \eqvClassMsmt{} \right).
\end{split}\end{align}
%{\color{purple}[I've REPAIRED/reverted \cref{eq:MNC,eq:MNCexpectationvalues,eq:PNC,eq:PNC15,eq:PNC2}. \~EW]}

%Therefore it suffices to deal with a single response function $\xi\left(\classicalOutputMsmtEqv{}|\lambda, \eqvClassMsmt{} \right)$ associated with the equivalence class $\eqvClassMsmt{}$ whose classical outcome is $\classicalOutputMsmtEqv{}$.

%Denoting the expectation value of outcome $\classicalOutputMsmt{}$ of measurement $\eqvClassMsmt{}$ for the ontic state
%$\lambda$ by
%\begin{align}\label{eq:onticExpectationX}
% \langle \classicalOutputMsmtEqv{} \rangle_{\lambda,\eqvClassMsmt{}}
%  \equiv
%  \sum_{\classicalOutputMsmt{}} \classicalOutputMsmt{}
%  \xi\left(\classicalOutputMsmt{}|\lambda,\eqvClassMsmt{}\right),
%\end{align}

We can also express measurement noncontextuality in terms of expectation values, using Eq.~\eqref{eq:onticExpectationX}, as
\begin{align}\label{eq:MNCexpectationvalues}
%  \text{If } \msmt{1},\msmt{2} \in \eqvClassMsmt{},
%   \nonumber\\
%   \text{Then } \forall \lambda  \in \Lambda:
%   \langle\classicalOutputMsmt{1} \rangle_{\lambda, \msmt{1}}&=    \langle\classicalOutputMsmt{2} \rangle_{\lambda, \msmt{2}}\nonumber\\
%    &\equiv    \langle \classicalOutputMsmtEqv{} \rangle_{\lambda,\eqvClassMsmt{}}
%   %\onticAvg{\classicalOutputMsmt{1}} &= \onticAvg{\classicalOutputMsmt{2}}
%  % \equiv \onticAvg{\classicalOutputMsmtEqv{}}.
\begin{split}&\text{If }
  \msmt{1},\msmt{2} \in \eqvClassMsmt{}\text{, then}
\;\;\forall\; {\lambda\in \Lambda}:
\\&\qquad
 \langle \classicalOutputMsmt{1} \rangle_{\lambda,\msmt{1}} =
 \langle \classicalOutputMsmt{2} \rangle_{\lambda,\msmt{2}} \equiv
  \langle \classicalOutputMsmtEqv{} \rangle_{\lambda,\eqvClassMsmt{}}.
%   \onticAvg{\classicalOutputMsmt{1}} = \onticAvg{\classicalOutputMsmt{2}}  \equiv \onticAvg{\classicalOutputMsmtEqv{}}.
\end{split}\end{align}

%An assignment to an equivalence class of measurements $\mathcal{M}$ by ontic state $\lambda$ will be said to be {\em deterministic}
An ontic state $\lambda$ will be said to be make a {\em deterministic} assignment to an equivalence class of measurements $\mathcal{M}$ if
\begin{align}\label{OD}
\xi\left( \classicalOutputMsmtEqv{} | \lambda, \eqvClassMsmt{} \right) \in \{0,1\},
\end{align}
or equivalently, if $\langle \classicalOutputMsmtEqv{} \rangle_{\lambda,\eqvClassMsmt{}} \in \textrm{Range}( \classicalOutputMsmtEqv{})$.
%the ontic state $\lambda$ makes a deterministic measurement assignment if and only if the expectation value $\onticAvg{ \classicalOutputMsmtEqv{}}$ is equal to either the extremal minimal possible value or the extremal maximal possible value of $\classicalOutputMsmtEqv{}$.
%\begin{align}\onticAvg{\classicalOutputMsmt{}} \in \{+1,-1\}.\end{align}

We pause to note the how the standard notion of measurement context that appears in proofs of the Kochen-Specker theorem is understood in our framework.
%sort of application of measurement noncontetuality that arises in our no-go theorem for quantum theory.
Consider a measurement associated to an observable $\obs{}$.  If $\obs{}$ is compatible with $\obs{1}$ and $\obs{}$ is compatible with $\obs{2}$, but $\obs{1}$ and $\obs{2}$ are not compatible with one another, then there are two operationally equivalent ways of implementing a measurement of $\obs{}$, namely, by measuring it jointly with $\obs{1}$ and by measuring it jointly with $\obs{2}$.  If $ \langle\classicalOutputMsmt{1} \rangle_{\lambda, \obs{} (\obs{1})}$ denotes the expectation value for the outcome of the $\obs{}$ measurement when it is measured jointly with $\obs{1}$ and $ \langle\classicalOutputMsmt{2} \rangle_{\lambda,\obs{}( \obs{2})}$ denotes the expectation value for the outcome of the $\obs{}$ measurement when it is measured jointly with $\obs{2}$, then measurement noncontextuality implies that $\forall \lambda  \in \Lambda:   \langle\classicalOutputMsmt{1} \rangle_{\lambda, \obs{}(\obs{1})}=    \langle\classicalOutputMsmt{2} \rangle_{\lambda, \obs{}(\obs{2})} \equiv    \langle \classicalOutputMsmtEqv{} \rangle_{\lambda,\obs{}}.$  The fact that the notion of measurement noncontextuality does not include the assumption of outcome determinism translates, in the case of quantum observables, to the fact that these expectation values are not assumed, a priori, to lie in the eigenspectrum of $\obs{}$.

The generalized notion of noncontextuality allows one to extend this analysis to the case of
%Finally, note that the same comments apply to
a quantum measurement that is associated to a POVM that is nonprojective.  If it can be measured jointly with either one of two other POVMs which are not compatible with one another, then the assumption of measurement noncontextuality implies that the expectation value of its outcome given the ontic state $\lambda$ should be independent of which of the two other POVMs it is measured jointly with.

%The expectation value assigned to the random variable $\classicalOutputMsmt{}$ by the ontic state
%$\lambda$ is
%\begin{align}\label{eq:onticExpectationX}
%  \onticAvg{\classicalOutputMsmt{}} =
%  \sum_{\classicalOutputMsmt{}} \classicalOutputMsmt{}
%  \xi\left(\classicalOutputMsmt{}|\lambda,\msmt{}\right)~.
%\end{align}
%We can express measurement noncontextuality in terms of such expectation values.
%\begin{align}\label{eq:MNC}
%  \text{If } \msmt{1},\msmt{2} \in \eqvClassMsmt{},
 %  \nonumber\\
 %  \text{Then } \forall \lambda  \in \Lambda:
 %  \onticAvg{\classicalOutputMsmt{1}} &= \onticAvg{\classicalOutputMsmt{2}} \nonumber \\
%   &\equiv \onticAvg{\classicalOutputMsmtEqv{}},
%\end{align}
%where $\onticAvg{\classicalOutputMsmtEqv{}}$ denotes the expectation value of the outcome $\classicalOutputMsmtEqv{}$ for ontic state $\lambda$.

\subsubsection{Preparation noncontextuality}

The assumption of preparation noncontextuality has previously been expressed in terms of individual preparations.  However, in this article we will be describing experiments in terms of sources, and so we will here express it in the language of sources.
% Because sources include preparations as a special case (when the outcome is a singleton set)

%, but we shall express it here in terms of sources.

The assumption of preparation noncontextuality stipulates that if two sources
$\src{1}$ and $\src{2}$ are operationally equivalent,  then the joint distributions over ontic states and outcomes that represent these sources in the ontological model are equal.  Equivalently, the joint distribution over ontic states and outcomes representing a source depends only on the operational equivalence class of that source.  Denoting the operational equivalence class of $\src{1}$ and $\src{2}$ by $\eqvClassSrc{}$ (with outcome denoted by $\classicalOutputSrcEqv{}$), and denoting the joint distribution over ontic states and outcomes for $\src{1}$ and $\src{2}$ and $\eqvClassSrc{}$ by $\mu\left(\classicalOutputSrc{1},\lambda| \src{1} \right)$, $\mu\left(\classicalOutputSrc{2},\lambda| \src{2} \right)$, and $\mu\left(\classicalOutputSrcEqv{},\lambda| \eqvClassSrc{} \right)$ respectively, the assumption of preparation noncontextuality can be formalized as follows:
\begin{align}\label{eq:PNC}\begin{split}
&\text{If }\src{1},\src{2} \in \eqvClassSrc{}\text{, then}\;\;\forall\;{\lambda\in \Lambda}:
\\&\qquad  \mu\left(\classicalOutputSrc{1},\lambda| \src{1} \right) =
  \mu\left(\classicalOutputSrc{2},\lambda| \src{2} \right)
  \equiv \mu\left( \classicalOutputSrcEqv{}, \lambda| \eqvClassSrc{} \right).
\end{split}\end{align}

Note that an individual preparation can be understood as a special kind of source, one wherein the outcome is trivial (i.e. taking a value in a singleton set), and for such sources the definition of noncontextuality provided above reduces to the standard one for preparations articulated in Ref.~\cite{Spekkens05}.

%Under the assumption of source noncontextuality, it suffices to deal with a single distribution $\mu\left(\classicalOutputSrcEqv{},\lambda| \eqvClassSrc{}\right)$ associated with the equivalence class $\eqvClassSrc{}$ with classical output $\classicalOutputSrcEqv{}$.

We can also express this assumption in terms of expectation values, as we did for the case of measurements, but with a critical difference, as noted above Eq.~\eqref{eq:onticExpectationY}: for sources, the relevant expectation values concern {\em retrodictions} rather than predictions.

If two sources, $\src{1}$ and $\src{2}$, are operationally equivalent then not only does preparation noncontextuality imply that the distributions $\mu\left(\classicalOutputSrc{1},\lambda| \src{1} \right)$ and $ \mu\left(\classicalOutputSrc{2},\lambda| \src{2} \right)$ are equal (and hence the marginals $\mu\left(\lambda| \src{1} \right)$ and $ \mu\left(\lambda| \src{2} \right)$ are equal as well),  it also implies, via Eq.~\eqref{Binversion}, that the conditional distributions $\mu\left(\classicalOutputSrc{1}|\lambda, \src{1} \right)$ and $ \mu\left(\classicalOutputSrc{2}|\lambda, \src{2} \right)$ are equal as well.
That is, applying Eq.~\eqref{Binversion} to Eq.~\eqref{eq:PNC}, we find that we can express the assumption of preparation noncontextuality as
\begin{align}\label{eq:PNC15}\begin{split}
&\text{If }\src{1},\src{2} \in \eqvClassSrc{}\text{, then}\;\;\forall\;{\lambda\in \Lambda}:
\\&\qquad \mu\left(s|\lambda, \src{1} \right) =
  \mu\left(s|\lambda, \src{2} \right)\equiv \mu\left( s |\lambda, \eqvClassSrc{} \right).
\end{split}\end{align}
or, translating this into expectation values using Eq.~\eqref{eq:onticExpectationY}, we can express it as
\begin{align}\label{eq:PNC2}\begin{split}
&\text{If }\src{1},\src{2} \in \eqvClassSrc{}\text{, then}\;\;\forall\;{\lambda\in \Lambda}:
\\&\qquad
 \langle \classicalOutputSrc{1} \rangle_{\lambda,\src{1}} =
 \langle \classicalOutputSrc{2} \rangle_{\lambda,\src{2}} \equiv
  \langle \classicalOutputSrcEqv{} \rangle_{\lambda,\eqvClassSrc{}}.
%   \onticAvg{\classicalOutputSrc{1}} = \onticAvg{\classicalOutputSrc{2}}
%   \equiv \onticAvg{\classicalOutputSrcEqv{}}.
\end{split}\end{align}
It is apparent, therefore, that the assumption of noncontextuality for sources is a kind of retrodictive analogue of the assumption of noncontextuality for measurements.

An ontic state $\lambda$ will be said to be make a {\em deterministic} assignment to an equivalence class of sources $\mathcal{S}$ if
\begin{align}
\mu\left( \classicalOutputSrcEqv{} | \lambda, \eqvClassSrc{} \right) \in \{0,1\},
\end{align}
or equivalently, if $ \langle \classicalOutputSrcEqv{} \rangle_{\lambda,\eqvClassSrc{}} \in  \textrm{Range}( \classicalOutputSrcEqv{} )$.
%the ontic state $\lambda$ makes a deterministic source assignment if and only if the expectation value $\onticAvg{ \classicalOutputSrcEqv{}}$ is equal to either the extremal minimal possible value or the extremal maximal possible value of $\classicalOutputSrcEqv{}$.

We again pause to illustrate these notions by specializing to the quantum case.
Consider a quantum source associated to the ensemble $\ens{} = \{ p_s \rho_s \}$.  Suppose that $\ens{}$ is compatible with an ensemble $\ens{1} = \{ p_{s_1} \rho_{s_1} \}$ and that it is also compatible with an ensemble  $\ens{2} = \{ p_{s_2} \rho_{s_2} \}$ but that $\ens{1}$ and $\ens{2}$ are not compatible with one another.   In this case, there are two operationally equivalent ways of implementing the source associated to $\ens{}$, namely, by implementing it jointly with $\ens{1}$ and by implementing it jointly with $\ens{2}$.  If $ \langle\classicalOutputSrc{1} \rangle_{\lambda, \ens{} (\ens{1})}$ denotes the expectation value for the outcome of the source $\ens{}$ when it is implemented jointly with $\ens{1}$ and $ \langle\classicalOutputSrc{2} \rangle_{\lambda,\ens{} (\ens{2})}$ denotes the expectation value for the outcome of the source $\ens{}$ when it is implemented jointly with $\ens{2}$, then preparation noncontextuality implies that $\forall \lambda  \in \Lambda:   \langle\classicalOutputSrc{1} \rangle_{\lambda, \ens{} (\ens{1}) }=    \langle\classicalOutputSrc{2} \rangle_{\lambda, \ens{} (\ens{2})} \equiv    \langle \classicalOutputSrcEqv{} \rangle_{\lambda,\ens{}}.$

\subsubsection{Universal noncontextuality}

An  operational theory
 will be said to admit of a universally noncontextual ontological
model if it admits of an ontological model that is noncontextual for all experimental procedures, and therefore for both the preparations and the measurements~\cite{Spekkens05}.

\section{Quantum no-go theorems based on the Peres-Mermin square}\label{NoGo}

KS-noncontextuality can be understood as the conjunction of the assumption of measurement noncontextuality defined above and the assumption that the ontic state assigns outcomes to projective measurements deterministically.  It is well known that quantum theory does not admit of a KS-noncontextual model.  The Peres-Mermin proof~\cite{Peres, Mermin} is particularly intuitive and has therefore become a paradigm example.

Ref.~\cite{Spekkens05} provided several reasons for adopting the notion of universal noncontextuality described in the previous section rather than KS-noncontextuality.
First of all, the reference to projective measurements in the definition of KS-noncontextuality makes it clear that the notion of KS-noncontextuality can only be applied to quantum theory and leaves open the question of how to apply it to other operational theories or to experimental data.  Furthermore, it has been argued in Refs.~\cite{Spekkens05,Spekkens14} that the notion of universal noncontextuality stands to the notion of KS-noncontextuality as the notion of local causality stands to that of local determinism (defined by Bell in Ref.~\cite{Bell2}).
The problem with both KS-noncontextuality and local determinism is that in the face of a contradiction, one can always salvage the spirit of noncontextuality or locality by simply abandoning determinism.  Such an option is not available if one derives a contradiction from universal noncontextuality or local causality.

It was shown in Sec. VIII. of Ref.~\cite{Spekkens05} that one can turn {\em any} no-go theorem for a KS-noncontextual model of quantum theory into a no-go theorem for a universally noncontextual model of quantum theory.\footnote{The proof of this result relies on preparation noncontextuality, together with two facts about quantum theory: (i) for every observable, there is a basis of quantum states each element of which makes its outcome perfectly predictable and (ii) the uniform mixture of any basis of quantum states is operationally equivalent to the uniform mixture of any other such basis. }
In this section, we carry out this translation for the proof based on the Peres-Mermin square.
This will serve to clarify the constrast between KS-noncontextuality and universal noncontextuality.  However, the main purpose of this section is to provide  the reader with some intuition about how the contradiction arises, so that she may better follow our technique for deriving noncontextuality inequalities from the Peres-Mermin construction.

\subsection{No-go theorem for KS-noncontextuality based on the Peres-Mermin square}\label{sec:NoGoKS}

The Peres-Mermin magic square construction \cite{Peres, Mermin}
consists of nine observables, each defined on two qubits and each expressible as a product of Pauli operators.  Denoting the four Pauli operators by:
\begin{align}\begin{split}
  \1 = \left(
  \begin{matrix}
    1&0\\
    0  &1\\
  \end{matrix}
  \right), \quad \quad
  &X = \left(
  \begin{matrix}
    0&1\\
    1  &0\\
  \end{matrix}
  \right),\\
  Y = \left(
  \begin{matrix}
    0&i\\
    -i  &0\\
  \end{matrix}
  \right), \quad\quad
  &Z = \left(
  \begin{matrix}
    1&0\\
    0  &-1\\
  \end{matrix}
  \right),
\end{split}\end{align}
the nine observables relevant to the construction are as follows:
  \begin{align}\label{fig:PMSquare}
   \begin{matrix}
        X \otimes \1 \;\;\;\;&  \1 \otimes X  \;\;\;\;\;& X \otimes X\\ \\
 \1 \otimes Z  \;\;\;\;\;& Z \otimes \1  \;\;\;\;\;& Z \otimes Z\\ \\
 X \otimes Z  \;\;\;\;\;&  Z \otimes X  \;\;\;\;\;& Y \otimes Y\\
 \end{matrix}
 \end{align}
These are organized into a $3\times 3$ grid (the ``square'') to visually represent their commutativity properties: the three observables on any row or column of the square commute and therefore are jointly measurable.

%six triples of observables on two qubits, organized into the
%table shown in fig. \ref{fig:PMSquare} where $\1$, $X$, $Y$ and
%$Z$ are the Pauli matrices defined as

%\begin{figure}[H]
 % \centering
 % \begin{align*}
  % \begin{matrix}
  %      X \otimes \1&  \1 \otimes X  & X \otimes X\\
% \1 \otimes Z  & Z \otimes \1  & Z \otimes Z\\
% X \otimes Z  &  Z \otimes X  & Y \otimes Y\\
% \end{matrix}
% \end{align*}
 % \caption{Observables of the Peres-Mermin Square}
 % \label{fig:PMSquare}
%\end{figure}

KS-noncontextuality implies that every observable in the square is assigned a value deterministically by the ontic state $\lambda$ and independently of whether that observable is measured together with the other observables in its row or whether it is measured together with the other observables in its column.
The deterministic value assigned to observable $O$ by the ontic state $\lambda$ we denote by $\lfloor O\rfloor_{\lambda}$.
Because an observable is only ever found to take values from the eigenspectrum of the associated operator, it follows that the deterministic assignments to $O$ by $\lambda$ can only take values in this set,
\begin{align}\label{spectrum}
\lfloor O\rfloor_{\lambda} \in {\rm spec}(O).
\end{align}

% Therefore, $\langle X \otimes \1 \rangle_{\lambda} \in \{ -1,+1\}$, $\langle  \1 \otimes X \rangle_{\lambda} \in \{ -1,+1\}$, etcetera.

Finally, for any set of observables that can be jointly measured, the functional relations that hold among the observables in the set must also hold among the values assigned to them by the ontic state.  This follows from the fact that if a given functional relation {\em failed} to hold for the values assigned by the ontic state, then the ontological model would predict that it failed to hold for the values obtained in a joint measurement.   In the Peres-Mermin square, one can show that the product of the observables along each of the rows and along each of the first two columns is $\1 \otimes \1$, while the product of the observables
on the last column is $-\1 \otimes \1$.  Therefore, the functional relations are
\begin{subequations}\begin{align}
&(X \otimes \1) (\1 \otimes X)( X \otimes X)   &=& \hphantom{-}\,\,\1\otimes \1,\\
&( \1 \otimes Z )( Z \otimes \1)( Z \otimes Z) &=& \hphantom{-}\,\,\1 \otimes \1,\\
&(X \otimes Z )( Z \otimes X ) ( Y \otimes Y)  &=& \hphantom{-}\,\,\1\otimes \1,\\
&(X \otimes \1)(\1 \otimes Z )(X \otimes Z )   &=& \hphantom{-}\,\,\1\otimes \1,\\
 &(\1 \otimes X)(Z \otimes \1)( Z \otimes X )  &=& \hphantom{-}\,\,\1\otimes \1,\\
 &( X \otimes X)(Z \otimes Z)(Y \otimes Y)     &=& -\1\otimes \1.
\end{align}\end{subequations}

Together with the fact, inferred from Eq.~\eqref{spectrum}, that
\begin{align}\label{identityrep}\begin{split}
\lfloor \1 \otimes \1 \rfloor_{\lambda} &= +1\\
\lfloor -\1 \otimes \1 \rfloor_{\lambda} &= -1,
\end{split}\end{align}
we conclude that the functional relations holding among the deterministic assignments to the nine observables in the Peres-Mermin square are
%and Eq.~\eqref{identityrep}, we obtain the constraints
%Consequently, we obtain constraints
\begin{subequations}\begin{align}\label{PM1}
&\lfloor X \otimes \1\rfloor_{\lambda} \lfloor \1 \otimes X\rfloor_{\lambda} \lfloor X \otimes X\rfloor_{\lambda} &=&+1,&\\
&\lfloor \1 \otimes Z \rfloor_{\lambda} \lfloor Z \otimes \1\rfloor_{\lambda}\lfloor Z \otimes Z\rfloor_{\lambda} &=&+1,&\\
&\lfloor X \otimes Z \rfloor_{\lambda} \lfloor Z \otimes X \rfloor_{\lambda} \lfloor Y \otimes Y\rfloor_{\lambda}&=&+1,&\\
&\lfloor X \otimes \1\rfloor_{\lambda}\lfloor \1 \otimes Z \rfloor_{\lambda}\lfloor X \otimes Z \rfloor_{\lambda}&=&+1,&\\
&\lfloor \1 \otimes X\rfloor_{\lambda} \lfloor Z \otimes \1\rfloor_{\lambda}  \lfloor Z \otimes X \rfloor_{\lambda} &=& +1,&\\
&\lfloor X \otimes X\rfloor_{\lambda} \lfloor Z \otimes Z\rfloor_{\lambda} \lfloor Y \otimes Y\rfloor_{\lambda}&=&-1.&\label{PM6}
\end{align}\end{subequations}
% such as $[X \otimes \1]_{\lambda} [\1 \otimes X]_{\lambda} [ X \otimes X]_{\lambda} =1$, and similarly for the other rows and columns.
%But there is no assignment of values to the nine observables that satisfies all six of these constraints.

By Eq.~\eqref{spectrum}, we also have that, for all $O,O' \in \{X,Y,Z\}$,
%$i,j\in \{1,2,3\}$,
\begin{align} \label{Paulirep}\begin{split}
\lfloor O \otimes \1\rfloor_{\lambda} \,\,\,&\in \{ -1,+1\},\\
\lfloor \1 \otimes O'\rfloor_{\lambda} \,&\in \{ -1,+1\},\\
\lfloor O \otimes O'\rfloor_{\lambda} &\in \{ -1,+1\}.
\end{split}\end{align}
%while
%\begin{align}\label{identityrep}
%[ \1 \otimes \1 ]_{\lambda} = 1\nonumber\\
%[ -\1 \otimes \1 ]_{\lambda} = -1.
%\end{align}
%From these constraints, we obtain
%Finally, one obtains a contradiction by noting that
However, given this constraint, the set of equations (\ref{PM1}-\ref{PM6}) has no solution. To see this, it suffices to note that the product of the left-hand sides of the six equations is +1 (because every term appears squared in this product), while the product of the right-hand sides of the six equations is -1.  We have thereby arrived at a contradiction.
%But famously these six constraints cannot all be satisfied because the products of the nine values of the observables, if multiplied row-wise must yield 1, while if multiplied column-wise must yield -1.

%\color{red} [Mention where the terminology of ``magic square'' comes from] \color{black}

\subsection{No-go theorem for universal noncontextuality based on the Peres-Mermin square}\label{NoGoUniversal}

As noted in Section~\ref{MNC}, unlike KS-noncontextuality, measurement noncontextuality allows for measurement outcomes to be assigned {\em indeterministically} by the ontic state.  Obtaining the contradiction in the previous section relied critically on this assumption of deterministic assignments.
%he assumption of determinism was critical in obtaining the contradiction derived in the previous section,
%This implies that the sort of contradiction from the previous section can no longer be derived.
Indeed, once one allows indeterministic assignments, one finds that there {\em are}, in fact, many noncontextual assignments to the observables in the Peres-Mermin square.  For instance, every quantum state defines such an assignment through the Born rule, and there are other valid assignments as well which do not arise from the Born rule (but could arise in some putative post-quantum theory).  At first glance, therefore, it may seem that by replacing the notion of KS-noncontextuality with the generalized notion of noncontextuality proposed in Ref.~\cite{Spekkens05}, one has lost the possibility of deriving a contradiction.
%the notion of noncontextuality has been weakened so much that it no longer contradicts with quantum theory.
However, although the generalized notion of noncontextuality does indeed weaken the constraints on how one represents {\em measurements} in a noncontextual ontological model,  it also introduces a novel constraint on how one represents {\em preparations}.  By availing oneself of the assumption of preparation noncontextuality, one can again derive a no-go theorem for a noncontextual model of quantum theory.

%\color{red}
%The standard Peres-Mermin no-go theorem proves the impossibility of a universally noncontextual model if one makes use of the result from my PRA and notes the perfect predictability of the measurements appearing therein for various ensembles that all average to the completely mixed state.  This is a special case of the connection noted in my PRA.  However, rather than going over the argument in its generic form, it is more pedagogically useful to see an explicit contradiction derived in the language of sources and measurements.  This is what we do now.
%\color{black}

%\color{red} These relations imply that when the measurements in a row are jointly implemented, the outcomes are related by...  Similarly for the measurements in a given column.

%We can also associate sources with these observables and then the relations imply ...
\color{black}

Each observable on a two-level quantum system represents an operational equivalence class of binary-outcome measurements.  We denote the observable in position $i,j$ of the Peres-Mermin square by $\obs{ij}$.
%and the associated equivalence class of binary-outcome measurements by $\eqvClassMsmt{ij}$. \color{red} [Avoid reference to equivalence classes here?] \color{black}
%, which we denote by $\eqvClassMsmt{ij}$, and
%To each observable $\obs{ij}$ in position $i,j$ of the square, we associate a binary-outcome measurement, $\eqvClassMsmt{ij}$,
%which
The projector-valued measure associated to the observable $\obs{ij}$ consists of the pair of orthogonal rank-2 projectors:
%The class $\eqvClassMsmt{ij}$ is represented by
%The resolution of identity into the pair of orthogonal projectors
\begin{align}\label{projmeetsobs}\begin{split}
  \Pi_{ij}^{+} &= \frac{1}{2}\left(\1 \otimes \1 + \obs{ij} \right) \\
  \Pi_{ij}^{-} &= \frac{1}{2}\left(\1 \otimes \1 - \obs{ij} \right),
\end{split}\end{align}
which correspond respectively to the $+1$ and $-1$ eigenspaces of $\obs{ij}$.

We also define an equivalence class of binary-outcome quantum sources for each of the observables as follows.
For each observable $\obs{ij}$, we consider the quantum source associated to the 2-element ensemble
\begin{align*}
\ens{ij} \equiv \{ \frac{1}{2} \rho_{ij}^{+}, \frac{1}{2} \rho_{ij}^{-} \},\end{align*} where
\begin{align}\label{statemeetsobs}\begin{split}
  \rho_{ij}^{+} &=  \frac{1}{2} \Pi_{ij}^{+} = \frac{1}{4}\left(\1 \otimes \1 + \obs{ij} \right) \\
  \rho_{ij}^{-} &= \frac{1}{2} \Pi_{ij}^{-} =  \frac{1}{4}\left(\1 \otimes \1 - \obs{ij} \right),
\end{split}\end{align}
are normalized density operators. Note that each of these quantum sources defines the same average state, namely, the  completely mixed state $\frac{1}{2}\1$.

%For observable $\obs{ij}$, we consider the quantum source that samples a
%quantum state from the ensemble of eigenstates of $\obs{ij}$, that is, from
%\begin{align}\label{statemeetsobs}
%  \rho_{ij}^{+} &= \frac{1}{4}\left(\1 \otimes \1 + \obs{ij} \right) \\
%  \rho_{ij}^{-} &= \frac{1}{4}\left(\1 \otimes \1 - \obs{ij} \right),
%\end{align}
%with sampling probability $1/2$ for each; it outputs the system prepared in this state and it outputs a classical binary variable specifying the state as its outcome.  We denote this ensemble by $$\ens{ij} \equiv \{ (\frac{1}{2},\rho_{ij}^{+}), (\frac{1}{2},\rho_{ij}^{-})\}.$$ \color{black}

%Recall from Sec. ??? that one has a notion of compatibility for quantum ensembles just as one does for quantum measurements.
If we arrange these nine quantum sources into a square, then they are compatible along the rows and the columns.
For example, in the first row of the square,
%the first source corresponds to the uniform ensemble of normalized projectors onto the eigenspaces of $\1 \otimes X$, the second source corresponds to the uniform ensemble of normalized projectors onto the eigenspaces of $\1 \otimes X$,
the three sources are seen to be compatible by virtue of the fact that they can all be obtained by post-processing of the outcome of a single 4-outcome source, namely, the one associated to the uniform ensemble of joint eigenstates of the set of commuting observables associated to that row.  The other rows and columns are  analogous.
We will speak of the {\em source version} of the Peres-Mermin square to refer to the compatibility relations among the sources.
%correspond to the uniform ensembles associated to the observable $\1 \otimes X, ...$, and these are compatible because they can all be simulated by the source corresponding to the uniform ensemble associated to eigenstates
%, with the same compatibility relations as for the measurements.

%\color{red} [Use observable and ensemble language in what follows, and reserve the measurement and source language for the next section.] \color{black}

%For $\alpha,\beta \in \{+1,-1\}$, it can be shown that
Given these definitions, it is clear that when the measurement associated to the observable $\obs{ij}$ is implemented on the source associated to the ensemble $\ens{ij}$ (i.e., the source at the same location in the square), the outcome $\classicalOutputMsmtEqv{}$ of the measurement is perfectly correlated with the outcome $\classicalOutputSrcEqv{}$ of the source and the marginal distribution over either outcome is uniform:
\begin{align}\label{perfcorrpr}
&\hspace{-\mathindent}\forall i,j: pr(\classicalOutputMsmtEqv{},\classicalOutputSrcEqv{}| \obs{ij}, \ens{ij}) =  \frac{1}{2} \tr{\Pi_{ij}^{\classicalOutputMsmtEqv{}}\rho_{ij}^{\classicalOutputSrcEqv{}}} = \frac{1}{2} \delta_{\classicalOutputMsmtEqv{},\classicalOutputSrcEqv{}}\!.
\shortintertext{This can be expressed equivalently  as}
%In terms of the correlator, this perfect correlation is expressed as
\label{perfcorrexp}
%\forall i,j: \langle \classicalOutputMsmtEqv{} \classicalOutputSrcEqv{} \rangle_{\eqvClassMsmt{ij},\eqvClassSrc{ij}} =  1.
&\forall i,j: \langle \classicalOutputMsmtEqv{} \classicalOutputSrcEqv{} \rangle_{\obs{ij}, \ens{ij}} =  1.
\end{align}

Now consider what this implies for any putative noncontextual ontological model of the experiment.
%\color{red} [Put assumptions of noncontextuality for these quantum measurements and sources up front and refer back to where we've described them?] \color{black}
  Denoting the expectation value
%of $\obs{ij}$
for the outcome $\classicalOutputMsmtEqv{}$ of the measurement of observable $\obs{ij}$
 given ontic state $\lambda$ by $\langle \classicalOutputMsmtEqv{} \rangle_{\lambda,\obs{ij}}$, and the expectation value for the outcome $\classicalOutputSrcEqv{}$ of the source associated to the ensemble $\ens{ij}$ given ontic state $\lambda$ by $\langle \classicalOutputSrcEqv{} \rangle_{\lambda, \ens{ij}}$ (note that the latter is a retrodictive expectation), then
%we have
%\begin{align}\label{correlatorPM}
 % \langle \classicalOutputMsmtEqv{} \classicalOutputSrcEqv{}\rangle_{\eqvClassMsmt{ij},\eqvClassSrc{ij}}
 % &=  \sum_{\classicalOutputMsmtEqv{}, \classicalOutputSrcEqv{}}   \classicalOutputMsmtEqv{}, \classicalOutputSrcEqv{}
%{\rm pr}(\classicalOutputMsmtEqv{},\classicalOutputSrcEqv{}| \eqvClassMsmt{ij},\eqvClassSrc{ij})  \nonumber\\
%  &=  \sum_{\classicalOutputMsmtEqv{}, \classicalOutputSrcEqv{}}   \classicalOutputMsmtEqv{}, \classicalOutputSrcEqv{}
%... \nonumber\\
%&=
%  \sum_{\lambda \in \Lambda} \left(
% \sum_{\classicalOutputMsmtEqv{}, \classicalOutputSrcEqv{}}  \langle \classicalOutputMsmtEqv{} \rangle_{\lambda,\eqvClassMsmt{ij}} \langle \classicalOutputSrcEqv{} \rangle_{\lambda,\eqvClassSrc{ij}} \right) \left( \sum_{\classicalOutputSrcEqv{}'} \mu(\lambda, \classicalOutputSrcEqv{}' |\eqvClassSrc{ij})\right).
%\end{align}
given Eq.~\eqref{correlationasexpectationvalues}, we have
\begin{align}\label{correlatorPM}
  \langle \classicalOutputMsmtEqv{} \classicalOutputSrcEqv{}\rangle_{\obs{ij},\ens{ij}}  =
  \sum_{\lambda \in \Lambda}
   \langle \classicalOutputMsmtEqv{} \rangle_{\lambda,\obs{ij}} \langle \classicalOutputSrcEqv{} \rangle_{\lambda,\ens{ij}}
    \mu(\lambda |\ens{ij})
   %\left( \sum_{\classicalOutputSrcEqv{}'} \mu(\lambda, \classicalOutputSrcEqv{}' |\eqvClassSrc{ij})\right).
\end{align}

An assumption of measurement noncontextuality has been made at this stage because we have assumed that the expectation value $ \langle \classicalOutputMsmtEqv{} \rangle_{\lambda,\obs{ij}}$  depends only on $\obs{ij}$ and not
%depends only on the equivalence class  $\eqvClassMsmt{ij}$ and not on which particular measurement procedure from that equivalence class was implemented, in particular, the assignment to a given observable is presumed to
on what other observables were measured together with it, that is, we have assumed that this expectation value is independent of whether we measure $\obs{ij}$ with the other observables in the same row of the Peres-Mermin square or with the other observables in the same column.
%\color{red} [Include the quantum realization business here?] \color{black}
%The fact that $ \langle \classicalOutputMsmtEqv{} \rangle_{\lambda,\eqvClassMsmt{ij}}$ depends only on the equivalence class  $\eqvClassMsmt{ij}$ and not on which particular measurement procedure from that equivalence class was implemented.
%identity of the binary-outcome measurement $\eqvClassMsmt{ij}$ and not on what other measurements are implemented jointly with it is an instance of measurement noncontextuality.
Similarly, an assumption of preparation noncontextuality has been made at this stage
%already been applied here
because we have assumed that $ \langle \classicalOutputSrcEqv{} \rangle_{\lambda,\ens{ij}}$ depends only on $\ens{ij}$ and not on which set of compatible ensembles are implemented jointly with it, those on the same row of the source version of the Peres-Mermin square or those on the same column.

%depends only on the equivalence class $\eqvClassSrc{ij}$ and not on which particular source from that class was implemented.
%\color{red} [use ensemble language, i.e., which fine-graining of a given ensemble was used]
%  In particular, for a given observable in the Peres-Mermin square, one can implement the binary-outcome source associated to it either by coarse-graining the four-outcome source associated to the joint eigenstates of the observables in its column or by coarse-graining the four-outcome source associated to the joint eigenstates of the observables in its row.   Preparation noncontextuality dictates that our expectation value for the outcome is independent of this choice of implementation.
 %  \color{black}

%The fact that $ \langle \classicalOutputSrcEqv{} \rangle_{\lambda,\eqvClassSrc{ij}}$ depends only on the identity of the binary-outcome source $\eqvClassSrc{ij}$ and not on what other sources are implemented jointly with it is an instance of preparation noncontextuality.

Finally, the assumption of preparation noncontextuality has an additional consequence that Eq.~\eqref{correlatorPM} does not yet fully incorporate.
%There is, however, an additional consequence of the assumption of preparation noncontextuality that Eq.~\eqref{correlatorPM} does not yet incorporate.
For each of the nine binary-outcome quantum sources, if one marginalizes over its oucome, one obtains the source that
  simply prepares $\frac{1}{2}\1$, the average state associated to the ensemble.  Recall that no quantum measurement can distinguish the different ensembles by which  the completely mixed state might have been prepared.  Therefore, for any given pair of quantum sources, $\ens{ij}$ and $\ens{i'j'}$, the pair of quantum sources one obtains by marginalizing over their outcomes are operationally equivalent.
% and associated to the preparation of $\frac{1}{2}\1$.
%We denote the equivalence class of the marginalization of all nine sources by $\eqvClassSrc{}^*$.

%\color{red} [avoid giving eq cladss a name.  jus call the distribution $\nu$.  then again...]\color{black}

%There is one additional consequence of the assumption of preparation noncontextuality.  Suppose one implements the source $\eqvClassSrc{ij}$ and then one marginalizes over its outcome $\classicalOutputSrcEqv{}$.  Call the resulting source, whose outcome is a singleton set, $\eqvClassSrc{ij}^{*}$.

If $\mu(\lambda, \classicalOutputSrcEqv{}| \ens{ij})$ is the representation in the ontological model of the quantum source $\ens{ij}$, then the quantum source that one obtains by marginalizing over its outcome $\classicalOutputSrcEqv{}$ is represented in the ontological model by
\begin{align}
\mu(\lambda |\ens{ij}) \equiv \sum_{\classicalOutputSrcEqv{ij}} \mu(\lambda, \classicalOutputSrcEqv{ij}|\ens{ij}).
\end{align}
%\color{blue} Note that the quantum state obtained by averaging the elements of each of the $\ens{ij}$ is always the completely mixed state, $\frac{1}{2}\1$, independent of the values of $i$ and $j$.  The outcome-marginalized versions of these quantum sources are therefore operationally equivalent. \color{black} %
Applying the assumption of preparation noncontextuality, Eq.~\eqref{eq:PNC}, to the operational equivalence of $\ens{ij}$ and $\ens{i'j'},$ we obtain
\begin{align}\begin{split}
%\forall (i,j),(i',j') : \sum_{\classicalOutputSrcEqv{ij}} \mu(\lambda, \classicalOutputSrcEqv{ij}|\eqvClassSrc{ij})
%&= \sum_{\classicalOutputSrcEqv{i'j'}} \mu(\lambda, \classicalOutputSrcEqv{i'j'}|\eqvClassSrc{i'j'})\nonumber\\
\hspace{-\mathindent}\forall (ij),(i'j') : \mu(\lambda|\ens{ij})
&=  \mu(\lambda|\ens{i'j'})
\equiv \mu\left( \lambda| \tfrac{1}{2}\1 \right).
\end{split}\end{align}

Substituting this into Eq.~\eqref{correlatorPM},  we finally obtain
\begin{align}\label{correlatorNC}
  \langle \classicalOutputMsmtEqv{} \classicalOutputSrcEqv{}\rangle_{\obs{ij},\ens{ij}}  =
\sum_{\lambda \in \Lambda}
 \langle \classicalOutputMsmtEqv{} \rangle_{\lambda,\obs{ij}} \langle \classicalOutputSrcEqv{} \rangle_{\lambda,\ens{ij}}  \mu(\lambda|\tfrac{1}{2}\1).
\end{align}

Now we are in a position to derive a contradiction.  The only way to reproduce the perfect correlations of Eq.~\eqref{perfcorrexp} is if for all $\lambda$ in the support of $ \mu(\lambda|\tfrac{1}{2}\1)$ and for all $i,j$,
\begin{align}\begin{split}
 &\langle \classicalOutputMsmtEqv{} \rangle_{\lambda,\obs{ij}}, \langle \classicalOutputSrcEqv{} \rangle_{\lambda,\ens{ij}} \in \{ +1,-1\},\\&
 \text{and}\quad \langle \classicalOutputMsmtEqv{} \rangle_{\lambda,\obs{ij}} =\langle \classicalOutputSrcEqv{} \rangle_{\lambda,\ens{ij}}.
 \end{split}\end{align}
In other words, every ontic state in the support of $\mu(\lambda|\tfrac{1}{2}\1)$ must assign {\em perfectly correlated outcomes} to the source and measurement when these are associated to the same observable, and the only way to achieve this is if it assigns these outcomes  {\em deterministically}.
However, any deterministic assignment to all of the measurements in the Peres-Mermin square must satisfy the functional relationships that hold among the outcomes of the compatible subsets of those measurements, that is, it must satisfy Eqs.~(\ref{PM1}-\ref{PM6}).  But following the standard argument (reviewed in the previous section), there are no such deterministic assignments, so we have arrived at our contradiction\footnote{Note that the contradiction could have been obtained equally well by considering the impossibility of finding deterministic assignments to all of the sources while respecting the functional relations that hold among compatible subsets of these.}.

%In a quantum language, the sorts of correlations we will be interested in are those that arise between preparations of the eigenstates of $O_{ij}$ and the outcomes of the measurements of the observable $O_{ij}$.

%[Prove a quantum no-go in the language of sources.  Start from observed perfect correlations at the nine loci.  Conclude that all of the ontic noncontextual expectation values need to be deterministic.  Then the standard contradiction can be derived, either from the nonexistence of a deterministic noncontextual assignment to the Peres-Mermin square of measurements or to the nonexistence of a {\em retrodictive} deterministic noncontextual assignment to the Peres-Mermin square of sources.  [drawback: it still some of the thunder of other articles on retrodictive proofs of Kochen-Specker]]

%The feature of the operational quantum predictions that the noncontextual ontological model fails to reproduce is the perfect correlation that exists between the outcome of the source and the outcome of the measurement, Eq.~\eqref{perfcorrexp}.  Therefore,
It follows that if one entertains the hypothesis that a given experiment {\em is}, in fact, described by a noncontextual ontological model, then one expects that for some subset of the nine source-measurement pairs, the correlations will be {\em imperfect}.  The noncontextuality inequalities that we derive for the Peres-Mermin scenario will capture the precise tradeoffs among the strengths of these nine correlations.

First, however, we must operationalize our description of the experiment, which is to say that we must purge it of any reference to the quantum formalism.

%We want to derive noncontextuality inequalities rather than a no-go theorem.  The idea is this: we can have noncontextual assignments, but they have to be indeterministic.  This then forces some of the nine correlations to be less than perfect.  Our inequalities will capture what are the precise tradeoffs among these nine correlations.

\section{From the quantum no-go theorem to noncontextuality inequalities}\label{sec:III}

\subsection{A purely operational description of the Peres-Mermin square}\label{sec:operationalPM}

%Our first objective is to cast this construction in an operational framework.

%\color{red} [We need to say it like this: \color{black}

%The operational description is as follows.

In Sec.~\ref{sec:Operational}, we defined operational equivalence relations and compatibility relations among experimental procedures in a manner that made reference only to experimental statistics, without appeal to the quantum formalism.  Here, we use these notions to express the relations must hold among a set of measurements and sources in an operational version of the quantum Peres-Mermin construction.  Any experiment satisfying all of these relations will be termed an {\em operational Peres-Mermin scenario}.

\begin{figure}[t!]
 \centering
{
 \includegraphics[scale=0.25]{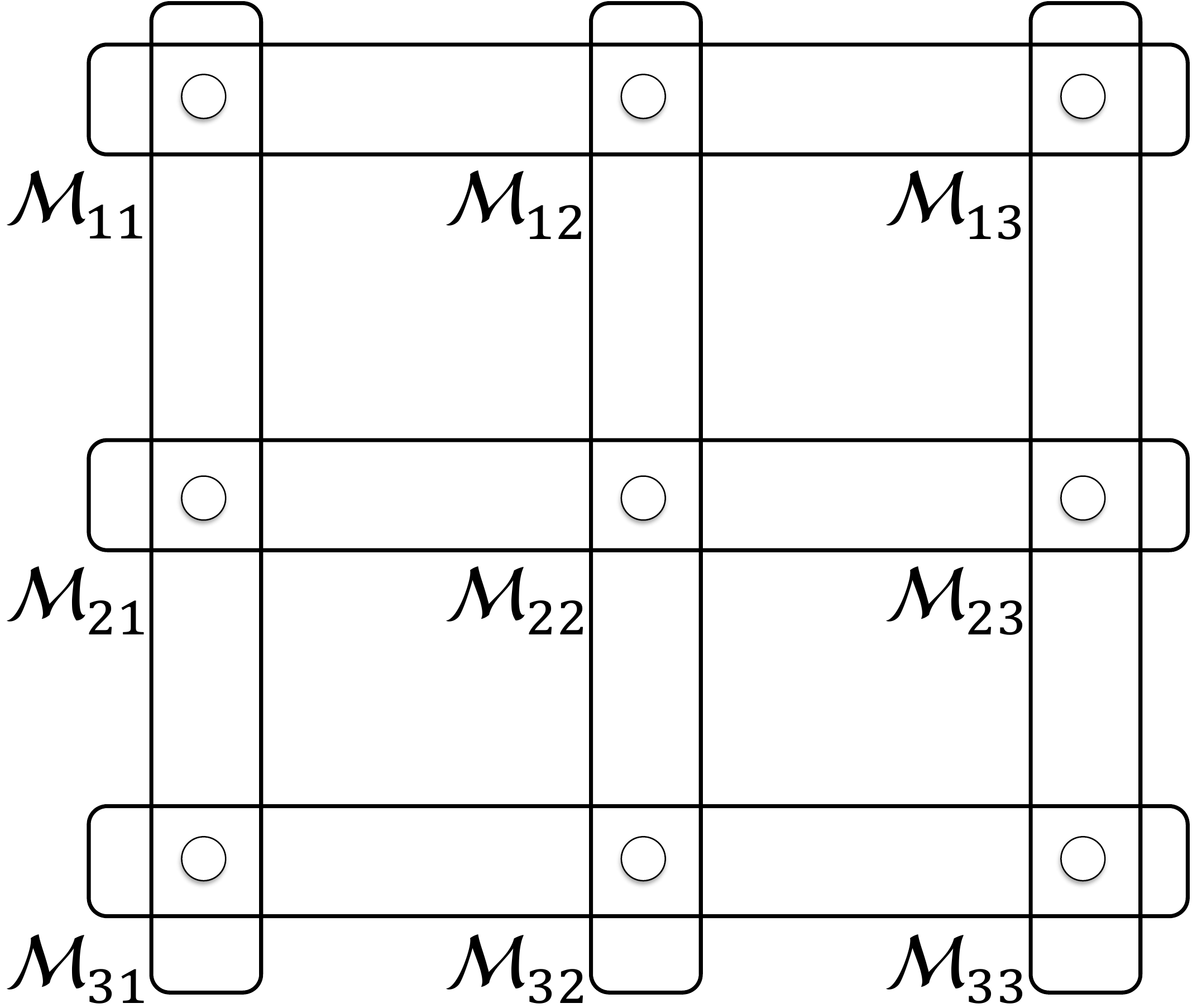}}
 \caption{The hypergraph representing compatibility relations among the nine binary-outcome measurements. }
 \label{CompatHypergraphMmts}
\end{figure}

We start with the measurements. There are 9 distinct equivalence classes of binary-outcome measurements, which we label by $\eqvClassMsmt{ij}$ where $i \in \{1,2,3\}$ and $j \in \{1,2,3\}$.  Laying these out in a $3\times 3$ square, where the measurement $\eqvClassMsmt{ij}$ appears at the $i$th row and $j$th column, each triple of measurements making up a row or a column of the square constitutes a compatible set of measurements.
%  compatibility relations that hold among these measurements can be described as follows: for each row and each column of the square, the triple of measurements are compatible.
This is depicted in the compatibility hypergraph of Fig.~\ref{CompatHypergraphMmts}.

By the definition of compatibility for measurements, Eq.~\eqref{compatibilitymsmts}, this implies that for every row and column there exists a measurement that simulates all the measurements on that row or column.  We denote the measurement that simulates the triple of measurements in row 1 by $\msmt{R_1}$, the one that simulates the triple in column 1 by $\msmt{C_1}$ and so forth.  We denote their outcomes by $m_{R_1}$, $m_{C_1}$, and so forth.

%the {\em particular} compatibility relation that holds among a set of measurements is specified by the conditional probabilities appearing in Eq.~\eqref{}.

% The particular compatibility relation for each row and column is defined as follows.  We consider the first row as an example.

We now turn to the nature of the {\em particular} relation that holds between the measurements on a given row or column and the measurement that simulates them.
Consider the measurements in the first row. The outcomes of the simulating measurement  $m_{R_1}$ is presumed to be 4-valued, such that it can
%The outcome of $\msmt{R_1}$ is presumed to be 4-valued.  It can therefore
be presented as an ordered pair of binary outcomes, which we denote by $m_{R_1,1}$ and $m_{R_1,2}$.  In terms of this notation, the three measurements in the first row are presumed to be obtained from the simulating measurement by the following identification of outcomes,
\begin{subequations}\label{cc}
\begin{align}
  \classicalOutputMsmtEqv{11} &= \classicalOutputMsmt{R_1,1}\\
  \classicalOutputMsmtEqv{12} &= \classicalOutputMsmt{R_1,2}\\
  \classicalOutputMsmtEqv{13} &= \classicalOutputMsmt{R_1,1}\cdot \classicalOutputMsmt{R_1,2}.
\end{align}
\end{subequations}
%This implies, in particular, that
%\begin{align}\label{eq:productRelation}
%  \classicalOutputMsmt{11}   \classicalOutputMsmt{12}  \classicalOutputMsmt{13} &= +1~.
%\end{align}
which in terms of the conditional probabilities in Eq.~\eqref{compatibilitymsmts} corresponds to the following post-processings of the simulating measurement:
\begin{subequations}
\begin{align}
{\rm pr}\{\classicalOutputMsmtEqv{11} |\classicalOutputMsmt{R_1} \} &=   \delta_{\classicalOutputMsmtEqv{11}, \classicalOutputMsmt{R_1,1}}\label{cc1}\\
{\rm pr}\{\classicalOutputMsmtEqv{12} |\classicalOutputMsmt{R_1} \} &=   \delta_{\classicalOutputMsmtEqv{12}, \classicalOutputMsmt{R_1,2}}\\
{\rm pr}\{\classicalOutputMsmtEqv{13} |\classicalOutputMsmt{R_1} \} &=   \delta_{\classicalOutputMsmtEqv{13},\classicalOutputMsmt{R_1,1}\cdot \classicalOutputMsmt{R_1,2}}.\label{cc3}
\end{align}
\end{subequations}

Analogous compatibility relations hold for the second and third rows and for the first and second column.  The relations are slightly different for the third column:
\begin{subequations}\label{cc33}
\begin{align}
  \classicalOutputMsmtEqv{13} &= \classicalOutputMsmt{C_3,1}\\
  \classicalOutputMsmtEqv{23} &= \classicalOutputMsmt{C_3,2}\\
  \classicalOutputMsmtEqv{33} &= - \classicalOutputMsmt{C_3,1}\cdot \classicalOutputMsmt{C_3,2},
\end{align}
\end{subequations}
or in terms of the conditional probabilities,
\begin{subequations}
\begin{align}
{\rm pr}\{\classicalOutputMsmtEqv{13} |\classicalOutputMsmt{C_3} \} &=   \delta_{\classicalOutputMsmtEqv{13}, \classicalOutputMsmt{C_3,1}}\label{comp1}\\
{\rm pr}\{\classicalOutputMsmtEqv{23} |\classicalOutputMsmt{C_3} \} &=   \delta_{\classicalOutputMsmtEqv{23}, \classicalOutputMsmt{C_3,2}}\\
{\rm pr}\{\classicalOutputMsmtEqv{33} |\classicalOutputMsmt{C_3} \} &=   \delta_{\classicalOutputMsmtEqv{33},-\classicalOutputMsmt{C_3,1}
\cdot \classicalOutputMsmt{C_3,2}}.\label{comp3}
\end{align}
\end{subequations}

\begin{figure}[t!]
 \centering
{
 \includegraphics[scale=0.25]{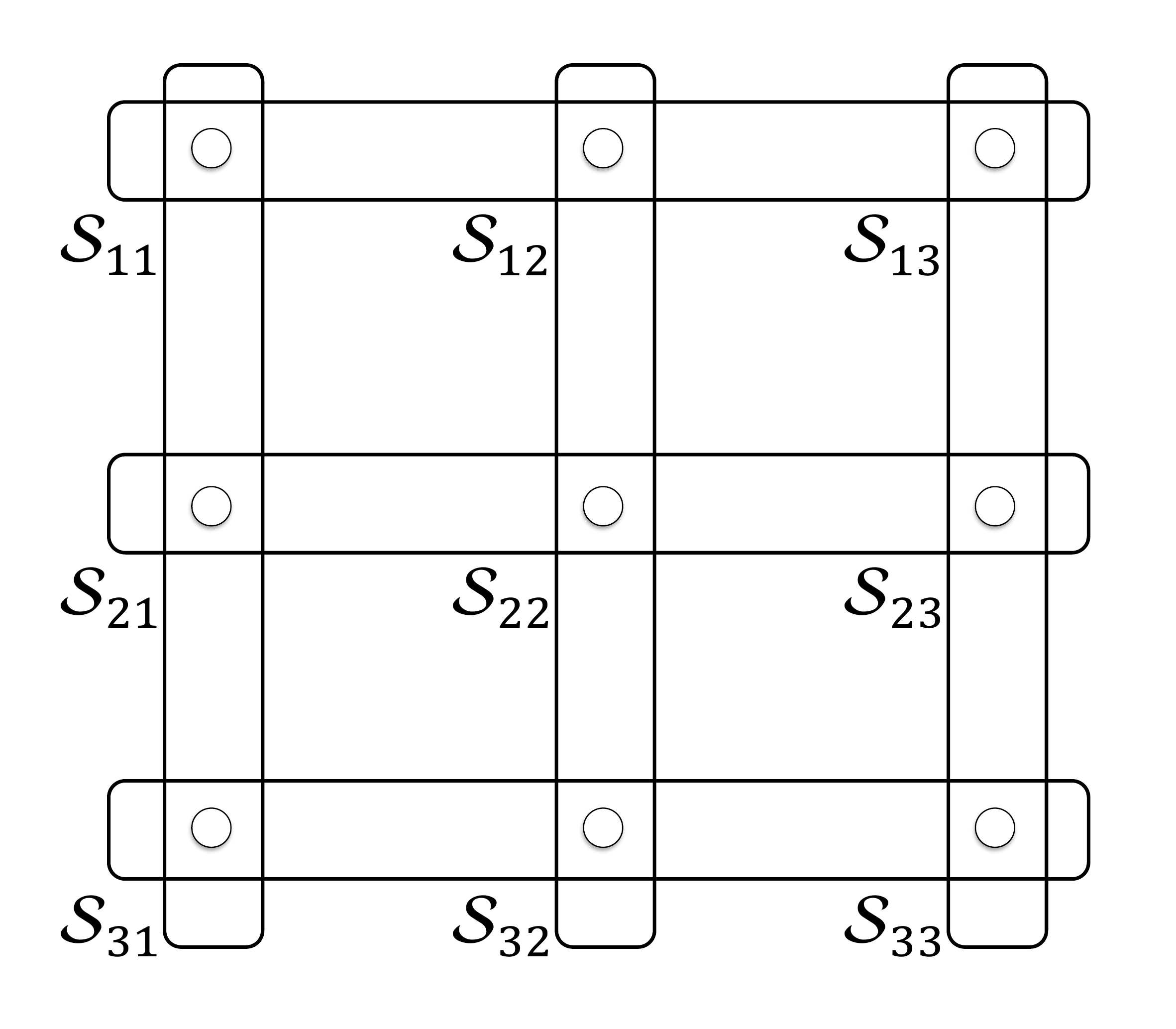}}
 \caption{The hypergraph representing compatibility relations among the nine binary-outcome sources. }
 \label{CompatHypergraphSources}
\end{figure}

A similar story holds for the sources.  There are 9 distinct equivalence classes of binary-outcome sources, which we label by $\eqvClassSrc{ij}$ where $i \in \{1,2,3\}$ and $j \in \{1,2,3\}$, with compatibility relations described by the hypergraph of Fig.~\ref{CompatHypergraphSources}.
% with extra constraint on the outcome-marginalized sources.

By the definition of compatibility for sources, Eq.~\eqref{compatibilitysources}, this implies that for every row and column there exists a source that simulates all the sources on that row or column.  We denote the source that simulates the triple of measurements in row 1 by $\src{R_1}$ and its outcome by $s_{R_1}$, the one that simulates the triple in column 1 by $\src{C_1}$ and its outcome by $s_{C_1}$, and so forth.  Each such outcome is presumed to be 4-valued, such that it can be presented as an ordered pair of binary variables, so that $s_{R_1} = (s_{R_1,1}, s_{R_1,2})$, etcetera. The conditional probabilities, which, by Eq.~\eqref{compatibilitysources}, define the precise nature of the compatibility relations are exactly the same as for the measurements.  For the first row, they are
\begin{subequations}
\begin{align}
{\rm pr}\{\classicalOutputSrcEqv{11} |\classicalOutputSrc{R_1} \} &=   \delta_{\classicalOutputSrcEqv{11}, \classicalOutputSrc{R_1,1}}\label{ss1}\\
{\rm pr}\{\classicalOutputSrcEqv{12} |\classicalOutputSrc{R_1} \} &=   \delta_{\classicalOutputSrcEqv{12}, \classicalOutputSrc{R_1,2}}\\
{\rm pr}\{\classicalOutputSrcEqv{13} |\classicalOutputSrc{R_1} \} &=   \delta_{\classicalOutputSrcEqv{13},\classicalOutputSrc{R_1,1}\cdot \classicalOutputSrc{R_1,2}}.\label{ss3}
\end{align}
\end{subequations}
%\begin{subequations}
%\begin{align}
%  \classicalOutputSrcEqv{11} &= \classicalOutputSrc{R_1,1}\\
%  \classicalOutputSrcEqv{12} &= \classicalOutputSrc{R_1,2}\\
 % \classicalOutputSrcEqv{13} &= \classicalOutputSrc{R_1,1}\cdot \classicalOutputSrc{R_1,2}.
%\end{align}
%\end{subequations}
with analogous relations holding for the other rows and the first and second column, while for the third column, they are
\begin{subequations}
\begin{align}
{\rm pr}\{\classicalOutputSrcEqv{13} |\classicalOutputSrc{C_3} \} &=   \delta_{\classicalOutputSrcEqv{13}, \classicalOutputSrc{C_3,1}}\label{sss1}\\
{\rm pr}\{\classicalOutputSrcEqv{23} |\classicalOutputSrc{C_3} \} &=   \delta_{\classicalOutputSrcEqv{23}, \classicalOutputSrc{C_3,2}}\\
{\rm pr}\{\classicalOutputSrcEqv{33} |\classicalOutputSrc{C_3} \} &=   \delta_{\classicalOutputSrcEqv{33},-\classicalOutputSrc{C_3,1}
\cdot \classicalOutputSrc{C_3,2}}.\label{sss3}
\end{align}
\end{subequations}

\subsubsection{Noiseless quantum realization}\label{noiselessQR}

It is straightforward to verify that the quantum measurements and quantum sources appearing in the no-go theorem described in Sec.~\ref{NoGoUniversal}  instantiate all of the compatibility relations that were described in the previous section.

%Since the observables in a given row or column of the Peres-Mermin square commute, they are compatible in the sense defined in Sec.~\ref{Compatibility}. \color{red} [This point has been made before] \color{black} Furthermore, one can explicitly verify that they satisfy the compatibility relations prescribed by Eqs.~\ref{comp1}-\ref{comp3}.
We begin with the quantum measurements.  Take, for example, the three observables in the first row of the Peres-Mermin square.
These  are associated to the projector-valued measures
$\{ \Pi_{11}^{(m_{11})} \}_{m_{11}}$,
$\{ \Pi_{12}^{(m_{12})} \}_{m_{12}}$ and $\{ \Pi_{13}^{(m_{13})} \}_{m_{13}}$ where $m_{11},m_{12},m_{13} \in \{-1,+1\}$, each corresponding to the projectors onto the pair of eigenspaces of the corresponding observables, as in Eq.~\eqref{projmeetsobs}.
The measurement that simulates all of these is, of course, the one associated to the joint eigenspaces of the three commuting observables, which as a projector-valued measure is
$\{  \Pi_{R_1}^{(m_{R_1,1} ,m_{R_1,2})}\}_{m_{R_1,1} , m_{R_1,2}}$ where ${(m_{R_1,1}, m_{R_1,2})\in \{-1,+1\}^2}$ and where
\begin{align}\begin{split}\label{simulatingmmt}
  \hspace{-\mathindent}\Pi_{R_1}^{(m_{R_1,1}, m_{R_1,2})} = \frac{1}{4} \Big( &\1 \otimes \1 +
  m_{R_1,1} \obs{11} + m_{R_1,2} \obs{12}\\
   &+ (m_{R_1,1} \cdot m_{R_1,2})\obs{13} \Big).
\end{split}\end{align}
The simulation of each of the three measurements in the row is achieved by implementing this PVM and then post-processing its outcome using the three conditional probability distributions specified in  Eqs.~(\ref{cc1}-\ref{cc3}), that is,
\begin{subequations}\begin{align}
 \Pi_{11}^{(m_{11})} &=  \smashoperator{\sum_{m_{R_1,1}, m_{R_1,2}}} \delta_{m_{11},m_{R_1,1}} \Pi_{R_1}^{(m_{R_1,1} ,m_{R_1,2})}\label{sim1}\\
  \Pi_{12}^{(m_{12})} &=  \smashoperator{\sum_{m_{R_1,1}, m_{R_1,2}}} \delta_{m_{12},m_{R_1,2}} \Pi_{R_1}^{(m_{R_1,1} ,m_{R_1,2})}\label{sim2}\\
    \Pi_{12}^{(m_{13})} &=  \smashoperator{\sum_{m_{R_1,1}, m_{R_1,2}}} \delta_{m_{13},m_{R_1,1}\cdot m_{R_1,2}} \Pi_{R_1}^{(m_{R_1,1}, m_{R_1,2})}.\label{sim3}
\end{align}\end{subequations}

In a similar fashion, one can verify that the other rows and columns of the Peres-Mermin square of quantum measurements have the compatibility relations described in the previous section.

The nine quantum sources appearing in the no-go theorem of Sec.~\ref{NoGoUniversal} also have the compatibility relations described in the previous section.  Consider the first row of the source version of the Peres-Mermin square as an example.  The three sources on this row are associated to the ensembles
 $\{ \tfrac{1}{2}\rho_{11}^{(s_{11})} \}_{s_{11}}$,
$\{ \tfrac{1}{2}\rho_{12}^{(s_{12})} \}_{s_{12}}$ and $\{ \tfrac{1}{2}\rho_{13}^{(s_{13})} \}_{s_{13}}$ where $s_{11},s_{12},s_{13} \in \{-1,+1\}$,  and $\rho_{ij}^{(+)}$ and $\rho_{ij}^{(-)}$ are the normalized projectors onto the eigenspaces of the observables associated to corresponding point on the Peres-Mermin square, as in Eq.~\eqref{statemeetsobs}.  The quantum source that simulates all of these is the one associated to the ensemble $\{ \tfrac{1}{4}  \rho_{R_1}^{(s_{R_1,1}, s_{R_1,2})}\}_{(s_{R_1,1}, s_{R_1,2})}$, where $\rho_{R_1}^{(s_{R_1,1}, s_{R_1,2}) } \equiv  \Pi_{R_1}^{(s_{R_1,1}, s_{R_1,2})}$ with $\Pi_{R_1}^{(s_{R_1,1}, s_{R_1,2})}$ the rank-1 projector defined in Eq.~\eqref{simulatingmmt}.  The conditional probabilities appearing in the simulation are precisely those given in Eqs.~(\ref{ss1}-\ref{ss3}).  This fact follows from Eqs.~(\ref{sim1}-\ref{sim3}).  The compatibility relations for the other rows and columns are verified similarly.

\subsubsection{Noisy quantum realization}\label{noisyQrealization}

In deriving noncontextuality inequalities, it is critical that one not base these on assumptions that are only valid when the measurements or the sources are noiseless because this ideal is never achieved in real experiments.
The compatibility relations outlined in Sec.~\ref{sec:operationalPM} satisfy this desideratum.
In the quantum case, for instance, they can be satisfied even if the measurements and sources are not sharp (i.e., not associated to an orthogonal set of projectors).

A specific example helps to clarify the point.

Suppose the nine sharp measurements appearing in the Peres-Mermin square are replaced by noisy versions thereof, that is, by the nine {\em unsharp} measurements that are the images of the sharp measurements under a partially depolarizing channel $\mathcal{D}$.  In this case, the projector-valued measures are replaced by POVMs that are not projective.   For instance, the three measurements in the first row of the Peres-Mermin square are associated to the binary-outcome POVMs
$\{ \mathcal{D} (\Pi_{11}^{(m_{11})}) \}_{m_{11}}$,
$\{  \mathcal{D} (\Pi_{12}^{(m_{12})}) \}_{m_{12}}$ and $\{  \mathcal{D} (\Pi_{13}^{(m_{13})}) \}_{m_{13}}$ where $m_{11},m_{12},m_{13} \in \{-1,+1\}$.  These can be jointly implemented using the 4-outcome POVM $\{ \mathcal{D}( \Pi_{R_1}^{(m_{R_1,1} ,m_{R_1,2})})\}_{m_{R_1,1} , m_{R_1,2}}$ where $(m_{R_1,1}, m_{R_1,2})\in \{-1,+1\}^2$.
The three binary-outcome POVMs are simulated by the 4-outcome POVM
using  the conditional probabilities in Eqs.~(\ref{cc1}-\ref{cc3}); to see this, it suffices to apply $\mathcal{D}$ to Eqs.~(\ref{sim1}-\ref{sim3}) and recall that it is a linear map).

Similarly, suppose that the nine sources appearing in the source version of the Peres-Mermin square are replaced by partially depolarized versions thereof.  (For simplicitly, we will assume that strength of the noise on the sources is equal to that on the measurements.)  In this case, the ensembles associated to the three sources in the first row are $\{\tfrac{1}{2} \mathcal{D} (\rho_{11}^{(s_{11})}) \}_{s_{11}}$,
$\{ \tfrac{1}{2} \mathcal{D} (\rho_{12}^{(s_{12})}) \}_{s_{12}}$ and $\{ \tfrac{1}{2} \mathcal{D} (\rho_{13}^{(s_{13})}) \}_{s_{13}}$ where $s_{11},s_{12},s_{13} \in \{-1,+1\}$.  The source that simulates all of these is then simply the partially depolarized version of the one that simulated the sharp sources, that is, $\{ \tfrac{1}{4}  \mathcal{D} ( \rho_{R_1}^{(s_{R_1,1}, s_{R_1,2})}) \}_{(s_{R_1,1}, s_{R_1,2})}$, which again follows from the linearity of Eqs.~(\ref{sim1}-\ref{sim3}).

Recall that a partial depolarization map $\mathcal{D}$ can be written as a convex mixture of the identity channel, $\mathcal{I}$, and the channel that traces over the system and reprepares the completely mixed state.  An element of the 1-parameter family of such maps is
\begin{align}
\mathcal{D}_r = r\, \mathcal{I} + (1-r)\, \frac{1}{4}\1 {\rm Tr},
\end{align}
where $r \in [0,1]$. The strength of depolarization is specified by the probability $r$ of realizing the identity map (with lower values of $r$ corresponding to stronger noise).

It follows that the degree of correlation that can be observed between sources and measurements is a function of $r$.  For $r<1$, one no longer achieves the perfect correlations of the noiseless quantum realization, \cref{perfcorrpr,perfcorrexp}, but rather imperfect correlations.  Denoting the $r$-depolarized versions of the observables $\obs{ij}$ and the ensembles $\ens{ij}$ by $\obs{ij}^{(r)}$ and $\ens{ij}^{(r)}$ respectively, we have  \begin{align}\label{imperfcorrpr}\begin{split}
\forall i,j: \;\;& pr(\classicalOutputMsmtEqv{ij},\classicalOutputSrcEqv{ij}| \obs{ij}^{(r)}, E^{(r)}_{ij})\\
& =  \tfrac{1}{2} \tr{\mathcal{D}_r\left(\Pi_{ij}^{(\classicalOutputMsmtEqv{ij})}\right)\mathcal{D}_r\left(\rho_{ij}^{(\classicalOutputSrcEqv{ij})}\right)}\\
& = \tfrac{1}{2} r^2 \delta_{\classicalOutputMsmtEqv{ij},\classicalOutputSrcEqv{ij}}~.
\end{split}\shortintertext{
This can be expressed equivalently  as
}
\label{imperfcorrexp}
%\forall i,j:\;\; \langle \classicalOutputMsmtEqv{} \classicalOutputSrcEqv{} \rangle_{\eqvClassMsmt{ij},\eqvClassSrc{ij}} =  1.
\forall i,j:&\;\; \langle \classicalOutputMsmtEqv{ij} \classicalOutputSrcEqv{ij} \rangle_{\obs{ij}^{(r)}, E^{(r)}_{ij}} =  r^2.
\end{align}

Because the no-go theorem of Sec.~\ref{NoGoUniversal} relied on having perfect correlations, it is not applicable to the noisy quantum realization of the operational Peres-Mermin scenario.  Nonetheless, one expects that for values of $r$ sufficiently close to 1, a noncontextual model should still be ruled out.  The noncontextuality inequalities that we derive confirm this expectation.  They are robust to  noise in the sense that they can be violated by values of $r$ strictly less than 1.  The lower bound on $r$ that they imply is determined in Sec.~\ref{sec:Robustness}.  This bound specifies how much noise one can tolerate in the noisy quantum realization of the Peres-Mermin scenario and still rule out a noncontextual model of the experiment.
\color{black}

%This establishes
%Indeed, noisy versions of the ideal quantum measurements continue to satisfy the compatibility relations.

%one must have an operational account that can contend with noise.
%requires the construction to  It is in this sense that the operational account of the scenario goes beyond the standard account.

%------------------------------------------------------------------------------------------------
%\subsection{Constraints on the ontological model}\label{sec:ontic}
\subsection{Expressing operational correlations in terms of noncontextual ontic assignments}\label{sec:ontic}
%\subsection{Where the assumption of noncontextuality gets applied}\label{sec:ontic}
%\section{The space of expectation values assigned by an ontic state}\label{sec:ontic}
%------------------------------------------------------------------------------------------------

Consider an experiment that can realize the nine equivalence classes of measurements  and the nine equivalence classes of sources having the compatibility structures of Figs.~\ref{CompatHypergraphMmts} and \ref{CompatHypergraphSources} respectively and having the compatibility relations specified in the text, such as Eqs.~(\ref{cc1}-\ref{cc3}) and Eqs.~(\ref{ss1}-\ref{ss3}).
There are 81 possible pairings of a source with a measurement.  For a given such pairing, say $\mathcal{S}_{ij}$ with $\mathcal{M}_{i'j'}$, the experiment yields a joint probability distribution over outcomes, $\textrm{pr} ( \mathcal{m}_{i'j'} \mathcal{s}_{ij}  | \mathcal{M}_{i'j'} \mathcal{S}_{ij})$.   Equivalently, the experimental data can be summarized by the expectation values $\langle \mathcal{s}_{ij} \rangle_{\mathcal{S}_{i'j'}}$, $\langle \mathcal{m}_{i'j'} \rangle_{\mathcal{M}_{ij}}$, and $\langle  \mathcal{m}_{i'j'}\mathcal{s}_{ij}  \rangle_{ \mathcal{M}_{i'j'} \mathcal{S}_{ij}}$.

In this article, we limit our focus to deriving constraints which do not refer to the marginal expectations $\langle \mathcal{s}_{ij} \rangle_{\mathcal{S}_{i'j'}}$ and $\langle \mathcal{m}_{i'j'} \rangle_{\mathcal{M}_{ij}}$, i.e. we focus on deriving inequalities which refer only to
the correlations $\langle \mathcal{s}_{ij}  \mathcal{m}_{i'j'} \rangle_{\mathcal{S}_{ij} \mathcal{M}_{i'j'}}$.
%rather than the marginal expectation values.
Furthermore, we consider only 9 of the 81 possible pairings of a source with a measurement, namely those wherein the source and the measurement are associated with a common label in their respective compatibility hypergraphs. That is, we hereafter consider only those correlations $\langle \mathcal{s}_{ij}  \mathcal{m}_{i'j'} \rangle_{\mathcal{S}_{ij} \mathcal{M}_{i'j'}}$ wherein $(i',j') = (i,j)$.  We will derive the necessary and sufficient conditions -- with respect to these nine correlations -- for an experiment to admit a noncontextual model.

For the equivalence class of measurements $\mathcal{M}_{ij}$, there are two associated measurement procedures, which we denote by $\msmt{ij}^{R}$ and $\msmt{ij}^{C}$, with outcomes denoted by $m^{R}_{ij}$ and $m^{C}_{ij}$, and which correspond to whether $\mathcal{M}_{ij}$ is implemented jointly with the other measurements in its row or with the other measurements in its column.  Similarly, the equivalence class of sources $\mathcal{S}_{ij}$ is associated with two sources, $S^{R}_{ij}$ and $S^{C}_{ij}$, with outcomes denoted by $s^{R}_{ij}$ and $s^{C}_{ij}$.  %The operational equivalences imply that $\langle s_{ij}  m_{i'j'} \rangle_{S_{ij} M_{ij}} = \langle s_{ij}  m'_{i'j'} \rangle_{S_{ij} M'_{ij}}=\langle s'_{ij}  m_{i'j'} \rangle_{S'_{ij} M_{ij}}=\langle s'_{ij}  m'_{i'j'} \rangle_{S'_{ij} M'_{ij}}$, which we can simply denote by $\langle \mathcal{s}_{ij}  \mathcal{m}_{i'j'} \rangle_{\mathcal{S}_{ij} \mathcal{M}_{i'j'}}$.
 The operational equivalences imply that
$\langle s^R_{ij}  m^R_{ij} \rangle_{S_{ij}^R M_{ij}^R} = \langle s^R_{ij}  m^C_{ij} \rangle_{S_{ij}^R M_{ij}^C}=\langle s^C_{ij}  m^R_{ij} \rangle_{S_{ij}^C M_{ij}^R}=\langle s^C_{ij}  m^C_{ij} \rangle_{S_{ij}^C M_{ij}^C}$, which we can simply denote by $\langle \mathcal{s}_{ij}  \mathcal{m}_{ij} \rangle_{\mathcal{S}_{ij} \mathcal{M}_{ij}}$.
%In particular, these relations hold when $(k,l)$ = $(i,j)$, the case of interest to us here.

Recall Eq.~\eqref{correlationasexpectationvalues}, which specifies how correlations such as $\langle s_{ij}  m_{ij} \rangle_{S_{ij} M_{ij}} $ are expressed in an ontological model.  Under the assumption of measurement noncontextuality, every measurement in the equivalence class $\mathcal{M}_{ij}$ is assigned the same expectation value by the ontic state.  Therefore, because $\msmt{ij}^R,\msmt{ij}^C \in \mathcal{M}_{ij}$, it follows that
\begin{align}\label{MNCexp}
\langle m^R_{ij} \rangle_{\lambda,M_{ij}^R} = \langle m^C_{ij} \rangle_{\lambda,M_{ij}^C} =\langle \mathcal{m}_{ij} \rangle_{\lambda,\mathcal{M}_{ij}}.
\end{align}
  In other words, measurement noncontextuality warrants the assumption that the expectation value for the outcome of $\eqvClassMsmt{ij}$ does not depend on the measurement context, that is, whether it is implemented with the measurements in the same row or in the same column of Fig.~\ref{CompatHypergraphMmts}.
 Similarly, under the assumption of preparation noncontextuality, every source in the equivalence class $\mathcal{S}_{ij}$ is assigned the same retrodictive expectation value by the ontic state, such that because $S^R_{ij},S^C_{ij} \in \mathcal{S}_{ij}$, it follows that
 \begin{align}\label{PNCexp}
 \langle s^R_{ij} \rangle_{\lambda,S_{ij}^R} = \langle s^C_{ij} \rangle_{\lambda,S_{ij}^C} =\langle \mathcal{s}_{ij} \rangle_{\lambda,\mathcal{S}_{ij}}.
 \end{align}  In other words, the expectation value for the outcome of $\eqvClassSrc{ij}$ does not depend on the source context, that is, whether it is implemented with the sources in the same row or in the same column of Fig.~\ref{CompatHypergraphSources}.
%\color{red} [Would it be better to specify the measurement procedures as $M_{11} \equiv \mathcal{M}_{11} (\mathcal{M}_{12},\mathcal{M}_{13})$ versus  $M'_{11} \equiv \mathcal{M}_{11} (\mathcal{M}_{21},\mathcal{M}_{31})$?]

We conclude that
%Recalling Eq.~\eqref{correlationasexpectationvalues}, it follows that if $M\in...$ and $S\in...$, then the expression in the ontological model for their correlation is
\begin{align}\label{zzz}
 \hspace{-3ex}\langle \mathcal{s}_{ij}  \mathcal{m}_{ij} \rangle_{\mathcal{S}_{ij} \mathcal{M}_{ij}}
    &=    \sum_{\lambda \in \Lambda}   \langle \mathcal{s}_{ij} \rangle_{\lambda, \mathcal{S}_{ij}}  \,\langle \mathcal{m}_{ij} \rangle_{\lambda, \mathcal{M}_{ij}}\, \mu(\lambda| \mathcal{S}_{ij}).
\end{align}
The operational equivalence relations among the sources and the assumption of preparation noncontextuality together imply one further simplification of this expression, namely, that $\mu(\lambda| \mathcal{S}_{ij})$ is independent of $(i,j)$,
\begin{align}\label{marginalizedsourceNC}
\forall (iji'j'): \mu(\lambda| \mathcal{S}_{ij})= \mu(\lambda| \mathcal{S}_{i'j'}) \equiv \mu(\lambda).
\end{align}
%The proof is as follows.
To see why this is the case, consider the triple of sources in the first row of Fig.~\ref{CompatHypergraphSources}, $\mathcal{S}_{11}, \mathcal{S}_{12}$ and $\mathcal{S}_{13}$.  By assumption, these are each simulatable by a single source, namely, $S_{R_1}$, by post-processing its outcome in the manner specified by the compatibility relations, Eqs.~(\ref{ss1}-\ref{ss3}).  Marginalizing over the outcome of $\mathcal{S}_{11}, \mathcal{S}_{12}$ or $\mathcal{S}_{13}$ is simply a {\em further} post-processing of $S_{R_1}$ and consequently the outcome-marginalized versions of these three sources are each operationally equivalent to the outcome-marginalized version of $S_{R_1}$ and therefore operationally equivalent to one another.
%, which we denoted by
  %are each simulated by the marginalization over the outcome of $S_{R_1}$.
   %Consequently, these three outcome-marginalized sources are operationally equivalent.
   The assumption of preparation noncontextuality then implies that the distributions over ontic states associated to these, namely that the three ontic state distributions
  % Because they are represented in the ontological model
  %by the marginalization over the outcome of the probability distribution that represents the unmarginalized source, that is,
%${\mu(\lambda|\mathcal{S}_{11}) := \sum_{\mathcal{s}_{11}}\mu(\mathcal{s}_{11},\lambda|\mathcal{S}_{11})}$, ${\mu(\lambda|\mathcal{S}_{12}) := \sum_{\mathcal{s}_{12}}\mu(\mathcal{s}_{12},\lambda|\mathcal{S}_{12})}$ and ${\mu(\lambda|\mathcal{S}_{13}) := \sum_{\mathcal{s}_{13}}\mu(\mathcal{s}_{13},\lambda|\mathcal{S}_{13})}$,
%their operational equivalence together with
%.   The operational equivalence of the three outcome-marginalized sources, together with
%the assumption of preparation noncontextuality, implies that these distributions
\begin{align*}
&\mu(\lambda|\mathcal{S}_{11}) := \sum_{\mathcal{s}_{11}}\mu(\mathcal{s}_{11},\lambda|\mathcal{S}_{11}) \\
&\mu(\lambda|\mathcal{S}_{12}) := \sum_{\mathcal{s}_{12}}\mu(\mathcal{s}_{12},\lambda|\mathcal{S}_{12}) \\
&\mu(\lambda|\mathcal{S}_{13}) := \sum_{\mathcal{s}_{13}}\mu(\mathcal{s}_{13},\lambda|\mathcal{S}_{13})
\end{align*}
are equal,
%Consequently, the operational equivalence of the three outcome-marginalized sources implies that
\begin{align}\label{ModelOfoutcomemarginaliedsources}
\mu(\lambda|\mathcal{S}_{11})& = \mu(\lambda|\mathcal{S}_{12})
=\mu(\lambda|\mathcal{S}_{13}).
%\sum_{s_{11}}\mu(s_{11},\lambda|\mathcal{S}_{11})& = \sum_{s_{12}}\mu(s_{12},\lambda|\mathcal{S}_{12})
%=\sum_{s_{13}}\mu(s_{13},\lambda|\mathcal{S}_{13})\nonumber\\
%&= \sum_{s_{R_1,1} ,s_{R_1,2}} \mu_{R_1}(s_{R_1,1} ,s_{R_1,2},\lambda).\nonumber\\
%&\equiv \mu_{R_1}(\lambda).
\end{align}
The same argument repeated for the other rows and the columns yields analogous equalities.  Together these imply Eq.~\eqref{marginalizedsourceNC}.
%\begin{align}
%\forall (ij),(i'j'): \sum_{s_{ij}}\mu(s_{ij},\lambda|\mathcal{S}_{ij})& = \sum_{s_{i'j'}}\mu(s_{i'j'},\lambda|\mathcal{S}_{i'j'}).
%\end{align}

%Recalling Eq.~\eqref{correlatorOPOM}, in a noncontextual ontological model with the operational equivalences we have described, each such degree of correlation can be expressed as
%\begin{align}
% \langle \mathcal{s}_{ij}  \mathcal{m}_{i'j'} \rangle_{\mathcal{S}_{ij} \mathcal{M}_{i'j'}}
%    &=    \sum_{\lambda \in \Lambda}   \langle \mathcal{s}_{ij} \rangle_{\lambda, \mathcal{S}_{ij}}  \langle \mathcal{m}_{ij} \rangle_{\lambda, \mathcal{M}_{ij}} \mu(\lambda| \mathcal{S}_{ij}).
%\end{align}

Denoting  $\langle \classicalOutputMsmtEqv{ij}\rangle_{\lambda,\mathcal{M}_{ij}}$ simply as $\langle \classicalOutputMsmtEqv{ij}\rangle_{\lambda}$ and $\langle \classicalOutputSrcEqv{ij}\rangle_{\lambda,\mathcal{S}_{ij}}$ simply as $\langle \classicalOutputSrcEqv{ij}\rangle_{\lambda}$, and using Eq.~\eqref{marginalizedsourceNC}, we find that Eq.~\eqref{zzz} becomes
\begin{align}\label{finalcorr}
 \langle \mathcal{s}_{ij}  \mathcal{m}_{ij} \rangle_{\mathcal{S}_{ij} \mathcal{M}_{ij}}
    &=    \sum_{\lambda \in \Lambda}   \langle \mathcal{s}_{ij} \rangle_{\lambda} \langle \mathcal{m}_{ij} \rangle_{\lambda} \mu(\lambda).
\end{align}

%All of the assumptions of noncontextuality have now been applied.  All that remains is to determine what constraints are satisfied by the  $\langle \mathcal{s}_{ij} \rangle_{\lambda}$ and the $\langle \mathcal{m}_{ij} \rangle_{\lambda}$.

We pause here to note that this expression for the correlation between the measurement outcome and the source outcome in a noncontextual model has the same form as the expression for the correlation between the measurement outcomes at the two wings of a Bell experiment in a locally causal model of the latter.  This provides a particularly intuitive demonstration of the isomorphism between the assumption of local causality and the assumption of preparation noncontextuality for the outcome-marginalized sources articulated in Eq.~\eqref{marginalizedsourceNC}. Note, however, that the assumptions of noncontextuality articulated in \cref{MNCexp,PNCexp} cannot be inferred from an assumption of local causality in the corresponding Bell scenario, so that the noncontextuality inequalities that we derive here are not isomorphic to Bell inequalities. 
\footnote{In particular, the noncontextuality inequalities we derive here are not isomorphic to the Bell inequality derived in Ref. \cite{CabelloPMBell} (and experimentally tested in Ref. \cite{CabelloPMBellExperiment}) even though the latter is inspired by a consideration of the Peres-Mermin construction (one such construction on each wing of the Bell experiment). This is because the inequality of Ref. \cite{CabelloPMBell} is based on the assumption of local causality alone. The analogue, for our prepare-and-measure scenario, of this inequality would be a constraint that follows from the assumption of preparation noncontextuality for the outcome-marginalized sources alone, Eq. (\ref{marginalizedsourceNC}).}
%additional assumptions of noncontextuality that we areare applied here.

The compatibility relations among the measurements imply constraints on the $\onticAvg{\classicalOutputMsmt{ij}}$.  We will refer to any 9-tuple of expectation values, $(\onticAvg{\classicalOutputMsmt{11}},\onticAvg{\classicalOutputMsmt{12}},\dots,\onticAvg{\classicalOutputMsmt{33}})$, satisfying these constraints as a {\em noncontextual ontic assignment} to the measurements.  We will see that the set of all such 9-tuples defines a polytope in a 9-dimensional space, which we term the \textbf{noncontextual measurement-assignment polytope}.
  Similarly, the compatibility relations among the sources imply constraints on the $\onticAvg{\classicalOutputSrc{ij}}$. We will refer to any 9-tuple of expectation values, $(\onticAvg{\classicalOutputSrc{11}},\onticAvg{\classicalOutputSrc{12}},\dots,\onticAvg{\classicalOutputSrc{33}})$, satisfying these constraints as a  (retrodictive) {\em noncontextual ontic assignment} to the sources.  These also form a polytope, which we term the \textbf{noncontextual source-assignment polytope}. The vertices of a polytope, i.e. the extremal noncontextual ontic assignments, can be deduced from that polytope's defining constraints using standard convex hull algorithms \cite{Fukuda1996,Zolotykh2012,avis_convexhull_2015}.

Every ontic state $\lambda$ specifies some noncontextual assignment to measurements, but not every noncontextual assignment corresponds to a vertex of the noncontextual measurement-assignment polytope. Nevertheless, those non-vertex noncontextual assignments to measurements can be \emph{simulated} by a distribution over ontic states that \emph{do} correspond to vertices:  Suppose $\kappa$ is a variable that runs over the vertices of the noncontextual measurement-assignment polytope.  Then, for any $\lambda$ in the polytope, there exists a distribution $p(\kappa|\lambda)$ such that
\begin{multline}\label{kappastuff}
  \left(\begin{matrix}
     \langle \mathcal{m}_{11} \rangle_{\lambda} &      \langle \mathcal{m}_{12} \rangle_{\lambda}  &      \langle \mathcal{m}_{13} \rangle_{\lambda}\\
          \langle \mathcal{m}_{21} \rangle_{\lambda} &      \langle \mathcal{m}_{22} \rangle_{\lambda}  &      \langle \mathcal{m}_{23} \rangle_{\lambda}\\
     \langle \mathcal{m}_{31} \rangle_{\lambda} &      \langle \mathcal{m}_{32} \rangle_{\lambda}  &      \langle \mathcal{m}_{33} \rangle_{\lambda}
 \end{matrix}\right)
 \\=
\sum_{\kappa}
  \left(\begin{matrix}
     \langle \mathcal{m}_{11} \rangle_{\kappa} &      \langle \mathcal{m}_{12} \rangle_{\kappa}  &      \langle \mathcal{m}_{13} \rangle_{\kappa}\\
          \langle \mathcal{m}_{21} \rangle_{\kappa} &      \langle \mathcal{m}_{22} \rangle_{\kappa}  &      \langle \mathcal{m}_{23} \rangle_{\kappa}\\
     \langle \mathcal{m}_{31} \rangle_{\kappa} &      \langle \mathcal{m}_{32} \rangle_{\kappa}  &      \langle \mathcal{m}_{33} \rangle_{\kappa}
 \end{matrix}\right)
 p(\kappa |\lambda)
\end{multline}
where we have presented the 9-tuple as a $3\times 3$ array.
% and the equality is to be understood as holding component-wise.
%For simplicity, assume that one is interested in solving for an expectation value of a measurement or a source alone, rather than a correlation. [.provide standard argument...]

A similar arguments holds for the noncontextual source-assignment polytope.
 %for the 9-tuple of noncontextual ontic assignments to the sources.
 Denoting its vertices by $\kappa'$, for any point $\lambda$ in the noncontextual source-assignment polytope, there exists a distribution $p(\kappa'|\lambda)$ such that
\begin{multline}\label{kappastuff2}
  \left(\begin{matrix}
     \langle \mathcal{s}_{11} \rangle_{\lambda} &      \langle \mathcal{s}_{12} \rangle_{\lambda}  &      \langle \mathcal{s}_{13} \rangle_{\lambda}\\
          \langle \mathcal{s}_{21} \rangle_{\lambda} &      \langle \mathcal{s}_{22} \rangle_{\lambda}  &      \langle \mathcal{s}_{23} \rangle_{\lambda}\\
     \langle \mathcal{s}_{31} \rangle_{\lambda} &      \langle \mathcal{s}_{32} \rangle_{\lambda}  &      \langle \mathcal{s}_{33} \rangle_{\lambda}
 \end{matrix}\right)
 \\=
\sum_{\kappa'}
  \left(\begin{matrix}
     \langle \mathcal{s}_{11} \rangle_{\kappa'} &      \langle \mathcal{s}_{12} \rangle_{\kappa'}  &      \langle \mathcal{s}_{13} \rangle_{\kappa'}\\
          \langle \mathcal{s}_{21} \rangle_{\kappa'} &      \langle \mathcal{s}_{22} \rangle_{\kappa'}  &      \langle \mathcal{s}_{23} \rangle_{\kappa'}\\
     \langle \mathcal{s}_{31} \rangle_{\kappa'} &      \langle \mathcal{s}_{32} \rangle_{\kappa'}  &      \langle \mathcal{s}_{33} \rangle_{\kappa'}
 \end{matrix}\right)
 p(\kappa' |\lambda)
\end{multline}

It is useful to introduce a simplified notation for the nine operational correlations in which we are interested, namely,
\begin{align}\label{defomega}
\omega_{ij}\equiv  \langle \mathcal{s}_{ij}  \mathcal{m}_{ij} \rangle_{\mathcal{S}_{ij} \mathcal{M}_{ij}}.
\end{align}
The set of 9-dimensional vectors $( \omega_{11}, \dots, \omega_{33})$ that can arise in an operational theory that admits of a noncontextual ontological model will be termed the \textbf{noncontextual correlation polytope}.   Recalling Eq.~\eqref{finalcorr}, and representing the $\omega_{ij}$ as a $3\times3$ array, it is defined in terms of the noncontextual measurement-assignment polytope and the noncontextual source-assignment polytope as follows:
%We now consider how to derive the \textbf{\em noncontextual correlation polytope} from the noncontextual measurement-assignment polytope and the noncontextual source-assignment polytope.
%For compactness, let us preliminarily introduce a simplified notation for the operational correlations in which we are interested, namely,
%\begin{align}\label{defomega}
%\omega_{ij}\equiv  \langle \mathcal{s}_{ij}  \mathcal{m}_{ij} \rangle_{\mathcal{S}_{ij} \mathcal{M}_{ij}}.
%\end{align}
%There are 9 such correlations.  We will present the 9-tuple of these as a $3\times3$ array.
%Recalling Eq.~\eqref{finalcorr}, and representing the $\omega_{ij}$ as a $3\times3$ array, we have
\begin{align}
&\hspace{-\mathindent}
\left(\begin{matrix}
     \omega_{11} & \omega_{12}  & \omega_{13}\\
     \omega_{21} & \omega_{22}  & \omega_{23}\\
     \omega_{31} & \omega_{32}  & \omega_{33}
 \end{matrix}\right)\nonumber\\
\begin{split}&\hspace{-\mathindent}=
\sum_{\lambda}
   \left(\begin{matrix}
     \langle \mathcal{m}_{11} \rangle_{\lambda} &      \langle \mathcal{m}_{12} \rangle_{\lambda}  &      \langle \mathcal{m}_{13} \rangle_{\lambda}\\
          \langle \mathcal{m}_{21} \rangle_{\lambda} &      \langle \mathcal{m}_{22} \rangle_{\lambda}  &      \langle \mathcal{m}_{23} \rangle_{\lambda}\\
     \langle \mathcal{m}_{31} \rangle_{\lambda} &      \langle \mathcal{m}_{32} \rangle_{\lambda}  &      \langle \mathcal{m}_{33} \rangle_{\lambda}
 \end{matrix}\right)\\
&\qquad\circ  \left(\begin{matrix}
     \langle \mathcal{s}_{11} \rangle_{\lambda} &      \langle \mathcal{s}_{12} \rangle_{\lambda}  &      \langle \mathcal{s}_{13} \rangle_{\lambda}\\
          \langle \mathcal{s}_{21} \rangle_{\lambda} &      \langle \mathcal{s}_{22} \rangle_{\lambda}  &      \langle \mathcal{s}_{23} \rangle_{\lambda}\\
     \langle \mathcal{s}_{31} \rangle_{\lambda} &      \langle \mathcal{s}_{32} \rangle_{\lambda}  &      \langle \mathcal{s}_{33} \rangle_{\lambda}
 \end{matrix}\right)
  \mu(\lambda),
\end{split}\end{align}
where $\;\circ\;$ denotes the entry-wise product of the arrays (also known as the Hadamard or Schur product).

Substituting Eqs.~\eqref{kappastuff} and \eqref{kappastuff2},
%its analogue for sources,
we have
\begin{align}\label{penultimate}
&\hspace{-\mathindent} \left(\begin{matrix}
     \omega_{11} & \omega_{12}  & \omega_{13}\\
     \omega_{21} & \omega_{22}  & \omega_{23}\\
     \omega_{31} & \omega_{32}  & \omega_{33}
 \end{matrix}\right)\nonumber\\
&\hspace{-\mathindent}=
\sum_{\kappa,\kappa'}
\left(\begin{matrix}
     \langle \mathcal{m}_{11} \rangle_{\kappa}  \langle \mathcal{s}_{11} \rangle_{\kappa'} &      \langle \mathcal{m}_{12} \rangle_{\kappa}  \langle \mathcal{s}_{12} \rangle_{\kappa'}  &      \langle \mathcal{m}_{13} \rangle_{\kappa}   \langle \mathcal{s}_{13} \rangle_{\kappa'}\\
          \langle \mathcal{m}_{21} \rangle_{\kappa}   \langle \mathcal{s}_{21} \rangle_{\kappa'} &      \langle \mathcal{m}_{22} \rangle_{\kappa}    \langle \mathcal{s}_{22} \rangle_{\kappa'}  &      \langle \mathcal{m}_{23} \rangle_{\kappa}  \langle \mathcal{s}_{23} \rangle_{\kappa'}\\
     \langle \mathcal{m}_{31} \rangle_{\kappa}   \langle \mathcal{s}_{31} \rangle_{\kappa'}  &      \langle \mathcal{m}_{32} \rangle_{\kappa}  \langle \mathcal{s}_{32} \rangle_{\kappa'}   &      \langle \mathcal{m}_{33} \rangle_{\kappa}   \langle \mathcal{s}_{33} \rangle_{\kappa'}
 \end{matrix}\right)\nonumber\\
& \times p(\kappa,\kappa'),\\
&\hspace{-\mathindent}\text{where}\quad p(\kappa,\kappa')\equiv  \sum_{\lambda}  p(\kappa |\lambda) p(\kappa' |\lambda) \mu(\lambda).
\end{align}
%We denote the $3\times 3$ array appearing in the summand by $A(\kappa,\kappa')$.
Therefore, the noncontextual correlation polytope is the convex hull of the correlations one obtains for all possible pairings $(\kappa,\kappa')$ of a vertex $\kappa$ from the noncontextual measurement-assignment polytope and a vertex $\kappa'$ from the noncontextual source-assignment polytope, that is, the convex hull of the 9-tuples $( \langle \mathcal{m}_{11} \rangle_{\kappa}  \langle \mathcal{s}_{11} \rangle_{\kappa'}, \ldots,  \langle \mathcal{m}_{33} \rangle_{\kappa}  \langle \mathcal{s}_{33} \rangle_{\kappa'})$, as one varies over $(\kappa,\kappa')$.

Not every pairing of $(\kappa,\kappa')$ corresponds to a unique vertex of the noncontextual correlation polytope: The fact that we consider correlations for only 9 source-measurement pairings and not the full set of 81 such pairings, and the fact that we do not consider any of the marginal expectations, implies that
%not ever pairing $(\kappa,\kappa')$ is extremal relative to the 9-tuple of correlations we consider. It can occur that
(i) more than one choice of  $(\kappa,\kappa')$ can yield the same 9-tuple of correlations, and (ii) one choice of $(\kappa,\kappa')$ can yield a 9-tuple of correlations that lies in the convex hull of the 9-tuples associated to several other choices of $(\kappa,\kappa')$.  It is therefore convenient to re-express Eq.~\eqref{penultimate} simply as
%an artifact of our considering correlations for only a strict subset of the possible source-measurement pairings (9 out of the 81 possibilities) is that occasionally more than one choice of  $(\kappa,\kappa')$ pairings will yield the same values for our nine correlations; additionally, sometimes one choice of $(\kappa,\kappa')$ will yield some 9-tuple which lies inside the convex hull of several other choices of pairings. It is therefore convenient to re-express Eq.~\eqref{penultimate} simply as
 \begin{align}\label{eq:tuplesfromverts}
&\hspace{-\mathindent}
\left(\begin{matrix}
     \omega_{11} & \omega_{12}  & \omega_{13}\\
     \omega_{21} & \omega_{22}  & \omega_{23}\\
     \omega_{31} & \omega_{32}  & \omega_{33}
 \end{matrix}\right)\\\nonumber
&\hspace{-\mathindent}
= \sum_{\gamma}
 \left(\begin{matrix}
     \langle \mathcal{m}_{11} \rangle_{\gamma}  \langle \mathcal{s}_{11} \rangle_{\gamma} &      \langle \mathcal{m}_{12} \rangle_{\gamma}  \langle \mathcal{s}_{12} \rangle_{\gamma}  &      \langle \mathcal{m}_{13} \rangle_{\gamma}   \langle \mathcal{s}_{13} \rangle_{\gamma}\\
          \langle \mathcal{m}_{21} \rangle_{\gamma}   \langle \mathcal{s}_{21} \rangle_{\gamma} &      \langle \mathcal{m}_{22} \rangle_{\gamma}    \langle \mathcal{s}_{22} \rangle_{\gamma}  &      \langle \mathcal{m}_{23} \rangle_{\gamma}  \langle \mathcal{s}_{23} \rangle_{\gamma}\\
     \langle \mathcal{m}_{31} \rangle_{\gamma}   \langle \mathcal{s}_{31} \rangle_{\gamma}  &      \langle \mathcal{m}_{32} \rangle_{\gamma}  \langle \mathcal{s}_{32} \rangle_{\gamma}   &      \langle \mathcal{m}_{33} \rangle_{\gamma}   \langle \mathcal{s}_{33} \rangle_{\gamma}
 \end{matrix}\right) p(\gamma).
%=\sum_{\gamma} A(\gamma) p(\gamma).
 \end{align}
where instead of ranging over all pairings of $(\kappa,\kappa')$ we restrict $\gamma$ to range over the vertices of the noncontextual correlation polytope without loss of generality.

We ultimately seek to derive noncontextuality inequalities, that is, the nontrivial facet inequalities of the noncontextual correlation polytope.  %Note that these facet inequalities are not those initially provided by the operational scenario, rather, the operational scenario initially determines the facet inequalities of the noncontextual measurement-assignment polytope and the noncontextual source-assignmnent polytope.
We begin by characterizing the noncontextual measurement-assignment polytope  and the noncontextual source-assignment polytope.  We will see that the nature of the compatibility relations among the measurements/sources determines their respective facet inequalities. From these, we infer the two set of vertices (measurements \& sources) using standard convex hull algorithms~\cite{Fukuda1996,barber1996quickhull,Zolotykh2012,avis_convexhull_2015}. Subsequently, by considering every possible pairing between those two sets of vertices, we determine the set of vertices of the noncontextual correlation polytope.  Finally, using standard convex hull algorithms again, we obtain all of the facet inequalities of the noncontextual correlation polytope.  The nontrivial facet inequalities define the set of noncontextuality inequalities for our problem.
 %The final step is to infer, from those deduced noncontextual correlation polytope vertices, the facets of the noncontextual correlation polytope (again using standard convex hull algorithms \cite{Fukuda1996,Zolotykh2012,avis_convexhull_2015}).
 In the following sections, we proceed through these various steps explicitly.

%We can obtain these given the {\em vertices}  of the noncontextual correlation polytope.  which are inferred from  The vertices of the noncontextual correlation polytope are easilly inferred from the vertices of the polytope of noncontextual ontic assignments to the measurements and the vertices of the polytope of noncontextual ontic assignments to the sources.
%Therefore, to determine the polytope of possibilities for the operational correlations, we must simply identify the $A(\gamma)$, that is, the vertices of the polytope of the $A(\kappa,\kappa')$, where {\em this} polytope is inferred by identifying the vertices of the polytope of noncontextual assignments to the measurements (which is equivalent to the polytope of noncontextual assignments to the sources).  From the vertices $A(\gamma)$ of the correlation polytope, we finally infer the facet inequalities, which constitute our noncontextuality inequalities.
 %We are now in a position to describe in detail how we derive our inequalities. Along the way, we will endeavour to provide some intuitions for their form.

\color{black}

\subsection{Facets of the noncontextual measurement-assignment and noncontextual source-assignment polytopes}

We will begin with the measurements. The compatibility relations holding among the measurements in a given row or column must also hold for the response functions representing these in the ontological model.  (This is a constraint on any ontological model, rather than one arising from the assumption of measurement noncontextuality.  See Sec.~\ref{sec:ExperimentalTest} for further discussion.)

Consider the response functions associated to the three equivalence classes of measurements in the first row of Fig.~\ref{CompatHypergraphMmts}.
%the Peres-Mermin square.
We denote the set of response functions associated to each of these by
$\{ \xi_{11}(m_{11}|\lambda) \}_{m_{11}}$,
$\{ \xi_{12}(m_{12}|\lambda) \}_{m_{12}}$ and $\{ \xi_{13}(m_{13}|\lambda) \}_{m_{13}}$ where $m_{11},m_{12},m_{13} \in \{-1,+1\}$, and we denote the set of response functions associated to the measurement that simulates these
$\{  \xi_{R_1}(m_{R_1,1} ,m_{R_1,2}|\lambda)\}_{m_{R_1,1} , m_{R_1,2}}$ where $(m_{R_1,1}, m_{R_1,2})\in \{-1,+1\}^2$.
The fact that the simulation is achieved
%each of the three measurements in the row can be simulated from this one by post-processing
with the three conditional probability distributions specified in  Eqs.~(\ref{cc1}-\ref{cc3}) implies that
\begin{subequations}\begin{align}
\hspace{-\mathindent} \xi_{11}(m_{11}|\lambda) &=  \smashoperator{\sum_{m_{R_1,1}, m_{R_1,2}}} \delta_{m_{11},m_{R_1,1}} \xi_{R_1}(m_{R_1,1} ,m_{R_1,2}|\lambda)\label{rf1}\\\hspace{-\mathindent}
  \xi_{12}(m_{12}|\lambda) &=  \smashoperator{\sum_{m_{R_1,1}, m_{R_1,2}}} \delta_{m_{12},m_{R_1,2}} \xi_{R_1}(m_{R_1,1} ,m_{R_1,2}|\lambda)\label{rf2}\\\hspace{-\mathindent}\begin{split}
    \xi_{12}(m_{13}|\lambda) &= \\ \smashoperator{\sum_{m_{R_1,1}, m_{R_1,2}}} &\delta_{m_{13},m_{R_1,1} m_{R_1,2}} \xi_{R_1}(m_{R_1,1}, m_{R_1,2}|\lambda).\label{rf3}\end{split}
\end{align}\end{subequations}
Recalling Eq.~\eqref{eq:onticExpectationX} and Eq.~\eqref{cc}, we infer that
\begin{subequations}\begin{align}\begin{split}
\langle m_{11} \rangle_{\lambda}  = &\,\xi_{R_1}(++|\lambda) + \xi_{R_1}(+-|\lambda) \\
&- \xi_{R_1}(-+|\lambda) - \xi_{R_1}(--|\lambda)\label{evintermsofrf1}\end{split}\\\begin{split}
\langle m_{12} \rangle_{\lambda}  = &\,\xi_{R_1}(++|\lambda) - \xi_{R_1}(+-|\lambda) \\
&+ \xi_{R_1}(-+|\lambda) - \xi_{R_1}(--|\lambda)\label{evintermsofrf2}\end{split}\\\begin{split}
\langle m_{13} \rangle_{\lambda}  = &\,\xi_{R_1}(++|\lambda) - \xi_{R_1}(+-|\lambda) \\
&- \xi_{R_1}(-+|\lambda) + \xi_{R_1}(--|\lambda).\label{evintermsofrf3}\end{split}
\end{align}\end{subequations}

Using the fact that $\xi_{R_1}$ is a normalized probability distribution,
\begin{align}\begin{split}\label{evintermsofrf4}
1  = &\xi_{R_1} (++|\lambda) + \xi_{R_1}(+-|\lambda) \\
&+ \xi_{R_1}(-+|\lambda) + \xi_{R_1}(--|\lambda),
\end{split}\end{align}
%normalization rule for probabilities,
the relations Eq.~(\ref{evintermsofrf1}-\ref{evintermsofrf3},\ref{evintermsofrf4})
can be inverted to write down the probabilities in terms of the expectation values: for $a,b\in \{-1,+1\}$,
\begin{align*}
%\hspace{-\mathindent}\xi_{R_1}(a, b|\lambda ) &= \tfrac{1}{4}(1 + m_{R_1,1}
% \langle m_{11} \rangle_{\lambda}  + m_{R_1,2} \langle m_{12} \rangle_{\lambda} \nonumber\\
 %&+ m_{R_1,1} m_{R_1,2} \langle m_{13} \rangle_{\lambda}  ).
 \hspace{-\mathindent}\xi_{R_1}(a, b|\lambda ) = \tfrac{1}{4}\left(1 + a
  \langle m_{11} \rangle_{\lambda}  + b \langle m_{12} \rangle_{\lambda}  + ab \langle m_{13} \rangle_{\lambda}  \right).
%\hspace{-\mathindent}\xi_{R_1}(\alpha, \beta|\lambda ) = \tfrac{1}{4}\left(1 + \alpha
 % \langle m_{11} \rangle_{\lambda}  + \beta \langle m_{12} \rangle_{\lambda}  + \alpha\beta \langle m_{13} \rangle_{\lambda}  \right).
\end{align*}

Finally, from the fact that
\begin{align}\label{jjj1}
&\forall a,b :\; \xi_{R_1}(a, b|\lambda )  \geq 0,
\shortintertext{we infer that \(\forall a,b :\)}
\label{jjj2}
& - a \onticAvg{ \classicalOutputMsmt{11} } - b \onticAvg{ \classicalOutputMsmt{12}} - ab \onticAvg{\classicalOutputMsmt{13} } \le   1.
\end{align}

This logic can be repeated for the other rows and the first two columns. For the third column, it is slightly different.  We find
\begin{align*}
\hspace{-\mathindent} \xi_{C_3}(a, b|\lambda ) = \tfrac{1}{4}\left(1 + a
  \langle m_{11} \rangle_{\lambda}  +b \langle m_{12} \rangle_{\lambda}  - ab \langle m_{13} \rangle_{\lambda}  \right)
%\hspace{-\mathindent} \xi_{C_3}(\alpha, \beta|\lambda ) = \tfrac{1}{4}\left(1 + \alpha
%  \langle m_{11} \rangle_{\lambda}  + \beta \langle m_{12} \rangle_{\lambda}  - \alpha\beta \langle m_{13} \rangle_{\lambda}  \right)
\end{align*}
from which we infer that \(\forall a,b :\)
\begin{align}
 - a \onticAvg{ \classicalOutputMsmt{11} } - b \onticAvg{ \classicalOutputMsmt{12}} + ab \onticAvg{\classicalOutputMsmt{13} } &\le   1.
\end{align}

%This logic can be repeated for the other rows and all three columns.
 In all then, we find that the 9 expectation values $\{ \onticAvg{\classicalOutputMsmt{ij}}: (i,j) \in \{ 1,2,3\}^2\}$ must satisfy
\begin{subequations}\begin{align}\begin{split}\label{mmtconstraints1}
\forall i &\in \{1,2,3\}:\\
 &\onticAvg{ \classicalOutputMsmt{i1} } + \onticAvg{ \classicalOutputMsmt{i2}} - \onticAvg{\classicalOutputMsmt{i3} } \le 1,
\\
& \onticAvg{ \classicalOutputMsmt{i1} } - \onticAvg{ \classicalOutputMsmt{i2}} + \onticAvg{\classicalOutputMsmt{i3} } \le   1,\\
-&  \onticAvg{ \classicalOutputMsmt{i1} } + \onticAvg{ \classicalOutputMsmt{i2}} + \onticAvg{\classicalOutputMsmt{i3} } \le   1,\\
-&  \onticAvg{ \classicalOutputMsmt{i1} } - \onticAvg{ \classicalOutputMsmt{i2}} -  \onticAvg{\classicalOutputMsmt{i3} } \le   1,
\end{split}
\shortintertext{and}
\begin{split}\label{mmtconstraints2}
\forall j  &\in \{1,2,3\}:\\
& \onticAvg{ \classicalOutputMsmt{1j} } + \onticAvg{ \classicalOutputMsmt{2j} } - \eta_{j}\onticAvg{ \classicalOutputMsmt{3j} }  \le 1,\\
& \onticAvg{ \classicalOutputMsmt{1j} } - \onticAvg{ \classicalOutputMsmt{2j} } + \eta_{j}\onticAvg{ \classicalOutputMsmt{3j} }  \le 1,\\
-& \onticAvg{ \classicalOutputMsmt{1j} } + \onticAvg{ \classicalOutputMsmt{2j} } + \eta_{j}\onticAvg{ \classicalOutputMsmt{3j} }  \le 1,\\
-& \onticAvg{ \classicalOutputMsmt{1j} } - \onticAvg{ \classicalOutputMsmt{2j} } -\eta_{j} \onticAvg{ \classicalOutputMsmt{3j} }  \le 1,
\end{split}\\\nonumber
%where $\eta_j=+1$ for $j\in \{1,2\}$ and $\eta_j=-1$ for $j=3$.
\hspace{-\mathindent}\text{where }&\eta_j=\begin{dcases*}
        +1  & for $j\in \{1,2\}$ \\
        -1 & for $j=3$
        \end{dcases*}.
\end{align}\end{subequations}

Note that these inequalities subsume the constraint that  $-1 \le \onticAvg{\classicalOutputMsmt{ij}} \le +1$ for all $i,j$.
%\color{red} [or is this implied by the others?] \color{black}
Eqs.~(\ref{mmtconstraints1}-\ref{mmtconstraints2}) capture all of the consequences of the compatibility relations for the ontic expectation values.
They are the facet inequalities for the noncontextual measurement-assignment polytope. \color{black}
% \color{red} We will refer to any 9-tuple of expectation values, $(\onticAvg{\classicalOutputMsmt{11}},\onticAvg{\classicalOutputMsmt{12}},\dots,\onticAvg{\classicalOutputMsmt{33}})$, satisfying these constrains as a {\em noncontextual ontic assignment} to the measurements.  The set of all such 9-tuples defines a polytope in a 9-dimensional space. \color{black}

Similar constraints hold for the nine retrodictive expectation values $\{ \onticAvg{\classicalOutputSrc{ij}}: (i,j) \in \{ 1,2,3\}^2\}$, as we now show.

Consider the the three equivalence classes of sources in the first row of Fig.~\ref{CompatHypergraphSources}.
%the source version of the Peres-Mermin square.
We will denote the probability distributions associated to each of these by $\mu_{11}(s_{11},\lambda)$, $\mu_{12}(s_{12},\lambda)$, and $\mu_{13}(s_{13},\lambda)$, where $s_{11},s_{12},s_{13} \in \{-1,+1\}$, and  the  probability distribution associated to the source that simulates these  by
$ \mu_{R_1}(s_{R_1,1} ,s_{R_1,2},\lambda)$ where $(s_{R_1,1}, s_{R_1,2})\in \{-1,+1\}^2$.
The fact that these sources obey the compatibility relations given in  Eqs.~(\ref{ss1}-\ref{ss3}) implies that
\begin{subequations}\begin{align}
\hspace{-3ex}\mu_{11}(s_{11},\lambda) &=  \smashoperator{\sum_{s_{R_1,1}, s_{R_1,2}}} \delta_{s_{11},s_{R_1,1}} \mu_{R_1}(s_{R_1,1} ,s_{R_1,2},\lambda)\label{muc1}\\\hspace{-3ex}
\mu_{12}(s_{12},\lambda) &=  \smashoperator{\sum_{s_{R_1,1}, s_{R_1,2}}} \delta_{s_{12},s_{R_1,2}} \mu_{R_1}(s_{R_1,1} ,s_{R_1,2},\lambda)\label{muc2}\\\begin{split}\hspace{-3ex}
\mu_{13}(s_{13},\lambda) &=  \\\smashoperator{\sum_{s_{R_1,1}, s_{R_1,2}}} & \delta_{s_{13},s_{R_1,1} s_{R_1,2}} \mu_{R_1}(s_{R_1,1}, s_{R_1,2},\lambda).\label{muc3}\end{split}
\end{align}\end{subequations}
Note that these expressions allow one to confirm the equality of
$\sum_{s_{11}}\mu(s_{11},\lambda|\mathcal{S}_{11})$, $\sum_{s_{12}}\mu(s_{12},\lambda|\mathcal{S}_{12})$ and
$\sum_{s_{13}}\mu(s_{13},\lambda|\mathcal{S}_{13})$, noted in Eq.~\eqref{ModelOfoutcomemarginaliedsources}, {\em via} their equality with
$\sum_{s_{R_1,1} ,s_{R_1,2}} \mu_{R_1}(s_{R_1,1} ,s_{R_1,2},\lambda)$.
Recalling Eq.~\eqref{marginalizedsourceNC},  these outcome-marginalized probability distributions are in fact equal to those associated to every other source in the problem, and we have denoted this unique distribution by $\mu(\lambda)$.
It follows that we can Bayesian invert all of the terms in Eqs.~(\ref{muc1}-\ref{muc3}) by dividing each equation by $\mu(\lambda)$.
We thereby obtain
\begin{subequations}\begin{align}
\hspace{-3ex}\mu_{11}(s_{11}|\lambda) &=  \smashoperator{\sum_{s_{R_1,1}, s_{R_1,2}}} \delta_{s_{11},s_{R_1,1}} \mu_{R_1}(s_{R_1,1} ,s_{R_1,2}|\lambda)\label{mucd1}\\\hspace{-3ex}
\mu_{12}(s_{12}|\lambda) &=  \smashoperator{\sum_{s_{R_1,1}, s_{R_1,2}}} \delta_{s_{12},s_{R_1,2}} \mu_{R_1}(s_{R_1,1} ,s_{R_1,2}|\lambda)\label{mucd2}\\\begin{split}\hspace{-3ex}
\mu_{13}(s_{13}|\lambda) &= \\ \smashoperator{\sum_{s_{R_1,1}, s_{R_1,2}}}& \delta_{s_{13},s_{R_1,1} s_{R_1,2}} \mu_{R_1}(s_{R_1,1}, s_{R_1,2}|\lambda).\label{mucd3}\end{split}
\end{align}\end{subequations}
Using these relations, together with Eq.~\eqref{eq:onticExpectationY}, we can express the expectation values $\{ \onticAvg{\classicalOutputSrc{ij}}: (i,j) \in \{ 1,2,3\}^2\}$ in terms of  the $\mu_{R_1}(s_{R_1,1} ,s_{R_1,2}|\lambda)$.  By appealing to the fact that $\mu_{R_1}$ is a normalized probability distribution, we can invert these equations and then use
$\forall a,b : \mu_{R_1}(a,b|\lambda) \ge 0$
%$\forall \alpha,\beta : \mu_{R_1}(\alpha,\beta|\lambda) \ge 0$
to obtain
inequality constraints on the $\{ \onticAvg{\classicalOutputSrc{ij}}: (i,j) \in \{ 1,2,3\}^2\}$.  The analysis proceeds precisely in analogy with the case of measurements, and we obtain:
\begin{subequations}\begin{align}\begin{split}\label{srcconstraints1}
\forall i &\in \{1,2,3\}:\\
 &\onticAvg{ \classicalOutputSrc{i1} } + \onticAvg{ \classicalOutputSrc{i2}} - \onticAvg{\classicalOutputSrc{i3} } \le 1,
\\
& \onticAvg{ \classicalOutputSrc{i1} } - \onticAvg{ \classicalOutputSrc{i2}} + \onticAvg{\classicalOutputSrc{i3} } \le   1,\\
-&  \onticAvg{ \classicalOutputSrc{i1} } + \onticAvg{ \classicalOutputSrc{i2}} + \onticAvg{\classicalOutputSrc{i3} } \le   1,\\
-&  \onticAvg{ \classicalOutputSrc{i1} } - \onticAvg{ \classicalOutputSrc{i2}} -  \onticAvg{\classicalOutputSrc{i3} } \le   1,
\end{split}
\shortintertext{and}\begin{split}
\label{srcconstraints2}
\forall j  &\in \{1,2,3\}:\\
& \onticAvg{ \classicalOutputSrc{1j} } + \onticAvg{ \classicalOutputSrc{2j} } - \eta_{j}\onticAvg{ \classicalOutputSrc{3j} }  \le 1,\\
& \onticAvg{ \classicalOutputSrc{1j} } - \onticAvg{ \classicalOutputSrc{2j} } + \eta_{j}\onticAvg{ \classicalOutputSrc{3j} }  \le 1,\\
-& \onticAvg{ \classicalOutputSrc{1j} } + \onticAvg{ \classicalOutputSrc{2j} } + \eta_{j}\onticAvg{ \classicalOutputSrc{3j} }  \le 1,\\
-& \onticAvg{ \classicalOutputSrc{1j} } - \onticAvg{ \classicalOutputSrc{2j} } -\eta_{j} \onticAvg{ \classicalOutputSrc{3j} }  \le 1.
\end{split}\\\nonumber
%where $\eta_j=+1$ for $j\in \{1,2\}$ and $\eta_j=-1$ for $j=3$.
\hspace{-\mathindent}\text{where }&\eta_j=\begin{dcases*}
        +1  & for $j\in \{1,2\}$ \\
        -1 & for $j=3$
        \end{dcases*}.
\end{align}\end{subequations}
These are the facet inequalities for the noncontextual source-assignment polytope.
%\color{red}We will refer to any 9-tuple of expectation values, $(\onticAvg{\classicalOutputSrc{11}},\onticAvg{\classicalOutputSrc{12}},\dots,\onticAvg{\classicalOutputSrc{33}})$, satisfying these constraints as a  (retrodictive) {\em noncontextual ontic assignment} to the sources.  \color{black}

Because Eqs.~(\ref{srcconstraints1}-\ref{srcconstraints2}) have the same form as Eqs.~(\ref{mmtconstraints1}-\ref{mmtconstraints2}), it follows that the noncontextual measurement-assignment polytope has precisely the same form as the noncontextual source-assignment polytope.
% polytope of noncontextual ontic assignments to the sources has the same form as the polytope of noncontextual ontic assignments to the measurements.
%We now turn to the problem of characterizing the latter.
It suffices, therefore, to characterize just one of them.  In the following, we consider the noncontextual measurement-assignment polytope for definiteness.

%We turn to characterizing the vertices of these two polytopes.

\subsection{Vertices of the noncontextual measurement-assignment and noncontextual source-assignment polytopes}

%As noted above, the polytope of ontic assignments to the measurements is the same as that for the sources.  It is sufficient, therefore, to characterize the polytope for the measurements.
%Because the polytope of ontic assignments to the measurements is the same as that for the sources, it is sufficient to characterize the polytope for the measurements.
%Because the two polytopes are equivalent, it is sufficient to solve the facet to vertex conversion for just one of them.  We consider the noncontextual measurement-assignment polytope.

In this section, we describe the conversion from the facet representation of the noncontextual measurement-assignment polytope, defined by the facet inequalities of Eqs.~(\ref{mmtconstraints1}-\ref{mmtconstraints2}), to its vertex represenation.
%We now describe all of the vertices of the polytope of ontic assignments to the measurements.  Recall that Eqs.~(\ref{mmtconstraints1}-\ref{mmtconstraints2}) are the facet inequalities of this polytope.
We use standard numerical algorithms to do so \cite{Fukuda1996,barber1996quickhull,Zolotykh2012,avis_convexhull_2015}, the details of which are provided in \cref{sec:repconversion}.  In addition to providing a description of this set of vertices, it is our aim here to provide some intuitions about their form.
%There are standard numerical algorithms for determining the vertices of a polytope from its facet inequalities\footnote{\color{red} [Elie: Do we need this given the lengthy discussion in the appendix?] \color{black} The authors utilized various freely-available software package for this purpose; initially \href{http://www.qhull.org}{\texttt{qhull}} by \citet{barber1996quickhull}, and later \href{http://www.uic.unn.ru/~zny/skeleton}{\texttt{skeleton}} and \href{http://sbastrakov.github.io/qskeleton}{\texttt{qskeleton}} by \citet{Bastrakov2015}.}, a computational task known as the convex hull problem \cite{Fukuda1996,Zolotykh2012,avis_convexhull_2015}; see \cref{sec:repconversion} for an extended discussion.

To begin with, note that all of the points within the noncontextual measurement-assignment polytope
%polytope of noncontextual ontic assignments to the measurements
are {\em indeterministic} assignments -- in the sense of violating Eq.~\eqref{OD} -- for one or more of the measurements.
To see that there are no noncontextual ontic assignments that are deterministic for all of the measurements, it suffices to note that for deterministic assignments, the constraints (\ref{mmtconstraints1}-\ref{mmtconstraints2}) simplify to
\begin{subequations}
\begin{align}
\lfloor \mathcal{m}_{11} \rfloor_{\lambda} \lfloor \mathcal{m}_{12} \rfloor_{\lambda} \lfloor \mathcal{m}_{13} \rfloor_{\lambda}&=+1,\label{gptPM1}\\
\lfloor \mathcal{m}_{21} \rfloor_{\lambda} \lfloor \mathcal{m}_{22} \rfloor_{\lambda} \lfloor \mathcal{m}_{23} \rfloor_{\lambda}&=+1,\label{gptPM2}\\
\lfloor \mathcal{m}_{31} \rfloor_{\lambda} \lfloor \mathcal{m}_{32} \rfloor_{\lambda} \lfloor \mathcal{m}_{33} \rfloor_{\lambda}&=+1,\label{gptPM3}\\
\lfloor \mathcal{m}_{11} \rfloor_{\lambda} \lfloor \mathcal{m}_{21} \rfloor_{\lambda} \lfloor \mathcal{m}_{31} \rfloor_{\lambda}&=+1,\label{gptPM4}\\
\lfloor \mathcal{m}_{12} \rfloor_{\lambda} \lfloor \mathcal{m}_{22} \rfloor_{\lambda} \lfloor \mathcal{m}_{23} \rfloor_{\lambda}&=+1,\label{gptPM5}\\
\lfloor \mathcal{m}_{13} \rfloor_{\lambda} \lfloor \mathcal{m}_{23} \rfloor_{\lambda} \lfloor \mathcal{m}_{33} \rfloor_{\lambda}&=-1.\label{gptPM6},
\end{align}
\end{subequations}
where $\lfloor \mathcal{m}_{11} \rfloor_{\lambda}$ denotes a deterministic assignment by ontic state $\lambda$,
and that these are equivalent to the constraints specified in Eqs.~(\ref{PM1}-\ref{PM6}), which, as noted in Sec.~\ref{sec:NoGoKS} admit no solution.
%\footnote{Here, we are considering deterministic noncontextual ontic assignments to {\em any} nine equivalence classes of measurements that satisfy the Peres-Mermin compatibility relations (having the structure of the hypergraph of Fig.~\ref{CompatHypergraphMmts} with simulation relations described near Eq.~\eqref{cc}), rather than simply the special case of the nine quantum observables in the Peres-Mermin square.}
%As noted in Sec.~\ref{sec:quantumPM}, these constraints admit of no solution, thereby establishing that there are no deterministic noncontextual ontic assignments to these measurements.
%\color{blue} [Say the following? If one wishes to maintain deterministic assignments these must be contextual which is to say that for some ontic state the outcomes assigned to operationally equivalent variables are not the same. [Measurements, not variables, are operationally equivalent] \color{black}

To get a feeling for how {\em in}deterministic noncontextual ontic assignments to the measurements escape contradiction, it is useful to see a concrete example (one that is a vertex of the noncontextual measurement-assignment polytope). We denote it by $\kappa_1$. We begin by describing it in terms of probabilistic assignments to the 4-outcome measurements associated to each row and column, rather than in terms of the expectation values for each the nine equivalence classes of binary-outcome measurements, because the correlations that hold between the different binary-outcome measurements are more transparent in this form.  Introducing the notation $p(m)=[\alpha]$ as a shorthand for $p(m)=\delta_{m,\alpha}$ and $p(m,m')=[\alpha,\beta]$ for $p(m,m')=\delta_{m,\alpha}\delta_{m',\beta}$, the vertex $\kappa_1$ is:
%The example of an indeterministic assignment corresponding to a vertex $\kappa$ of the noncontextual measurement-assignment polytope is:
%that an ontic state $\lambda$ might make is:
\begin{subequations}\begin{align}
\hspace{-\mathindent}\xi_{R_1}(m_{R_1,1},m_{R_1,2}|\kappa_1) &= \tfrac{1}{2}[+1,-1]+\tfrac{1}{2}[-1,+1]\\
\hspace{-\mathindent}\xi_{R_2}(m_{R_2,1},m_{R_2,2}|\kappa_1) &= \tfrac{1}{2}[+1,+1]+\tfrac{1}{2}[-1,-1]\\
\hspace{-\mathindent}\xi_{R_3}(m_{R_3,1},m_{R_3,2}|\kappa_1) &= [+1,+1]\\
\hspace{-\mathindent}\xi_{C_1}(m_{C_1,1},m_{C_1,2}|\kappa_1) &= \tfrac{1}{2}[+1,+1]+\tfrac{1}{2}[-1,-1]\\
\hspace{-\mathindent}\xi_{C_2}(m_{C_2,1},m_{C_2,2}|\kappa_1) &= \tfrac{1}{2}[+1,+1]+\tfrac{1}{2}[-1,-1]\\
\hspace{-\mathindent}\xi_{C_3}(m_{C_3,1},m_{C_3,2}|\kappa_1) &= [-1,+1]
\end{align}\end{subequations}
Using Eqs.~\eqref{rf1}-\eqref{rf3} and analogues thereof,
%From these,
one can compute from these the response functions for each of the nine equivalence classes of binary-outcome measurements.  They are
\begin{align}
&\hspace{-\mathindent}
\left(\begin{matrix}
     \xi_{11}( \mathcal{m}_{11} |\kappa_1)  &        \xi_{12}( \mathcal{m}_{12} |\kappa_1)   &        \xi_{13}( \mathcal{m}_{13} |\kappa_1) \\
     \xi_{21}( \mathcal{m}_{21} |\kappa_1)  &        \xi_{22}( \mathcal{m}_{22} |\kappa_1)   &        \xi_{23}( \mathcal{m}_{23} |\kappa_1) \\
     \xi_{31}( \mathcal{m}_{31} |\kappa_1)  &        \xi_{32}( \mathcal{m}_{32} |\kappa_1)   &        \xi_{33}( \mathcal{m}_{33} |\kappa_1)
 \end{matrix}\right)
\\&\nonumber =
 \left(\begin{matrix}
          \tfrac{1}{2}[+1]+\tfrac{1}{2}[-1] &\;\; \tfrac{1}{2}[+1]+\tfrac{1}{2}[-1] \;\; & [-1]\\
           \tfrac{1}{2}[+1]+\tfrac{1}{2}[-1]  &\;\; \tfrac{1}{2}[+1]+\tfrac{1}{2}[-1]\;\;  & [+1]\\
           [+1]  & [+1]  & [+1]\\
           \end{matrix}\right)
\end{align}
It is easy to verify that the two ways of defining the value of the response function for $\mathcal{M}_{ij}$  at $\kappa_1$  (via simulation by $M_{R_i}$ or via simulation by $M_{C_j}$)
 %with simulation relations given in \cref{cc,cc33}
  yield the same result, so that this is indeed a noncontextual assignment satisfying the compatibility relations.
In terms of expectation values, this assignment corresponds to
\begin{align}\label{examplevertex}
  \hspace{-\mathindent}\left(\begin{matrix}
     \langle \mathcal{m}_{11} \rangle_{\kappa_1} &      \langle \mathcal{m}_{12} \rangle_{\kappa_1}  &      \langle \mathcal{m}_{13} \rangle_{\kappa_1}\\
          \langle \mathcal{m}_{21} \rangle_{\kappa_1} &      \langle \mathcal{m}_{22} \rangle_{\kappa_1}  &      \langle \mathcal{m}_{23} \rangle_{\kappa_1}\\
     \langle \mathcal{m}_{31} \rangle_{\kappa_1} &      \langle \mathcal{m}_{32} \rangle_{\kappa_1}  &      \langle \mathcal{m}_{33} \rangle_{\kappa_1}
 \end{matrix}\right)
 =
 \left(\begin{matrix}
             0& 0  &-1\\
           0  & 0  &+1\\
           +1  & +1  &+1\\
           \end{matrix}\right).
\end{align}
%\color{red} [We should make note of the fact that one can extract the response functions from the matrix of expectation values] \color{black}
Note that it makes four of the nine measurements outcome-indeterministic.

A second concrete example of a vertex of the polytope, denoted $\kappa_2$, is
\begin{align}\label{examplevertex2}
  \hspace{-\mathindent}\left(\begin{matrix}
     \langle \mathcal{m}_{11} \rangle_{\kappa_2} &      \langle \mathcal{m}_{12} \rangle_{\kappa_2}  &      \langle \mathcal{m}_{13} \rangle_{\kappa_2}\\
          \langle \mathcal{m}_{21} \rangle_{\kappa_2} &      \langle \mathcal{m}_{22} \rangle_{\kappa_2}  &      \langle \mathcal{m}_{23} \rangle_{\kappa_2}\\
     \langle \mathcal{m}_{31} \rangle_{\kappa_2} &      \langle \mathcal{m}_{32} \rangle_{\kappa_2}  &      \langle \mathcal{m}_{33} \rangle_{\kappa_2}
 \end{matrix}\right)
 =
 \left(\begin{matrix}
             0& 0  &+1\\
           +1 & 0  &0\\
           0  & -1  &0\\
           \end{matrix}\right),
\end{align}
where six of the nine measurements are outcome-indeterministic.

% , as in \cref{examplevertex}, whereas the other 48 vertices make six of the nine measurements indeterministic, such as the following vertex

By considering the set of all deterministic processings of the measurements that preserve the compatibility relations holding among these, defined in \cref{symmetries}, one can determine the symmetries of the noncontextual measurement-assignment polytope.  Specifically, each such deterministic processing induces a bijective mapping of the set of vertices to itself.
%The deterministic processings of the measurements that preserve the compatibility relations among them, discussed in \cref{symmetries}, induce symmetries of the noncontextual measurement-assignment polytope in the sense of  bijectively mapping the set of vertices of this polytope to itself.
The full symmetry group is specified in \cref{symmetries}.
%A particularly efficient description of it is this group is in
It is straightforward to verify that it can be generated by the following three deterministic processings:
% are sufficient to generate the full symmetry group:
% \color{purple} As proven in \cref{symmetries}, the three generating symmetries that take a vertex of the polytope of noncontextual-ontic-assignments-to-the-measurements to another vertex (using the established matrix notation) are
\begin{align}\begin{split}\label{eq:msymgens}\hspace{-3ex}\underbrace{
\begin{array}{c}
\langle \mathcal{m}_{11} \rangle_{\kappa}\leftrightarrow\langle \mathcal{m}_{12} \rangle_{\kappa}\\
\langle \mathcal{m}_{21} \rangle_{\kappa}\leftrightarrow\langle \mathcal{m}_{22} \rangle_{\kappa}\\
\langle \mathcal{m}_{31} \rangle_{\kappa}\leftrightarrow\langle \mathcal{m}_{32} \rangle_{\kappa}\\
\end{array}}_{\mathclap{\text{Swap columns }1\leftrightarrow 2}}&,
\quad\text{and}\quad
\underbrace{
\begin{array}{c}
\langle \mathcal{m}_{21} \rangle_{\kappa}\leftrightarrow\langle \mathcal{m}_{31} \rangle_{\kappa}\\
\langle \mathcal{m}_{22} \rangle_{\kappa}\leftrightarrow\langle \mathcal{m}_{32} \rangle_{\kappa}\\
\langle \mathcal{m}_{23} \rangle_{\kappa}\leftrightarrow\langle \mathcal{m}_{33} \rangle_{\kappa}\\
\end{array}}_{\mathclap{\text{Swap rows }2\leftrightarrow 3}},\\
\text{and}\quad&
\underbrace{
\begin{array}{l}
\langle \mathcal{m}_{12} \rangle_{\kappa}\leftrightarrow\langle \mathcal{m}_{21} \rangle_{\kappa}\\
\langle \mathcal{m}_{31} \rangle_{\kappa}\leftrightarrow\langle \mathcal{m}_{31} \rangle_{\kappa}\\
\langle \mathcal{m}_{23} \rangle_{\kappa}\leftrightarrow\langle \mathcal{m}_{32} \rangle_{\kappa}\\
\langle \mathcal{m}_{33} \rangle_{\kappa}\leftrightarrow-\langle \mathcal{m}_{33} \rangle_{\kappa}\\
\end{array}}_{\mathclap{\text{Modified transpose}}}.
\end{split}\end{align}

Note that the number of measurements that are assigned outcomes deterministically is preserved by these symmetry operations.  Consequently, our two examples above, Eq.~\eqref{examplevertex} and Eq.~\eqref{examplevertex2}, are in different symmetry classes. In fact, we find that there are only these two symmetry classes.

The symmetry class wherein six of the nine measurements are indeterministic contains 48 vertices.  As $3\times 3$ matrices, they correspond to those with elements in $\{-1,0,+1\}$ having the property that every row and every column contains precisely one non-zero element.
The symmetry class wherein four of the nine measurements are indeterministic contains 72 vertices, and corresponds to those $3\times 3$ matrices with elements in $\{-1,0,+1\}$ having a single row of nonzero elements and a single column of nonzero elements such that the overall parity of the row is +1, and the overall parity of the column is $\eta$, where $\eta=-1$ if it is the third column and $\eta =+1$ otherwise.

%\color{black}We find that there are two symmetry classes of vertices under the action of these symmetries, where the classes are distinguished solely by the number of measurements that are assigned outcomes deterministically.  The 48 vertices that make six measurements indeterministic correspond to the set of $3\times 3$ matrices having the property that every row and every column contains precisely one non-zero element. The 72 vertices that make four measurements indeterministic correspond to the set of $3\times 3$ matrices with a single row of nonzero elements and a single column of nonzero elements such that the overall parity of the row is +1, and the overall parity of the column is $\eta$, where $\eta=-1$ if it is the third column and $\eta =+1$ otherwise.
%also +1, except when the column is the third column of the square, in which case its parity is -1.

%We find that the noncontextual measurement-assignment polytope has 120 vertices.  Of the 120 vertices, 72 make four of the nine measurements indeterministic, as in \cref{examplevertex}, whereas the other 48 vertices make six of the nine measurements indeterministic, such as the following vertex

\subsection{Vertices of the noncontextual correlation polytope}\label{sec:verticesCpolytope}

%Next, we compute the polytope of possibilities formed by the $A(\kappa,\kappa')$ defined in Eq.~\eqref{Akk}, which we term the {\em noncontextual correlation polytope}.

To determine the vertices of the noncontextual correlation polytope from the vertices of the noncontextual measurement-assignment polytope and those of the noncontextual source-assignment polytope, we preserve only those pairings which lead to extremal 9-tuples, as noted above Eq.~\eqref{eq:tuplesfromverts}.  Specifically, for each of the $120^2\!\!=\!\!14,\!400$ pairings $(\kappa,\kappa')$,
 %of a vertex $\kappa$ of the noncontextual measurement-assignment polytope with a vertex $\kappa'$ of the noncontextual source-assignment polytope,
 one computes the 9-tuple $(\langle \mathcal{m}_{11} \rangle_{\kappa}  \langle \mathcal{s}_{11} \rangle_{\kappa'}, \ldots,  \langle \mathcal{m}_{33} \rangle_{\kappa} \langle \mathcal{s}_{33} \rangle_{\kappa'})$.  By eliminating duplicate and non-extremal points from this set, we obtain the vertices of the noncontextual correlation polytope.
% \color{red} This is done using standard computational algorithms such as [Elie to fill in]. See \cref{sec:repconversion} for more details. \color{black}

A concrete example of a vertex of the noncontextual correlation polytope is obtained by pairing the vertex $\kappa_1$ of the noncontextual measurement-assignment polytope, described in Eq.~\eqref{examplevertex}, with a vertex $\kappa'_1$ of the noncontextual source-assignment polytope having precisely the same components as $\kappa_1$. This pairing yields
\begin{align}\label{eq:noncontexvertex1}
&\hspace{-\mathindent}
\left(\begin{matrix}
     \langle \mathcal{m}_{11} \rangle_{\kappa_1}  \langle \mathcal{s}_{11} \rangle_{\kappa'_1} &      \langle \mathcal{m}_{12} \rangle_{\kappa_1}  \langle \mathcal{s}_{12} \rangle_{\kappa'_1}  &      \langle \mathcal{m}_{13} \rangle_{\kappa_1}   \langle \mathcal{s}_{13} \rangle_{\kappa'_1}\\
          \langle \mathcal{m}_{21} \rangle_{\kappa_1}   \langle \mathcal{s}_{21} \rangle_{\kappa'_1} &      \langle \mathcal{m}_{22} \rangle_{\kappa_1}    \langle \mathcal{s}_{22} \rangle_{\kappa'_1}  &      \langle \mathcal{m}_{23} \rangle_{\kappa_1}  \langle \mathcal{s}_{23} \rangle_{\kappa'_1}\\
     \langle \mathcal{m}_{31} \rangle_{\kappa_1}   \langle \mathcal{s}_{31} \rangle_{\kappa'_1}  &      \langle \mathcal{m}_{32} \rangle_{\kappa_1}  \langle \mathcal{s}_{32} \rangle_{\kappa'_1}   &      \langle \mathcal{m}_{33} \rangle_{\kappa_1}   \langle \mathcal{s}_{33} \rangle_{\kappa'_1}
 \end{matrix}\right)\nonumber\\
&= \left(\begin{matrix}
             0& 0  &+1\\
           0  & 0  &+1\\
           +1  & +1  &+1\\
           \end{matrix}\right) .
\end{align}
Note that this vertex can also be constructed by pairing $\kappa_3$ with a corresponding $\kappa'_3$, where $\kappa_3$ is defined as $\kappa_{1}$ per Eq.~\eqref{examplevertex} but with the first two rows permuted, i.e. such that the $-1$ appears in the second row instead of the first. The distinct pairings $(\kappa_1, \kappa'_1)$ and $(\kappa_3, \kappa'_3)$ therefore yield duplicate noncontextual correlation points under entry-wise product.

Another vertex of the noncontextual correlation polytope is obtained by pairing the vertex $\kappa_2$ of the noncontextual measurement-assignment polytope, desribed in Eq.~\eqref{examplevertex2}, with a vertex $\kappa'_2$ of the noncontextual source-assignment polytope having precisely the same components as $\kappa_2$:
\begin{align}
&\hspace{-\mathindent}
\left(\begin{matrix}
     \langle \mathcal{m}_{11} \rangle_{\kappa_2}  \langle \mathcal{s}_{11} \rangle_{\kappa'_2} &      \langle \mathcal{m}_{12} \rangle_{\kappa_2}  \langle \mathcal{s}_{12} \rangle_{\kappa'_2}  &      \langle \mathcal{m}_{13} \rangle_{\kappa_2}   \langle \mathcal{s}_{13} \rangle_{\kappa'_2}\\
          \langle \mathcal{m}_{21} \rangle_{\kappa_2}   \langle \mathcal{s}_{21} \rangle_{\kappa'_2} &      \langle \mathcal{m}_{22} \rangle_{\kappa_2}    \langle \mathcal{s}_{22} \rangle_{\kappa'_2}  &      \langle \mathcal{m}_{23} \rangle_{\kappa_2}  \langle \mathcal{s}_{23} \rangle_{\kappa'_2}\\
     \langle \mathcal{m}_{31} \rangle_{\kappa_2}   \langle \mathcal{s}_{31} \rangle_{\kappa'_2}  &      \langle \mathcal{m}_{32} \rangle_{\kappa_2}  \langle \mathcal{s}_{32} \rangle_{\kappa'_2}   &      \langle \mathcal{m}_{33} \rangle_{\kappa_2}   \langle \mathcal{s}_{33} \rangle_{\kappa'_2}
 \end{matrix}\right)\nonumber\\
&= \left(\begin{matrix}
             0& 0  &+1\\
           +1 & 0  &0\\
           0  & +1  &0\\
           \end{matrix}\right).
\end{align}
%, corresponding to a subset of extremal points among the thousands of possible products of ontic-assignments-to-the-measurements with possible ontic-assignments-to-the-sources
%{\color{purple}The raw construction of (ontic) correlation vectors yielded $120^2\!\!=\!\!14,\!400$ points $\lfloor\bm{\omega}\rfloor_{\kappa,\kappa'}$, formed by pairing every vertex of the polytope of ontic assignments to the measurements (indexed by $\kappa$) with every vertex of the polytope of ontic assignments to the sources (indexed by $\kappa'$). To obtain the $120$ vertices of the noncontextual correlation polytope one must filter out duplicate and non-extremal points from that initial set of more than fourteen thousand correlation vectors; see \cref{sec:repconversion}.}
By contrast, if we pair $\kappa_1$ with $\kappa'_2$, we obtain a point that is nonextremal in the noncontextual correlation polytope.
We find that the noncontextual correlation polytope has
%\emph{also} has
120 vertices.\footnote{To be clear, the 120 vertices of the noncontextual correlation polytope should not be confused with the 120 vertices of the noncontextual measurement-assignment polytope; they are distinct sets, subject to different symmetry classifications.
%For example, some vertices of the  noncontextual correlation polytope satisfy ${\omega_{13}} {\omega_{23}} {\omega_{33}}=+1$, counter to the universality of $\lfloor \mathcal{m}_{13} \mathcal{m}_{23}\mathcal{m}_{33} \rfloor_{\lambda}\neq +1$.
}
%Among such vertices of the noncontextual correlation polytope one will find the correlation vector
%\begin{align}
% &\left(\begin{matrix}
 %    \langle \mathcal{m}_{11} \rangle_{\gamma}  \langle \mathcal{s}_{11} \rangle_{\gamma} &      \langle \mathcal{m}_{12} \rangle_{\gamma}  \langle \mathcal{s}_{12} \rangle_{\gamma}  &      \langle \mathcal{m}_{13} \rangle_{\gamma}   \langle \mathcal{s}_{13} \rangle_{\gamma}\\
%          \langle \mathcal{m}_{21} \rangle_{\gamma}   \langle \mathcal{s}_{21} \rangle_{\gamma} &      \langle \mathcal{m}_{22} \rangle_{\gamma}    \langle \mathcal{s}_{22} \rangle_{\gamma}  &      \langle \mathcal{m}_{23} \rangle_{\gamma}  \langle \mathcal{s}_{23} \rangle_{\gamma}\\
%     \langle \mathcal{m}_{31} \rangle_{\gamma}   \langle \mathcal{s}_{31} \rangle_{\gamma}  &      \langle \mathcal{m}_{32} \rangle_{\gamma}  \langle \mathcal{s}_{32} \rangle_{\gamma}   &      \langle \mathcal{m}_{33} \rangle_{\gamma}   \langle \mathcal{s}_{33} \rangle_{\gamma}
% \end{matrix}\right)\nonumber\\
%& =
% \left(\begin{matrix}
%             0& 0  &-1\\
%           0  & 0  &-1\\
%           +1  & +1  &+1\\
%           \end{matrix}\right) \,,
%\end{align}
%in contrast to the odd-parity final column dictum illustrated in \cref{examplevertex}.

By considering the deterministic processings of the measurements and sources which preserve the operational Peres-Mermin scenario---that is, the processings which preserve the compatibility relations among the measurements, the compatibility relations among the sources, and the manner in which the sources and the measurements are paired---we can infer the symmetries of the noncontextual correlation polytope, as discussed in \cref{symmetries}.  Specifically, every such symmetry bijectively maps the set of vertices of this polytope to itself.  The full symmetry group is described in \cref{symmetries}, and it is straightforward to verify that it
%the symmetry group described in \cref{symmetries}
can be generated by the following three processings (which we also describe as operations on the $3\times3$ matrix):
%As proven in \cref{symmetries}, the three generating symmetries which take any valid vertex of the noncontextual correlation polytope to another are the following operations:
\begin{align}\label{eq:mssymgens}\hspace{-\mathindent}\underbrace{
\begin{array}{c}
  \langle \mathcal{m}_{11} \rangle_{\gamma}  \langle \mathcal{s}_{11} \rangle_{\gamma} \leftrightarrow \langle \mathcal{m}_{12} \rangle_{\gamma}  \langle \mathcal{s}_{12} \rangle_{\gamma}\\
  \langle \mathcal{m}_{21} \rangle_{\gamma}  \langle \mathcal{s}_{21} \rangle_{\gamma} \leftrightarrow \langle \mathcal{m}_{22} \rangle_{\gamma}  \langle \mathcal{s}_{22} \rangle_{\gamma}\\
  \langle \mathcal{m}_{31} \rangle_{\gamma}  \langle \mathcal{s}_{31} \rangle_{\gamma} \leftrightarrow \langle \mathcal{m}_{32} \rangle_{\gamma}  \langle \mathcal{s}_{32} \rangle_{\gamma}\\
\end{array}}_{\mathclap{\text{Columns }1\leftrightarrow 2}}
\nonumber \\ \text{ and }
\underbrace{
\begin{array}{c}
  \langle \mathcal{m}_{21} \rangle_{\gamma}  \langle \mathcal{s}_{21} \rangle_{\gamma} \leftrightarrow \langle \mathcal{m}_{31} \rangle_{\gamma}  \langle \mathcal{s}_{31} \rangle_{\gamma}\\
  \langle \mathcal{m}_{22} \rangle_{\gamma}  \langle \mathcal{s}_{22} \rangle_{\gamma} \leftrightarrow \langle \mathcal{m}_{32} \rangle_{\gamma}  \langle \mathcal{s}_{32} \rangle_{\gamma}\\
  \langle \mathcal{m}_{23} \rangle_{\gamma}  \langle \mathcal{s}_{23} \rangle_{\gamma} \leftrightarrow \langle \mathcal{m}_{33} \rangle_{\gamma}  \langle \mathcal{s}_{33} \rangle_{\gamma}\\
\end{array}}_{\mathclap{\text{Rows }2\leftrightarrow 3}}
 \\ \text{ and }
\underbrace{
\begin{array}{l}
  \langle \mathcal{m}_{11} \rangle_{\gamma}  \langle \mathcal{s}_{11} \rangle_{\gamma} \leftrightarrow -\langle \mathcal{m}_{11} \rangle_{\gamma}  \langle \mathcal{s}_{11} \rangle_{\gamma}\\
  \langle \mathcal{m}_{12} \rangle_{\gamma}  \langle \mathcal{s}_{12} \rangle_{\gamma} \leftrightarrow \langle \mathcal{m}_{21} \rangle_{\gamma}  \langle \mathcal{s}_{21} \rangle_{\gamma}\\
  \langle \mathcal{m}_{13} \rangle_{\gamma}  \langle \mathcal{s}_{13} \rangle_{\gamma} \leftrightarrow -\langle \mathcal{m}_{31} \rangle_{\gamma}  \langle \mathcal{s}_{31} \rangle_{\gamma}\\
  \langle \mathcal{m}_{23} \rangle_{\gamma}  \langle \mathcal{s}_{23} \rangle_{\gamma} \leftrightarrow \langle \mathcal{m}_{32} \rangle_{\gamma}  \langle \mathcal{s}_{32} \rangle_{\gamma}\\
  \langle \mathcal{m}_{33} \rangle_{\gamma}  \langle \mathcal{s}_{33} \rangle_{\gamma} \leftrightarrow -\langle \mathcal{m}_{33} \rangle_{\gamma}  \langle \mathcal{s}_{33} \rangle_{\gamma}\\
\end{array}}_{\mathclap{\text{Modified transpose}}}\nonumber
\end{align}
%\par\vbox{
%$\bullet$ Permutation of any pair of rows\par
%$\bullet$ Permutation of any pair of columns\par
%$\bullet$ For any pair of row and column, multiplying the four coefficients not contained therein by  -1\par
%Multiplying any column AND any row by -1}\par
Note that the modified transpose operation appearing in \cref{eq:wsymgens} is distinct from the modified transpose appearing in \cref{eq:msymgens}.

%For many applications one may be satisfied with enumeration of the extremal correlations, which we have now completed. If one wishes to proceed to determine the facet inequalities for the correlation polytope from knowledge of these vertices, one proceeds by (again) solving the convex hull problem \cite{Fukuda1996,Zolotykh2012,avis_convexhull_2015}, discussed more fully in \cref{sec:repconversion}.

%We are now finally in a position to present our noncontextuality inequalities.

\subsection{Facets of the noncontextual correlation polytope: noncontextuality inequalities}\label{sec:NCInequalities}
%\subsection{The noncontextuality inequalities for the Peres-Mermin square}\label{sec:NCInequalities}

To determine the facet inequalities for the noncontextual correlation polytope from its vertices, we proceed (again) by solving the convex hull problem \cite{Fukuda1996,barber1996quickhull,Zolotykh2012,avis_convexhull_2015}, discussed more fully in \cref{sec:repconversion}.

Facet inequalities of the noncontextual correlation polytope
have the form
\begin{align}\label{eq:genericIneq}
  \sum_{i,j = 1}^{3} \alpha_{ij} \omega_{ij} \leq \beta~.
\end{align}
where $\{ \alpha_{ij}\}_{ij}$ and $\beta$ are integers.
%real numbers (which in the example considered here are integers).
%We denote such an inequality by $I(\{\alpha_{ij}\},\beta)$.
Arranging the $\alpha_{ij}$ and the $\omega_{ij}$ into $3\times 3$ matrices and denoting the entry-wise matrix product by $\;\circ\;$ and the sum of the elements of a matrix $A$ by $\textrm{su}(A)$, we can express this as
%the $I(\{\alpha_{ij}\},\beta)$
%the inequality of Eq.~\eqref{eq:genericIneq} as
\begin{align}\label{eq:ineqmatform}
\hspace{-3ex}\textrm{su}\left[ \left(\begin{matrix}
     \alpha_{11} & \alpha_{12}  & \alpha_{13}\\
     \alpha_{21} & \alpha_{22}  & \alpha_{23}\\
     \alpha_{31} & \alpha_{32}  & \alpha_{33}\\
 \end{matrix}\right)
 \circ
 \left(\begin{matrix}
     \omega_{11} & \omega_{12}  & \omega_{13}\\
     \omega_{21} & \omega_{22}  & \omega_{23}\\
     \omega_{31} & \omega_{32}  & \omega_{33}
 \end{matrix}\right)\right] \leq \beta.
\end{align}
We will refer to the matrix of $\alpha_{ij}$'s as the {\em coefficient matrix} for the inequality.
%This representation is convenient for enumerating inequalities since these can be classified by the upper bound $\beta$  and by the matrix of $\alpha_{ij}$'s, which we term the {\em coefficient matrix}.

We find that there are 184 inequalities, all of which are expressed using coefficient matrices where $\forall i,j : \alpha_{ij} \in \{0,-1,+1\}$ and where $\beta \in \{1,3,5\}$.

%Recalling the definition of $\omega_{ij}$, Eq.~\eqref{}, we see that any matrix of $\omega_{ij}$s satisfying the inequalities is mapped to another such matrix under the symmetries described in Eq.~\eqref{eq:mssymgens}.
%In the previous section, we described the symmetry operations that map an extremal matrix of $\omega_{ij}$ within the noncontextual correlation polytope to another such matrix. These same symmetry operations are such that they will map a valid coefficient matrix to another.
Note that the deterministic processings of the experiment that bijectively map the set of vertices of the noncontextual correlation polytope to itself also bijectively map the set of facet inequalities of the noncontextual correlation polytope to itself, and vice versa.   Consequently, the symmetry group of the set of facet inequalities is the same as the symmetry group of the set of vertices.  The value of $\beta$ in a facet inequality is invariant under the symmetry group, so that only the matrix of $\alpha$ coefficients transforms nontrivially.  Given that the facet inequalities can be expressed in the form of Eq.~\eqref{eq:ineqmatform}, any map on the $\omega$ matrix can be transferred onto the matrix of $\alpha$ coefficients.  Consequently, the action of the symmetry group on the coefficient matrix is precisely parallel to that described in Eq.~\eqref{eq:mssymgens}, namely, the group generated by
% the action of the group on the matrix of $\alpha$ coefficients is precisely the same as its action on the $\omega$ matrix.
%A processing of the experiment can be understood as a map on the $\omega$ matrix.  However, we can transfer this action onto the matrix of $\alpha$ coefficients.
 %It is not difficult to see that if we express this symmetry group by its action on the coefficient matrix of each facet inequality, it can be  generated by:
%As proven in \cref{symmetries}, the three generating symmetries which take any valid facet-inequality-coefficient-matrix of the noncontextual correlation polytope to another are the following operations:
\begin{align}\label{eq:wsymgens}\hspace{-\mathindent}\underbrace{
\begin{array}{c}
\alpha_{11}\leftrightarrow\alpha_{12}\\
\alpha_{21}\leftrightarrow\alpha_{22}\\
\alpha_{31}\leftrightarrow\alpha_{32}\\
\end{array}}_{\mathclap{\text{Columns }1\leftrightarrow 2}}
\text{ and }
\underbrace{
\begin{array}{c}
\alpha_{21}\leftrightarrow\alpha_{31}\\
\alpha_{22}\leftrightarrow\alpha_{32}\\
\alpha_{23}\leftrightarrow\alpha_{33}\\
\end{array}}_{\mathclap{\text{Rows }2\leftrightarrow 3}}
\text{ and }
\underbrace{
\begin{array}{l}
\alpha_{11}\leftrightarrow-\alpha_{11}\\
\alpha_{12}\leftrightarrow\alpha_{21}\\
\alpha_{13}\leftrightarrow-\alpha_{31}\\
\alpha_{23}\leftrightarrow\alpha_{32}\\
\alpha_{33}\leftrightarrow-\alpha_{33}\\
\end{array}}_{\mathclap{\text{Modified transpose}}}
\end{align}
%\par\vbox{
%$\bullet$ Permutation of any pair of rows\par
%$\bullet$ Permutation of any pair of columns\par
%$\bullet$ For any pair of row and column, multiplying the four coefficients not contained therein by  -1\par
%Multiplying any column AND any row by -1}\par
%Note that the modified transpose operation appearing in \cref{eq:wsymgens} is distinct from the previous modified transpose appearing in \cref{eq:msymgens}.\color{black}

%The symmetries apply to both facet-defining inequalities as well as vertices, in that those operations of the correlation vectors which bijectively map vertices to vertices will also bijectively map facet inequalities to facet inequalities, and vice versa. As such, the operations of \cref{eq:mssymgens}, which preserve the facets of the noncontextual correlation polytope, also generate the symmetry group for this polytope's vertices. Although the operations in \cref{eq:mssymgens} are expressed in terms of inequality coefficients, they may be interpreted as applying to the correlation vectors directly; the vertices of the noncontextual correlation polytope fall into two distinct orbits under the closure of those operations.

%Thus, for two facet inequalities that share the same value of $\beta$,
Thus, if the $\alpha$ coefficient matrix of one inequality is related to that of another inequality by one of the symmetry operations we have identified, then these two inequalities are in the same symmetry class. An efficient description of all of the facet inequalities is achieved by describing representatives of each of the symmetry classes, and closing under the action of the symmetries.  We find that there are just three symmetry classes of facet inequalities, conveniently distinguished by their values of $\beta$.
%The vertices of the noncontextual correlation polytope fall into two distinct orbits under the closure of those operations.
%Under these operations, the facet inequalities are classified into three symmetry classes.

The first symmetry class is trivial in the sense that the facet inequalities therein hold for all correlations that are logically possible in the operational Peres-Mermin scenario, and consequently they are not sensitive to whether or not the correlations admit of a noncontextual model.

\noindent\textbf{Trivial Class}:
%$\beta=1$}\par
A representative of this class is
%An example of an inequality from this class is
    \begin{align}\label{trivialineq}
    \hspace{-3ex}\textrm{su}\left[
         \left(\begin{matrix}
             -1& +1  &+1\\
           0  & 0  & 0\\
           0  & 0  & 0\\
           \end{matrix}\right) \circ
            \left(\begin{matrix}
     \omega_{11} & \omega_{12}  & \omega_{13}\\
     \omega_{21} & \omega_{22}  & \omega_{23}\\
     \omega_{31} & \omega_{32}  & \omega_{33}
 \end{matrix}\right)\right]
            \leq 1\,.
           \end{align}
Closing under the symmetries, one finds that there are 24 such inequalities, corresponding to coefficient matrices where only one row or column has all nonzero elements, and such that the overall parity of these is $-1$.
%\color{red} PROBLEM: The inequality should be of a different form for the third column! \color{blue}
We justify the claim that these inequalities are trivial in Appendix~\ref{TrivialIneqs}.

The term {\em noncontextuality inequality} is reserved for those facet inequalities of the noncontextual correlation polytope that are nontrivial.   There are two symmetry classes of these.

%           \item
\noindent\textbf{Nontrivial Class I}:
% $\beta=3$}\par
A representative of this class is
%n example of a noncontextuality inequality from this class is
\begin{align}\label{exampleclassII}
    \hspace{-3ex}\textrm{su} \left[
         \left(\begin{matrix}
             -1 & 0  & +1\\
           +1  & 0   & +1\\
           0  & -1  & 0\\
           \end{matrix}\right) \circ
            \left(\begin{matrix}
     \omega_{11} & \omega_{12}  & \omega_{13}\\
     \omega_{21} & \omega_{22}  & \omega_{23}\\
     \omega_{31} & \omega_{32}  & \omega_{33}
 \end{matrix}\right)\right]
            \leq 3\,.
           \end{align}
Closing under the symmetries, one finds that there are 144 such inequalities, all of which can be constructed as follows: Choose a special position in the matrix of coefficients, say element $\alpha_{ij}$, and make it +1 or $-1$.
Let all other elements in the same row or column of the coefficients matrix be zero. Finally, choose any assignment of $\pm 1$ for the remaining four elements such that the overall parity of the five nonzero elements is $+1$. The example inequality of Eq.~\eqref{exampleclassII}  is one of the eight inequalities that follow by starting with $\alpha_{32}=-1$ as the special element.

An example of an inequality from this class that is maximally violated by the noiseless quantum realization of the operational Peres-Mermin scenario (described in Sec.~\ref{noiselessQR}) is
 %An example of an inequality from this class which is violated by quantum theory is
      \begin{align}\label{QVI1}
    \hspace{-3ex}\textrm{su} \left[
         \left(\begin{matrix}
             +1 & +1  & 0\\
           +1  & +1   & 0\\
           0  & 0  & +1\\
           \end{matrix}\right) \circ
            \left(\begin{matrix}
     \omega_{11} & \omega_{12}  & \omega_{13}\\
     \omega_{21} & \omega_{22}  & \omega_{23}\\
     \omega_{31} & \omega_{32}  & \omega_{33}
 \end{matrix}\right)\right]
            \leq 3\,,
           \end{align}
as we will demonstrate in the next section.

%\item
\noindent\textbf{Nontrivial Class II}:
%: $\beta=5$}\par
A representative of this class is
%An example of a noncontextuality inequality from this class is
%The third class of noncontextuality inequalities are upper bounded by $\beta=5$, such as
\begin{align}
    \hspace{-3ex}\textrm{su}\left[
         \left(\begin{matrix}
             +1 & -1  & -1\\
           +1  & +1   & +1\\
           +1  & -1  & -1\\
           \end{matrix}\right) \circ
            \left(\begin{matrix}
     \omega_{11} & \omega_{12}  & \omega_{13}\\
     \omega_{21} & \omega_{22}  & \omega_{23}\\
     \omega_{31} & \omega_{32}  & \omega_{33}
 \end{matrix}\right)\right]
            \leq 5\,.
           \end{align}
Closing under the symmetries, one finds that there are 16 inequalities in this class, all of which have only nonzero elements in the coefficient matrix. The 16 inequalities are precisely those whose coefficient matrices have $+1$ overall parity for every row and every column.

      An example of an inequality from this class that is maximally violated by the noiseless quantum realization of the operational Peres-Mermin scenario (described in Sec.~\ref{noiselessQR}) is
      \begin{align}\label{QVI2}
    \hspace{-3ex}\textrm{su} \left[
         \left(\begin{matrix}
             +1 & +1  & +1\\
           +1  & +1   & +1\\
           +1 & +1  & +1\\
           \end{matrix}\right) \circ
            \left(\begin{matrix}
     \omega_{11} & \omega_{12}  & \omega_{13}\\
     \omega_{21} & \omega_{22}  & \omega_{23}\\
     \omega_{31} & \omega_{32}  & \omega_{33}
 \end{matrix}\right)\right]
            \leq 5\,,
           \end{align}
           as we will demonstrate in the next section.
%\end{enumerate}

\subsubsection{Quantum violation of the inequalities}\label{sec:QViolationInequalities}

We have already seen in Sec.~\ref{NoGoUniversal} that one can identify sharp quantum sources and sharp quantum measurements that satisfy the Peres-Mermin compatibility structure and whose statistics are inconsistent with a universally noncontextual model.   These quantum sources and measurements must therefore violate our inequalities.
Indeed, for these quantum sources and measurements, it follows from Eq.~\eqref{perfcorrexp} that $\forall i,j : \omega_{ij}=1$, that is, every one of the nine source-measurement pairs exhibits perfect correlation.  This implies that the right-hand side of the noncontextuality inequality of Eq.~\eqref{QVI1} evaluates to 5 for this quantum realization, thereby exceeding the noncontextual bound of 3.  It also implies that the right-hand side of the noncontextuality inequality of Eq.~\eqref{QVI1} evaluates to 9, which exceeds the noncontextual bound of 5.

\subsubsection{Robustness of the inequalities to noise}\label{sec:Robustness}
%\subsection{Quantum noise limit}

We now demonstrate explicitly how the noncontextuality inequalities we have derived are robust to noise by showing that the noisy quantum realization of the operational Peres-Mermin scenario, described in Sec.~\ref{noisyQrealization}, can still lead to a violation.  Recall that this consisted of quantum sources and quantum measurements that were the image under a partial depolarization map of those appearing in the no-go theorem of Sec.~\ref{NoGoUniversal}.

For these noisy sources and measurements, the value of the correlation for each of the nine source-measurement pairs was computed, as a function of  the weight $r$ of the identity map in the partial depolarization, in Eq.~\eqref{imperfcorrexp}. Translating into the $\omega_{ij}$ notation of Eq.~\eqref{defomega}, the result is
\begin{align}
\forall i,j: \omega_{ij} = r^2.
\end{align}

Substituting this expression into the noncontextuality inequality of Eq.~\eqref{QVI1}, we obtain
\begin{align}
5 r^2 \le 3,
\end{align}
Consequently, as long as the level of noise is such that ${r > \frac{\sqrt{3}}{\sqrt{5}}\simeq 0.77460}$, one has a violation of the noncontextuality inequality.  (Lower values of $r$ correspond to stronger noise, so this is an upper bound on the noise.)

Similarly, from the noncontextuality inequality of Eq.~\eqref{QVI2}, we obtain
\begin{align}
9 r^2 \le 5.
\end{align}
implying that we require $r > \frac{\sqrt{5}}{\sqrt{9}}\simeq 0.73536$ to see a violation.

Because $\frac{\sqrt{5}}{9} < \frac{\sqrt{3}}{\sqrt{5}}$, the noncontextuality inequality of Eq.~\eqref{QVI2} can tolerate more noise than that of Eq.~\eqref{QVI1}.

We have shown that our noncontextuality inequalities still admit a violation even in the presence of significant depolarizing noise relative to the ideal quantum realization.  As such, the fact that any attempt at an experimental implementation of the ideal quantum sources and measurements inevitably only realizes noisy versions of these is not an obstacle to demonstrating an experimental failure of noncontextuality.

\section{How to implement an experimental test of these inequalities}\label{sec:ExperimentalTest}
%\subsection{What an experiment must probe}

%\color{red} [Perhaps the compatibility-equivalence hypergraphs should be presented first in this section.  Their role is to make it clear what one has to do experimentally.  They don't really appear in the analysis of the derivation of noncontextuality inequalities.]

%\color{red} [We must emphasize that the operational equivalences must be checked experimentally.  This requires the secondary procedures trick from our other paper.]
% \color{black}

%\color{red} One implements {\em primary} versions of the eighteen measurements and the eighteen sources.  One characterizes each of the measurements (sources) by a tomographically complete set of sources (measurements).  One then defines secondary procedures as mixtures of these and others (that one implements to help move in any direction in the space of possible measurements and sources) and these secondary procedures are constrained to satisfy the operational equivalence relations.  It is the secondary procedures, therefore, that represent the nine equivalence classes of measurements and the nine equivalence classes of sources.  We then plug in the correlations for pairs of {\em these} into our noncontextuality inequalities.
\color{black}

The assumption of noncontextuality only has nontrivial consequences if one has experimentally verified that certain operational equivalence relations hold among the measurements and among the sources.  This creates a problem for experimental tests of noncontextuality because the definition of operational equivalence for sources is in terms of equivalence of statistics for {\em all} measurements, and for measurements it is in terms of equivalence of statistics for {\em all} sources, and, strictly speaking, one can never experimentally implement {\em all} possible procedures of either type.    The problem may be summarized as the physical impossibility of verifying any criterion that involves a universal quantifier. We here explain our view on what is the appropriate attitude to take towards this problem.

A tomographically-complete set of measurements  is defined as a set of measurements whose statistics are sufficient to infer the statistics of any other measurement.  A tomographically-complete set of sources is defined as a set of sources whose statistics are sufficient to infer the statistics of any other source.  It follows that to judge operational equivalence relations among sources, it is sufficient to consider their statistics on a tomographically-complete set of measurements rather than on all measurements, and  to judge operational equivalence relations among measurements, it is sufficient to consider their statistics on a tomographically-complete set of sources rather than on all sources.  The problem therefore reduces to identifying tomographically complete sets for each.
%, which we shall denote by $\mathfrak{M}_{\rm tomo}$ and $\mathfrak{S}_{\rm tomo}$ respectively.

 It is well known that in quantum theory, the set of observables obtained by taking all products of Pauli operators corresponds to a tomographically complete set of measurements for a pair of qubits. Therefore, quantum theory dictates that one must {\em supplement} the products of Pauli operators appearing in the Peres-Mermin square with all the other nontrivial products of Pauli operators, that is, with $\{Y \otimes X, Y \otimes Z, Y \otimes \1, \1\otimes Y,  X\otimes Y, Z\otimes Y \}$, in order to obtain a tomographically complete set.  Consequently, by the lights of quantum theory, in order to test operational equivalence relations among the sources, it is necessary to do more measurements than appear in the Peres-Mermin square construction.

% In the case of an experiment seeking to realize the operational Peres-Mermin scenario, therefore, quantum theory leads us to expect that the  nine binary-outcome measurements appearing therein are   {\em not}, by themselves, a tomographically complete of measurements, but that they will need to be supplemented by six more binary-outcome measurements.
%\begin{align}
%\mathfrak{M}_{\rm tomo} = \{ M_{R_1}, M_{R_2}, M_{R_3}, M_{C_1}, M_{C_2}, M_{C_3}\}.
%\end{align}

Furthermore, if one seeks to implement a {\em direct} experimental test of noncontextuality, then one does not want to presume the correctness of quantum theory.  As such, it is inappropriate to presume that the fifteen binary-outcome measurements that one expects to be tomographically complete by the lights of quantum theory are {\em in fact}
%given set of measurements are
tomographically complete.  Instead, one should accumulate experimental evidence for this hypothesis by implementing the greatest diversity of measurements on the system that one can and by verifying that the statistics for each such measurement can be inferred from the statistics of the hypothetical tomographically complete set.

Similar comments apply for the problem of identifying a tomographically complete set of sources for the purpose of evaluating operational equivalence relations among the measurements.
%Similarly, for the sources, quantum theory leads us to expect that one can take a tomographically complete set to be,
%\begin{align}
%\mathfrak{S}_{\rm tomo} = \{ S_{R_1}, S_{R_2}, S_{R_3}, S_{C_1}, S_{C_2}, S_{C_3}\},
%\end{align}
%but, again, one can and should accumulate experimental evidence for this hypothesis.

%Similarly for the sources.
We refer the reader to Ref.~\cite{bootstraptomography} for more details on how to acquire experimental evidence for a given set of measurements (preparations) being tomographically complete.

Even given good evidence of tomographic completeness, one faces another problem with experimentally verifying operational equivalences, namely, that if one aims to implement a particular set of procedures (termed the {\em target} procedures),  the unavoidable imprecision of experimental implementations implies that one inevitably fails to do so precisely, and one's experiment instead realizes a set of procedures that deviate from the target procedures, termed the {\em primary} procedures.  Given the failure of the primary procedures to coincide with the target procedures, operational equivalence relations that hold among the target procedures need not hold among the primary procedures.  The primary procedures therefore do not generally realize the operational equivalence relations that are the starting point for derivations of noncontextuality inequalities.
%achieve these precise procedures
%and consequently the operational equivalence relations that hold among the target procedures will not hold for their actually-implemented counterparts.
This has been termed the problem of ``no strict operational equivalences'' in Ref.~\cite{MazurekEtAl}.

Before describing how to resolve it, we specify how it arises in the context of the Peres-Mermin construction.

 \begin{figure}[t!]
 \centering
 {
 \includegraphics[scale=0.25]{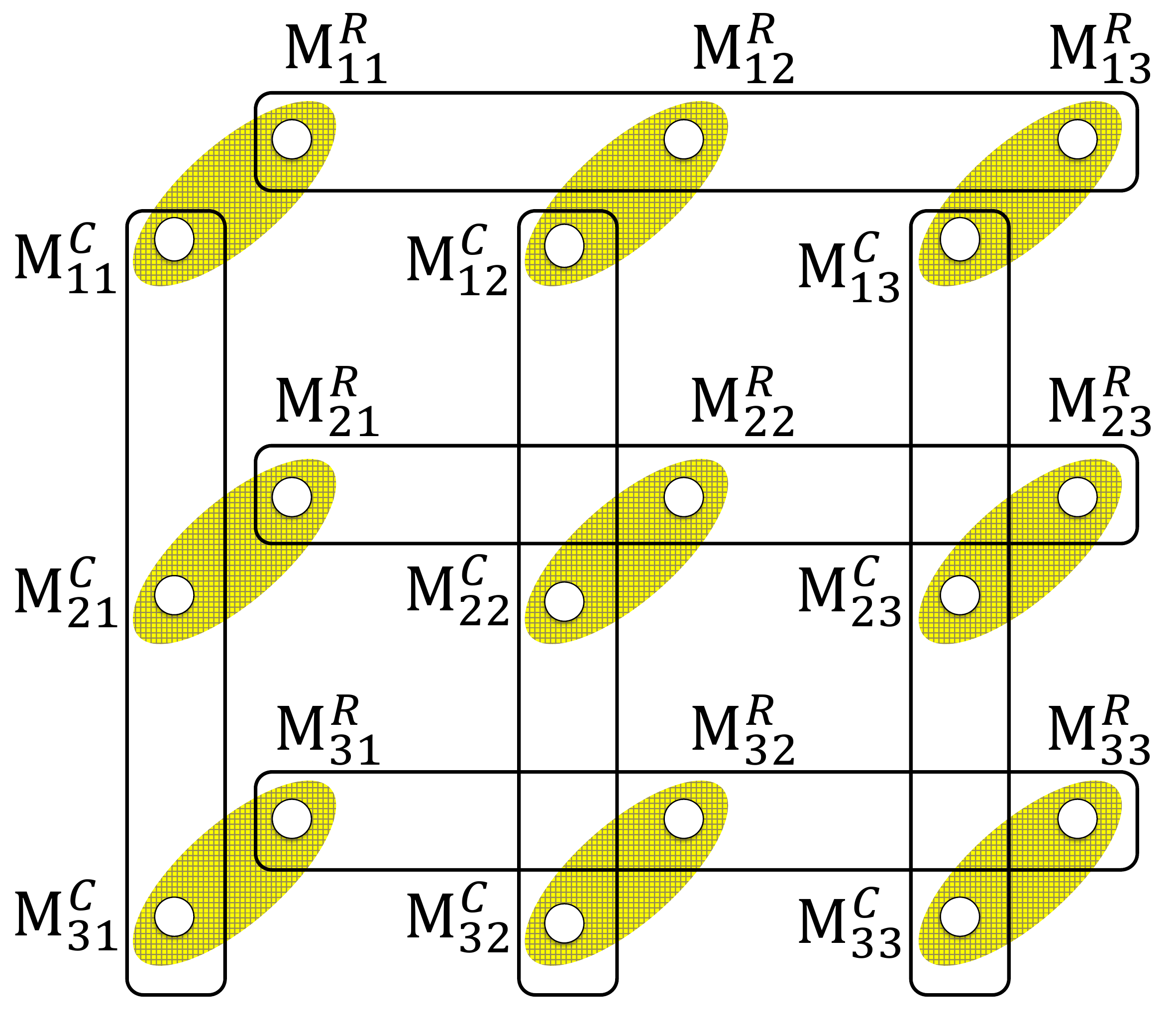}}
 \caption{The hypergraph representing compatibility and operational equivalence relations among the eighteen measurements $\{ \msmt{ij}^R\}_{ij} \cup \{ \msmt{ij}^C\}_{ij}$ defined in the text. }
 \label{CompatEquivHypergraphMmts}
\end{figure}

%First, consider Here are the measurement procedures that are to be targeted in an experiment.  We wish to implement the
There are eighteen target measurement procedures, corresponding to six compatible triples of binary-outcome measurements, one triple for each row and column of the square.  Recall that we denoted the 4-outcome measurement that simulates all of the binary-outcome measurements in the first row by $\msmt{R_1}$ and its outcome by the pair of binary-outcome variables $(m_{R_1,1},m_{R_1,2})$.   Recall also that we defined  $\msmt{11}^{R}$ to be the binary-outcome measurement procedure one obtains by implementing $\msmt{R_1}$ and outputting the single binary variable $m^R_{11} =  m_{R_1,1}$ (i.e., by marginalizing over  $m_{R_1,2}$).
Similarly, we defined $\msmt{11}^{C}$ to be the binary-outcome measurement procedure that one obtains by implementing $M_{C_1}$ and outputing  the single binary variable $m^C_{11} =  m_{C_1,1}$ (i.e., by marginalizing over  $m_{C_1,2}$).
$\msmt{11}^R$ and $\msmt{11}^C$ are two of the target procedures.  Although they are distinct in the sense of involving different physical operations in the laboratory, they are requried to be operationally equivalent in order for the assumption of noncontextuality to have any nontrivial consequences.  Similar comments hold for the pair of procedures associated to each of the other points of the Peres-Mermin square.  Consequently, if instead of considering the hypergraph where the nodes are operational equivalence classes of measurement procedures and the hyperedges are compatibility relations as we did in Fig.~\ref{CompatHypergraphMmts}, we consider the hypergraph where the nodes are individual measurements, and there are two types of hyperedges, one denoting compatibility relations and the other denoting operational equivalence relations, then we can represent the set of target measurement procedures by Fig.~\ref{CompatEquivHypergraphMmts}.

%{\bf Note on what relations need to be tested.}
Note that all of the relationships of compatibility that hold among the measurements
%or among the sources
 are guaranteed by the manner in which those measurements
%and sources
 are implemented.  Specifically, for every set of compatible measurements
%(sources)
 in the experiment, these are implemented by various coarse-grainings of
a single measurement.
% (source).
As such, the compatibility relations are ensured by
construction, and no further evidence must be accumulated to confirm their presence.
Only the operational equivalence relations among these different coarse-grained
measurements
%(sources)
 must be tested explicitly.

This distinguishes our approach
from other approaches to experimental tests of noncontextuality \cite{Kirchmair}
wherein the compatibility relations must be tested explicitly.  In the latter
approaches, the experiment seeks to implement a single measurement procedure
$\msmt{}$ before or after another measurement $\msmt{}'$ which is drawn from a set of possibilities that are
compatible with $\msmt{}$ but incompatible with one another.  It is then critical to
demonstrate that the procedures $\msmt{}$ and $\msmt{}'$ that are actually realized in the
experiment are indeed compatible.
%A criticism of the latter approach is provided in appendix \ref{subsec:prevProposals}.

\begin{figure}[t!]
 \centering
 {
 \includegraphics[scale=0.25]{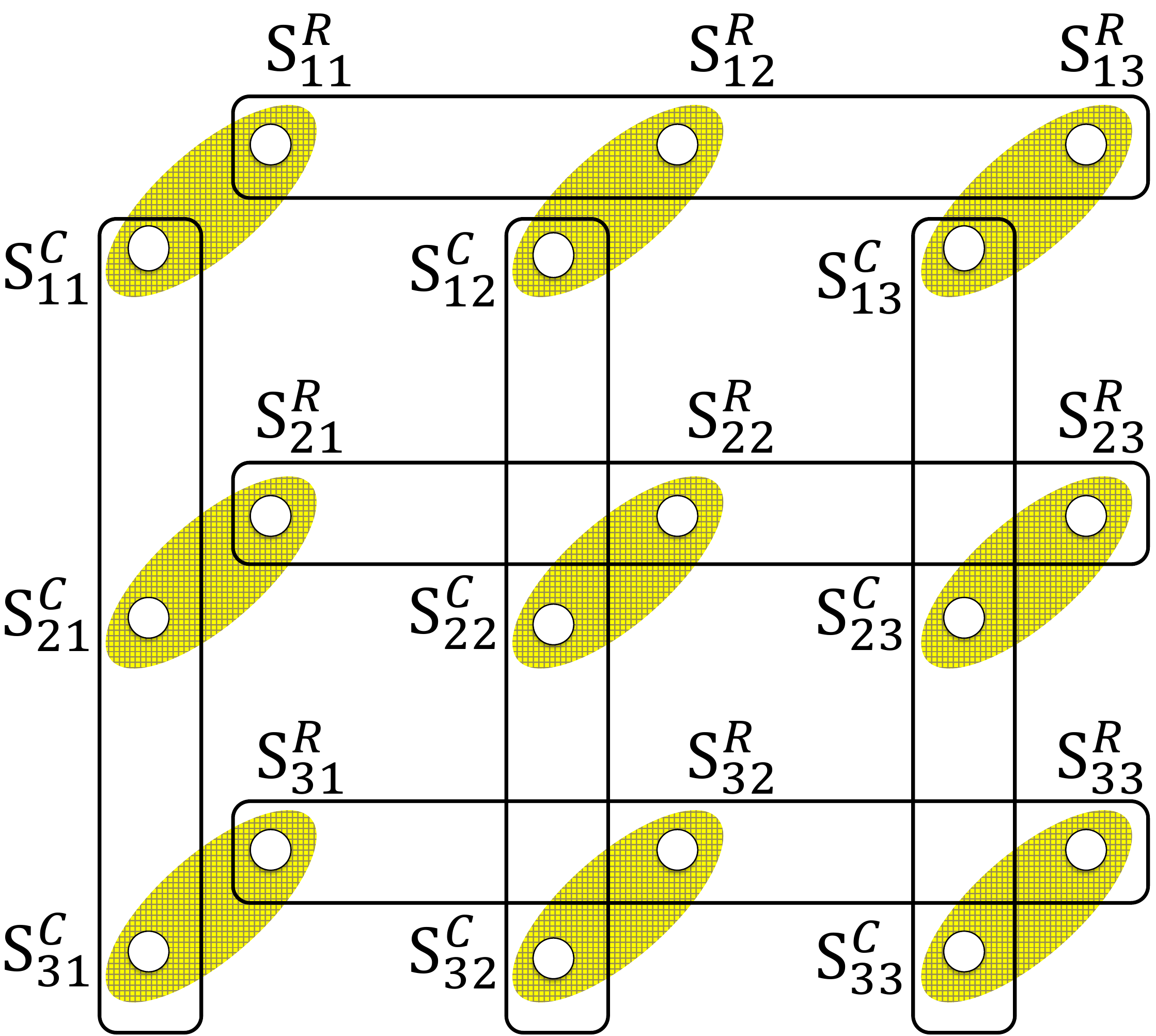}}
 \caption{The hypergraph representing compatibility and operational equivalence relations among the eighteen sources $\{ \src{ij}^R\}_{ij} \cup \{ \src{ij}^C\}_{ij}$ defined in the text. }
 \label{CompatEquivHypergraphSrcs}
\end{figure}

Similar comments apply to sources.  In a hypergraph where the nodes are individual sources, and there are two types of hyperedges, one denoting compatibility relations and the other denoting operational equivalence relations, we represent the set of target sources by Fig~\ref{CompatEquivHypergraphSrcs}.  If every set of compatible sources is implemented by a coarse-graining of
a single source, then the compatibility relations among the sources are achieved by construction and only the operational equivalence relations need to be tested.

We are now in a position to describe the problem of no strict operational equivalences in the case of Peres-Mermin.
Given the unavoidable deviation of the actually-realized procedures (the primary procedures) from the target procedures, one expects that
%the operational equivalences will generally not hold for the primary procedures that were realized in the experiment.
for each $ij$, the actually-implemented versions of $\msmt{ij}^R$ and of $\msmt{ij}^C$ will not be strictly operationally equivalent, nor will the actually-implemented versions of $\src{ij}^R$ and of $\src{ij}^C$.

The resolution to this problem was provided in Ref.~\cite{MazurekEtAl} and proceeds as follows.  When one's experiment realizes a finite set of primary procedures, it simultaneously provides a characterization of an infinite set of procedures, namely, those that can be obtained by classical post-processing of the set of primary procedures, for instance, any procedure in the convex hull of these.
%.  For instance, one can infer the characterization of any convex mixture of the primary procedures.
One can then choose a set of {\em secondary} procedures from among this infinite set  under the constraint of {\em exactly satisfying} the desired operational equivalence relations.

Specifically, we select secondary versions of the 4-outcome measurements and 4-outcome sources, under the constraint that the binary-outcome measurements and sources that they define precisely satisfy the operational equivalence relations depicted in the hypergraphs of Figs.~\ref{CompatEquivHypergraphMmts} and \ref{CompatEquivHypergraphSrcs} respectively.

Once this is done, one uses the secondary versions of the nine binary-outcome measurements and nine binary-outcome sources to compute the correlations $\{\omega_{ij}\}_{ij}$.  Because the operational equivalences hold for these secondary measurements and sources, the assumption of noncontextuality implies a constraint on their ontological representation, and consequently the operational correlations exhibited by these secondary measurements and sources are expected to satisfy the noncontextuality inequalities if the experiment admits of a noncontextual model.  A violation of these inequalities by the secondary procedures, therefore, witnesses a failure of noncontextuality.

\section{Conclusions}\label{sec:Conclusions}

In this paper, we have derived a set of noncontextuality inequalities based on the
Peres-Mermin square proof of the impossibility of a KS-noncontextuality in quantum theory.
These inequalities are robust to noise and consequently they can be tested directly by experiment.
If they are found to be violated by experiment, then not only quantum theory, but any physical theory that can do justice to the experimental data must fail to admit of a noncontextual model.
The procedure we have outlined for deriving such inequalities is quite general and can be applied to other state-independent proofs of the Kochen-Specker theorem, particularly those that are expressed algebraically.

In \cref{subsec:prevProposals}, we contrast our approach with previous attempts at deriving operationally-meaningful noncontextuality
inequalities based on the Peres-Mermin square and we criticize the latter.
%highlight some of their shortcomings.

The technique we have described here can be applied to deriving inequalities for correlations in the full set of 81 source-measurement pairings, or even to deriving inequalities on these together with the nine marginal expectation values for the measurements alone and the nine marginal expectation values for the sources alone.
What prevented us from doing so here was the computational infeasibility of solving the associated convex hull problems, using the best algorithms for this problem that we could identify and given the computational resources we devoted to the task.  It is conceivable, however, that by leveraging our knowledge of the symmetries of the polytopes involved in the problem, one might render it computationally feasible using the same algorithms and computational resources.
%The only obstacle to extending our technique to deriving constraints on all of the experimental data (both the full set of correlations between preparations and measurements and the marginal expectations on the preparation side or the measurement side alone) is computational infeasibility.
We are also hopeful that the graph-theoretic techniques for analyzing contextuality scenarios, described in Refs.~\cite{CSW, AFLS, KunjwalJointMeasurability}, might suggest better algorithms for deriving these inequalities.
%provide a path forward.

It is worth noting that the technique for deriving noncontextuality inequalities we have introduced here, insofar as it reduces to a convex hull problem, is an instance of the problem of quantifier elimination. Recent work in quantum foundations has seen increasing use of quantifier elimination algorithms.  Fourier-Motzkin elimination, which is appropriate for problems wherein the dependence on the variables to be eliminated is linear, has been used to derive Bell inequalities \cite{Budroni2012}, and also recently, to derive Bell-like inequalities for novel causal scenarios~\cite{ChavesGrossLuft, Chaves,inflation}.  In Ref.~\cite{inflation}, where the problem is reduced to what is known as the classical marginals problem---that of determining whether a given set of distributions on various subsets of a set of variables can arise as the marginals of a single joint distribution over all of the variables---and this problem can be solved by performing quantifier elimination on the probabilities in the joint distributions using convex hull algorithms. Nonlinear quantifier elimination using computational algebraic geometry has also found application in deriving Bell-like inequalities in simple scenarios~\cite{LeeSpekkens}.  We anticipate that these more general techniques for quantifier elimination will ultimately also find applications to the derivation of noncontextuality inequalities.

\begin{acknowledgments}
The authors thank D. Schmid for helpful comments.  This research was supported in part by Perimeter Institute for Theoretical Physics. Research at Perimeter Institute is supported by the Government of Canada through the Department of Innovation, Science and Economic Development Canada and by the Province of Ontario through the Ministry of Research, Innovation and Science.
\end{acknowledgments}

%\clearpage
\setlength{\bibsep}{2pt plus 1pt minus 2pt}
\bibliographystyle{apsrev4-1}
\nocite{apsrev41Control}
\bibliography{references}%{}

%\onecolumngrid
%\clearpage
\appendix

\renewcommand{\theequation}{\Alph{section}.\arabic{equation}}
\setcounter{equation}{0}

\section{Polytope Representation Conversion}\label{sec:repconversion}

%\section{Polytope Representation Conversion}\label{sec:repconversion}
The problem of converting between polytope representations is essential to our technique for deriving noncontextuality inequalities. We describe the abstract mathematical problem here, and refer the reader to various software packages which are capable of solving it efficiently.
%performing polytope representation conversion.
A polytope may be uniquely characterized in terms of two dual descriptions: (1) as a collection of facet-defining inequalities, which is called the halfspace representation of the polytope, denoted \emph{H-rep}, or (2) as a collection of vertices, which is called the vertex representation of the polytope, denoted \emph{V-rep}. It turns out that facet enumeration from vertices and vertex enumeration from facets are exactly the same computational problem. To understand the representation conversion problem in general, it is imperative to first understand the computational data structures used to represent a collection of points or a collection of inequalities pertaining to $d$-dimensional vector spaces.
%coordinate systems. In other words, we answer the questions in what form should one input a V-rep or H-rep polytope to a computer as input, and in what form should the computer return the dual representation as output.

We elect to represent the inequality
\begin{align*}
0\leq c_0 + \sum\limits_{i=1}^d c_i x_i
\end{align*}
by the $(d+1)$-dimensional vector
\begin{align*}
\bm{h}:=\,\left\{\frac{c_0}{g},\frac{c_1}{g},...,\frac{c_d}{g}\right\}
\end{align*}
where $g$ is chosen to efficiently store $\bm{h}$ in computer memory. For rational coefficients, taking $g=\mbox{GCD}[c_0,...,c_d]$ allows $\bm{h}$ to be stored as a vector of integers\footnote{GCD stands for Greatest Common Divisor; here we are applying GCD to the field of rational numbers, see \href{http://functions.wolfram.com/introductions/PDF/GCD.pdf}{functions.wolfram.com/IntegerFunctions/GCD} and \href{http://math.stackexchange.com/questions/151081/gcd-of-rationals/151431}{math.stackexchange.com/q/151081}.}. Indeed, for the sort of contextuality problems that involve linear rational compatibility relationships (such as the operational Peres-Mermin scenario),
 %considered in the main text here, or that of Ref. \cite{KunjwalSpekkens}),
 we can always represent the pertinent inequalities by vectors of integers.

We elect to represent the point
\begin{align*}
%x_1 = p_1,\;x_2 = p_3,\;...,\;x_d = p_d
&\bm{x}:=\,\left\{x_1,...,x_d\right\}
%\end{align*}
\shortintertext{by the $(d+1)$-dimensional vector}
%\begin{align*}
&\bm{v}:=\,\left\{\frac{1}{g'},\frac{x_1}{g'},...,\frac{x_d}{g'}\right\}
\end{align*}
where $g'$ is chosen to efficiently store $\bm{v}$ in computer memory. For rational components, taking ${g'=\mbox{GCD}[x_1,...,x_d]}$ allows $\bm{v}$ to be stored as a vector of integers.
%Assuming that the point ${\bm x}$ has rational components, we can represent it as a vector $\bm{v}$ of integers by taking $g'=\mbox{GCD}[x_1,...,x_d]$.
%We can any represent $d$-dimensional point $\bm{x}$ with rational coefficients as a $(d+1)$-dimensional vector $\bm{v}$ of integers in this manner.

If a polytope has $n$ facet inequalities, then given that each of these can be represented by a ${(d\!+\!1)}$-dimensional vector, its halfspace representation can be given as an ${n\!\times\!(d\!+\!1)}$ matrix
\begin{align}
\bm{H}:=\begin{pmatrix}
\bm{h}^{(1)}\\
\vdots\\
\bm{h}^{(n)}
\end{pmatrix}.
\end{align}
%An H-rep matrix $\bm{H}$ (either input or output) is a list of inequalities, where each inequality is represented by a $(d+1)$-dimensional vector.
We refer to $\bm{H}=\{ {\bm h}^{(i)} \}_i$ as the H-rep matrix.

Similarly, if a polytope has $m$ vertices, then given that each of these can be represented by a ${(d\!+\!1)}$-dimensional vector, its vertex representation can be given as an ${m\!\times\!(d\!+\!1)}$ matrix
%The same is true for polytopes in vertex representation: a collection of $m$ points is an $m\times (d+1)$ matrix denoted
\begin{align}
\bm{V}:=\begin{pmatrix}
\bm{v}^{(1)}\\
\vdots\\
\bm{v}^{(m)}
\end{pmatrix}.
\end{align}
We refer to $\bm{V}=\{ {\bm v}^{(j)} \}_j$ as the V-rep matrix.
%An V-rep matrix $\bm{V}$ (either input or output) is a list of points, where each point is represented by a $(d+1)$-dimensional vector.
%To reiterate, for these problems wherein the compatibility relations involve only rational coefficients we are effectively concerned with integer matrices without loss of generality.

It follows that a facet enumeration algorithm takes a V-rep matrix as input and returns an H-rep matrix as output, while a vertex enumeration algorithm takes an H-rep matrix as input and returns a V-rep matrix as output.
%So, the algorithms for facet enumeration and vertex enumeration both take a collection of $(d+1)$-dimensional vectors and return a collection of $(d+1)$-dimensional vectors as output; we can
We now show that the algorithms are equivalent.

A set of point-vectors ${\bm V}=\{ {\bm v}^{(j)} \}_j$ defines the \mbox{V-rep} of a polytope relative to some \mbox{H-rep} \mbox{inequalities-matrix} ${\bm H}=\{ {\bm h}^{(i)} \}_i$ if and only if the following conditions hold:
\begin{compactdesc}
\item[Consistent] Every element of $\{ {\bm v}^{(j)} \}_j$ must satisfy all of the facet inequalities,\par
\hspace{\mathindent}\hspace{\parindent}${\forall i,j: \bm{h}^{(i)}\cdot\bm{v}^{(j)}\geq 0}$.
%$\forall j:\,{\bm H} \bm{v}^{(j)} \geq 0$, where the inequality is evaluated component-wise.
%${\forall_{i}:\, \bm{h}^{(i)}\cdot\bm{v}^{(j)}\geq 0}$.
\item[Complete] Every point ${\bm x}$ which cannot be expressed as a positive combination of the elements of $\{ {\bm v}^{(j)} \}_j$ must violate some facet inequality,\par
\hspace{\mathindent}\hspace{\parindent}$\forall {\bm{x}\not\in \operatorname{PosLinSpan}\left[  \{ {\bm v}^{(j)} \}_j  \right]}\,, \exists_i :\, \bm{h}^{(i)} \cdot \bm{x} < 0$.
%\exists i :\,\bm{h}^{(i)} \cdot \tilde{\bm{v}} <0}$.
%${\forall_{\bm{v}}\not\exists_{\bm{w}\geq\bm{0}\text{ s.t. }\bm{v}=\bm{w}\bm{V}}\,:\,\exists_i \bm{H}_i \cdot \bm{v} <0}$.
\item[Concise] No element of $\{ {\bm v}^{(j)} \}_j$ is
%No point in the (output) collection of vertices may be
a positive combination of the others,\par
\hspace{\mathindent}\hspace{\parindent}${\forall j :\, \bm{v}^{(j)} \not\in \operatorname{PosLinSpan}\left[ \{ {\bm v}^{(k)} \}_{k\ne j} \right]}$.
%${\forall_j \not\exists_{\bm{w}\geq\bm{0},w_j=0}:\bm{v}^{(j)}=\bm{w}\bm{V}}$.\par
\end{compactdesc}

%Starting from some H-rep, the vertex enumeration task is to find the vertices, i.e. the set of points satisfying all three following properties:
%\begin{compactdesc}
%\item[Consistent] Every point in the (output) collection of vertices satisfies all the (input)  inequalities, ${\forall_{i,j}:\, \bm{h}^{(i)}\cdot\bm{v}^{(j)}\geq 0}$.
%\item[Complete] Every point which cannot be expressed as a positive combination of the points in the (output) collection of vertices should violate some (input) inequality, ${\forall_{\tilde{\bm{v}}\not\in \operatorname{PosLinSpan}\left[\bm{V}\right]}:\,\exists_i :\,\bm{h}^{(i)} \cdot \tilde{\bm{v}} <0}$.
%\item[Extremal] No point in the (output) collection of vertices may be a positive linear combination of the others, ${\forall_j :\, \bm{v}^{(j)} \not\in \operatorname{PosLinSpan}\left[\bm{V}\setminus \bm{v}^{(j)}\right]}$.
%\end{compactdesc}
\smallskip
A set of inequality-vectors ${\bm H}=\{ {\bm h}^{(i)} \}_i$ defines the \mbox{H-rep} of a polytope relative to some \mbox{V-rep} \mbox{vertices-matrix} ${\bm V}=\{ {\bm v}^{(j)} \}_j$ if and only if the following conditions hold:
\begin{compactdesc}
\item[Consistent]
%All of the elements of $\{ {\bm h}^{(i)} \}_j$ must satisfy the facet inequalities, $\forall j:\,{\bm H} \bm{v}^{(j)} \geq 0$, where the inequality is evaluated component-wise.
Every inequality in the set $\{ {\bm h}^{(i)} \}_i$ is satisfied by all of the vertices of the polytope,\par
\hspace{\mathindent}\hspace{\parindent}
${\forall i,j: \bm{h}^{(i)}\cdot\bm{v}^{(j)}\geq 0}$.
\item[Complete] Any inequality ${\bm c}$ which cannot be expressed as a positive combination of the inequalities $\{ {\bm h}^{(i)} \}_i$ must be violated by at least one vertex of the polytope,\par
\hspace{\mathindent}\hspace{\parindent} ${\forall {\bm{c}\not\in \operatorname{PosLinSpan}\left[ \{ {\bm h}^{(i)} \}_i \right]},\,\exists j :\,\bm{c}\cdot\bm{v}^{(j)}  <0}$.
\item[Concise] No inequality in $\{ {\bm h}^{(i)} \}_i$ is a positive combination of the others,\par
\hspace{\mathindent}\hspace{\parindent} ${\forall i :\, \bm{h}^{(i)} \not\in \operatorname{PosLinSpan}\left[ \{ {\bm h}^{(k)} \}_{k\ne i} \right]}$.
%${\forall_i\not\exists_{\bm{w}\geq\bm{0},w_i=0}:\bm{h}^{(i)}=\bm{w}\bm{H}}$.\par
\end{compactdesc}

%\noindent Starting from some V-rep, the facet enumeration task is to find the \emph{complete} set of facets, i.e. the set of inequalities satisfying all three following properties
%\begin{compactdesc}
%\item[Consistent] Every (input) vertex satisfies each inequality in our (output) collection of facets, ${\forall_{i,j}: \bm{h}^{(i)}\cdot\bm{v}^{(j)}\geq 0}$.
%\item[Complete] Any inequality which cannot be expressed as a positive combination of the inequalities in the (output) collection of facets should be violated by at least one (input) vertex, ${\forall_{\tilde{\bm{h}}\not\in \operatorname{PosLinSpan}\left[\bm{H}\right]}:\,\exists_j :\,\tilde{\bm{h}}\cdot\bm{v}^{(j)}  <0}$.%$\item[Extremal] No inequality in the (output) collection of facets may be a positive linear combination of the others, ${\forall_i :\, \bm{h}^{(i)} \not\in \operatorname{PosLinSpan}\left[\bm{H}\setminus \bm{h}^{(i)}\right]}$.
%\end{compactdesc}

It should now be clear that the conditions which define the dual representation are symmetric with respect to the interchange of $\bm{H}$ and $\bm{V}$. Consequently, the polytope representation conversion task is a single computational problem, oblivious to the subjective interpretation of the task as facet enumeration or vertex enumeration. The common underlying computational task is conventionally referred to as the \emph{convex hull} problem.

In the general procedure for deriving robust operational noncontextuality inequalities, one encounters the convex hull problem three times:
\begin{asparaenum}
\item When one enumerates the vertices of the noncontextual measurement-assignment polytope starting from its facet inequalities (\ref{mmtconstraints1}-\ref{mmtconstraints2}), one is performing H-rep to V-rep conversion.
%For the Peres-Mermin problem, we are converting a list of 24 inequalities into list of 120 vertices, both lists being equivalent representations of the polytope.
\item When one enumerates the vertices of the noncontextual source-assignment polytope starting from its facet inequalities (\ref{srcconstraints1}-\ref{srcconstraints2}), one is performing H-rep to V-rep conversion.
%For this problem, once again, the polytope's 24 facet inequalities are converted into its 120 vertices.
\item When one enumerates the facets of the noncontextual correlation polytope starting from its vertices, one is performing V-rep to H-rep conversion.
%The 120 vertices of the polytope of noncontextual correlations (per \cref{sec:verticesCpolytope}, not to be confused with the 120 vertices of either ontic-assignments polytope) are dual to the 184 operational noncontextuality inequalities, which comprise the ultimate goal of \cref{sec:verticesCpolytope}).
\end{asparaenum}
For the problem of interest in this article, the Peres-Mermin scenario, the noncontextual measurement-assignment polytope coincides with the noncontextual source-assignment polytope, so the first two instances of the convex hull problem are the same, converting  24 facet inequalities into 120 vertices.  The third instance concerns
%In the last instance of the convex hull problem,
 the noncontextual correlation polytope, and converts the 120 vertices thereof into 184 facet inequalities.
% of noncontextual correlations (per \cref{sec:verticesCpolytope}, not to be confused with the 120 vertices of either ontic-assignments polytope) are dual to the 184 operational noncontextuality inequalities, which comprise the ultimate goal of \cref{sec:verticesCpolytope}).

Solving the convex hull problem can, in practice, be accomplished by a variety of algorithms; the most prominent implementation is the dual description method. For details on the various algorithms used to solve the convex hull problem see Refs.~\cite{Fukuda1996,barber1996quickhull,Zolotykh2012,avis_convexhull_2015}. Efficient convex hull solvers are abundant in current software packages.\footnote{\textbf{\href{http://comopt.ifi.uni-heidelberg.de/software/PORTA}{\texttt{PORTA}}:}
 [Executable]  Older software, limited to moderate dimension polytopes $(d\lesssim 100)$, nevertheless \href{http://comopt.ifi.uni-heidelberg.de/software/PORTA}{\texttt{PORTA}} is quite fast.
\textbf{\href{https://www.inf.ethz.ch/personal/fukudak/cdd_home}{\texttt{CDD}}, \href{http://cgm.cs.mcgill.ca/~avis/C/lrs.html}{\texttt{LRS}}} \& \textbf{\href{http://www.math.uni-rostock.de/~rehn/software/sympol.html}{\texttt{SymPol}}:}
 [Executables \& C-libraries] All accept the same style of input file. The \href{https://www.inf.ethz.ch/personal/fukudak/cdd_home}{\texttt{CDD}} binary is no longer being updated, but it's quite fast. \href{http://cgm.cs.mcgill.ca/~avis/C/lrs.html}{\texttt{LRS}} uses a different internal algorithm than \href{https://www.inf.ethz.ch/personal/fukudak/cdd_home}{\texttt{CDD}}; sometimes it is much faster, sometimes much slower. \href{http://www.math.uni-rostock.de/~rehn/software/sympol.html}{\texttt{SymPol}} can automatically discover symmetries to reduce the size of the computation, and is the authors' preferred tool for solving large convex hull problems in the presence of high symmetry, such as is very often the case in noncontextuality polytopes.
%\textbf{{http://www.cs.unb.ca/~bremner/software/pd}{\texttt{PD}}:} [Standalone binary]  Accepts \href{https://www.inf.ethz.ch/personal/fukudak/cdd_home}{\texttt{CDD}}-style input files. No longer actively maintained.
\textbf{\href{http://www.qhull.org/}{\texttt{Qhull}}:} [Multiple interfaces] Widely used, but suboptimal for high dimensional polytopes. %Native implementation in \href{http://www.mathworks.com/help/matlab/ref/convhulln.html}{\texttt{MATLAB}}  and \href{https://octave.sourceforge.io/octave/function/convhulln.html}{\texttt{Octave}}.
\textbf{\href{http://www.uic.unn.ru/~zny/skeleton}{\texttt{Skeleton}}} \& \textbf{\href{http://sbastrakov.github.io/qskeleton}{\texttt{Qskeleton}}:} [Executables] The fastest convex hull solvers in the authors' informal benchmarking; both accept the same style of input file.
%\textbf{\href{http://people.ee.ethz.ch/~mpt/3}{\texttt{MPT3}}:} [MATLAB addon] Extremely efficient.
\textbf{\href{https://pypi.python.org/pypi/polytope}{\texttt{polytope}}:} [Python package] Polyhedral geometry software with good convex hull performance if used with \href{https://scaron.info/blog/linear-programming-in-python-with-cvxopt.html\#comparing-solver-performances}{\texttt{GLPK}} bindings.
%\item[\href{http://www.sagemath.org}{\texttt{SAGE}}:} [Python-based] Built-in class ``polyhedron.representation'' efficiently handles rational polytope representation conversion.
\textbf{\href{https://polymake.org}{\texttt{polymake}}:} [Software for Linux \& Mac] Versatile polyhedral geometry software, capable of much more than just convex hull. \href{https://polymake.org}{\texttt{polymake}} offers selecting amongst multiple internal convex hull algorithms.
%\textbf{\href{http://mathieudutour.altervista.org/Polyhedral}{\texttt{Polyhedral}}:} [GAP package]  Automatically exploits symmetries to reduce the size of the computation.
%\textbf{\href{http://www.math.uni-rostock.de/~rehn/software/sympol.html}{\texttt{SymPol}}:} [standalone binary] Can automatically discover symmetries to reduce the size of the computation. Accepts \href{https://www.inf.ethz.ch/personal/fukudak/cdd_home}{\texttt{CDD}}-style input files. \href{http://www.math.uni-rostock.de/~rehn/software/sympol.html}{\texttt{SymPol}} is the authors' preferred tool for solving large convex hull problems in the presence of high symmetry, such as is very often the case in noncontextuality polytopes.
\textbf{\href{http://comopt.ifi.uni-heidelberg.de/software/PANDA}{\texttt{PANDA}}:} [Executable \& C-library] Given a description of known symmetries, \href{http://comopt.ifi.uni-heidelberg.de/software/PANDA}{\texttt{PANDA}} is capable of converting polytope representations extremely efficiently; the symmetries are specified relative to the polytope's coordinate system.
%\end{asparadesc}
}

Note that when we infer the vertices of the noncontextual correlation polytope from the superset of all possible pairings of vertices from the noncontextual measurement-assignment polytope and the noncontextual source-assignment polytope, the computational problem is one of redundancy removal. Redancy removal is the task of isolating the vertices from a set of non-extremal points, or, computationally equivalently, isolating the irredundant linear constraints from some inconcise collection of inequalities. Standard algorithms have been developed for redundancy removal \cite{RedRomoval1989Caron,RedRemoval1993Matousek}, and a variety of current software tools are also available.\footnote{The authors customized a redundancy removal script to utilize \href{http://www.mosek.com}{\texttt{MOSEK}}'s efficient linear programming features; code available upon request. Filtering for extremality is also available natively in many polyhedral geometry software tools. A single-purpose freely-available tool is the \href{http://cgm.cs.mcgill.ca/~avis/C/lrslib/USERGUIDE.html\#redund}{\texttt{redund}} binary which ships with \href{http://cgm.cs.mcgill.ca/~avis/C/lrs.html}{\texttt{LRS}}.}
%Another computational task in polyhedral geometry that is related to our discussion in this article is redundancy removal.

%When we construct the vertices of the correlation polytope, initially the set of points we construct includes non-extremal points interspersed alongside the genuine vertices, i.e. the non-extremal points can be expressed as a positive linear combinations of the vertices. This is not terribly problematic in any way, but it is optimal to remove the non-extremal points before submitting the V-rep matrix to a convex hull algorithm\footnote{The authors used a script which relied on \href{http://www.mosek.com}{\texttt{MOSEK}}'s linear programming features to filter out the non-extremal points so as to isolate the vertices. Many polyhedral geometry software tools can natively filter for extremality. A single-purpose freely-available tool is the \texttt{redund} binary which ships with \href{http://cgm.cs.mcgill.ca/~avis/C/lrs.html}{\texttt{LRS}}.}.

\section{Deterministic processings of the experimental procedures that preserve the compatiblity relations of the Peres-Mermin scenario}\label{symmetries}

When we express our noncontextuality inequalities in Sec.~\ref{sec:NCInequalities}, the description of the full set of such inequalities is simplified by appealing to the symmetries of the operational construction.   These symmetries correspond to processings of the experiment that permute source-measurement pairs (i.e., deterministically process the setting variables in a correlated fashion), and that deterministically process the outcomes of the measurements and (independently) the outcomes of the sources.  It is therefore useful to explicitly characterize these symmetries.

Suppose one has an experiment satisfying the operational features of the Peres-Mermin scenario.  This means that there are nine equivalence classes of binary-outcome measurements, $\{\mathcal{M}_{11},\dots,\mathcal{M}_{33}\}$, satisfying the compatibility hypergraph of Fig.~\ref{CompatHypergraphMmts} and the compatibility relations defined
around Eqs.~(\ref{cc1}-\ref{cc3}).
Similarly, there are nine equivalence classes of binary-outcome sources, $\{\mathcal{S}_{11},\dots,\mathcal{S}_{33}\}$, satisfying the compatibility hypergraph of Fig.~\ref{CompatHypergraphSources} and the compatibility relations defined around Eqs.~(\ref{ss1}-\ref{ss3}).

Now suppose that from these one defined a new pair of sets, $\{\mathcal{M}'_{11},\dots,\mathcal{M}'_{33}\}$
%a new set $\{\mathcal{M}'_{11},\dots,\mathcal{M}'_{33}\}$ as a processing of $\{\mathcal{M}_{11},\dots,\mathcal{M}_{33}\}$, and
and $\{\mathcal{S}'_{11},\dots,\mathcal{S}'_{33}\}$, as processings of $\{\mathcal{M}_{11},\dots,\mathcal{M}_{33}\}$ and $\{\mathcal{S}_{11},\dots,\mathcal{S}_{33}\}$ as follows.  To implement the source-measurement pair $(\mathcal{S}'_{ij},\mathcal{M}'_{ij})$, one begins by implementing the source-measurement pair $(\mathcal{S}_{\pi(ij)},\mathcal{M}_{\pi(ij)})$, where $\pi$ is some permutation of the nine positions in the Peres-Mermin square, and then for each $\mathcal{M}_{\pi(ij)}$, one multiplies its outcome by $\zeta_{ij} \in \{-1,+1\}$ (thereby either flipping its value or leaving it the same) and for each $\mathcal{S}_{\pi(ij)}$, one multiplies its outcome by $\gamma_{ij} \in \{-1,+1\}$.

The deterministic processings by which the outcomes of the original measurements and sources are mapped to those of the primed measurements and sources can be expressed as follows.  Denoting component-wise product of matrices by $\circ$,
\begin{align}\label{relabelmmts}
&\left(\begin{matrix}
  \mathcal{m}'_{11}&       \mathcal{m}'_{12}   &       \mathcal{m}'_{13} \\
           \mathcal{m}'_{21}  &       \mathcal{m}'_{22}   &       \mathcal{m}'_{23} \\
      \mathcal{m}'_{31}  &       \mathcal{m}'_{32}   &       \mathcal{m}'_{33}
 \end{matrix}\right) \nonumber\\
 &\hspace{-\mathindent}=
 \left(\begin{matrix}
  \zeta_{11}&       \zeta_{12}   &       \zeta_{13} \\
           \zeta_{21}  &       \zeta_{22}   &       \zeta_{23} \\
      \zeta_{31}  &       \zeta_{32}   &       \zeta_{33}
 \end{matrix}\right)
 \circ \left(\begin{matrix}
  \mathcal{m}_{\pi(11)}&       \mathcal{m}_{\pi(12)}   &       \mathcal{m}_{\pi(13)} \\
           \mathcal{m}_{\pi(21)}  &       \mathcal{m}_{\pi(22)}   &       \mathcal{m}_{\pi(23)} \\
      \mathcal{m}_{\pi(31)}  &       \mathcal{m}_{\pi(32)}   &       \mathcal{m}_{\pi(33)}
 \end{matrix}\right)
\end{align}
 and
 \begin{align}\label{relabelsrcs}
&\left(\begin{matrix}
  \mathcal{s}'_{11}&       \mathcal{s}'_{12}   &       \mathcal{s}'_{13} \\
           \mathcal{s}'_{21}  &       \mathcal{s}'_{22}   &       \mathcal{s}'_{23} \\
      \mathcal{s}'_{31}  &       \mathcal{s}'_{32}   &       \mathcal{s}'_{33}
 \end{matrix}\right) \nonumber\\
 &\hspace{-\mathindent}=
 \left(\begin{matrix}
  \gamma_{11}&       \gamma_{12}   &       \gamma_{13} \\
           \gamma_{21}  &       \gamma_{22}   &       \gamma_{23} \\
      \gamma_{31}  &       \gamma_{32}   &       \gamma_{33}
 \end{matrix}\right)
 \circ \left(\begin{matrix}
  \mathcal{s}_{\pi(11)}&       \mathcal{s}_{\pi(12)}   &       \mathcal{s}_{\pi(13)} \\
           \mathcal{s}_{\pi(21)}  &       \mathcal{s}_{\pi(22)}   &       \mathcal{s}_{\pi(23)} \\
      \mathcal{s}_{\pi(31)}  &       \mathcal{s}_{\pi(32)}   &       \mathcal{s}_{\pi(33)}
 \end{matrix}\right)
\end{align}

% $\pi$ is a permutation of the nine measurements, and $\zeta_{11},\zeta_{12},\dots,\zeta_{33} \in \{ -1,+1\}$ are parameters specifying whether each outcome is flipped or not.

We must determine what constraints on $\pi$, $\{ \zeta_{ij}\}$ and $\{ \gamma_{ij}\}$ are implied by the requirement that the primed measurements have the compatibility relations of the Peres-Mermin scenario.

We begin by deriving some necessary conditions on the permutation $\pi$.
%considering what constraints on the permutation is implied by the requirement that the primed measurements have the appropriate compatibility relations.
To have the right compatibility structure, it is necessary that each of the rows and columns of primed measurements in the Peres-Mermin square  (and those of the primed sources) must constitute a compatible triple.  It is necessary, therefore, that each must be the image, under the permutation $\pi$, of a compatible triple in the original set of nine measurements (respectively sources), and given that the only triples that are compatible in the original set are the rows and columns, it follows that each must be the image of a row or column of the original.  The overall permutation must therefore be generated by permutations of the rows, permutations of the columns, and the transpose of the Peres-Mermin square.

But this condition is not sufficient.  We also require the particular compatibility {\em relations} among the nine primed measurements (and sources) to be those specified in the operational Peres-Mermin scenario. That is, we require that there exist six 4-outcome measurements $\msmt{R_1}', \msmt{R_2}',\dots, \msmt{C_3}'$ such that primed versions of Eqs.~(\ref{cc1}-\ref{cc3}) hold for the first row, and its analogues hold for the other rows and the first two columns, and primed versions of Eqs.~(\ref{comp1}-\ref{comp3}) hold for the third column, and we require that there exist six 4-outcome sources $\src{R_1}', \src{R_2}',\dots, \src{C_3}'$ such that primed versions of Eqs.~(\ref{ss1}-\ref{ss3}) hold for the first row, and its analogues hold for the other rows and the first two columns, and primed versions of Eqs.~(\ref{sss1}-\ref{sss3}) hold for the third column.

We require that for any triple corresponding to a row or a column in the primed Peres-Mermin square, the product of the outcomes in any given run of the experiment must be the same as the product of the outcomes from the corresponding triple in the original square.  In other words, the deterministic processing is required to satisfy
\begin{align}
\classicalOutputMsmtEqv{a}' \classicalOutputMsmtEqv{b}'\classicalOutputMsmtEqv{c}' = \classicalOutputMsmtEqv{a}\classicalOutputMsmtEqv{b}\classicalOutputMsmtEqv{c}.
\end{align}
% where $\zeta_a,\zeta_b,\zeta_c \in \{ -1,+1\}$.  A relabelling preserves the compatibility relations among the measurements, Eq.~\eqref{tripleproducts} if and only if it preserves the product of the the three variables, consequently, if and only if,
for all $(a,b,c)$ corresponding to compatible triples. Given Eq.~\eqref{relabelmmts},
it follows that
\begin{align}
 \zeta_a  \zeta_b \zeta_c = \frac{\classicalOutputMsmtEqv{a}\classicalOutputMsmtEqv{b}\classicalOutputMsmtEqv{c}}{\classicalOutputMsmtEqv{\pi(a)} \classicalOutputMsmtEqv{\pi(b)}\classicalOutputMsmtEqv{\pi(c)}},
\end{align}

Now note that the compatibility relations imply that for each of the nine compatible triples of measurements, the associated triple of outcomes that one can obtain in any run of the experiment has a fixed product,
 %variables one obtains always has a particular product in every run of the experiment,
 namely,
\begin{align}\begin{split}\label{tripleproducts}
 \classicalOutputMsmtEqv{11} \classicalOutputMsmtEqv{12} \classicalOutputMsmtEqv{13} &=+1,\\
 \classicalOutputMsmtEqv{21} \classicalOutputMsmtEqv{22} \classicalOutputMsmtEqv{23} &=+1,\\
 \classicalOutputMsmtEqv{31} \classicalOutputMsmtEqv{32} \classicalOutputMsmtEqv{33} &=+1,\\
 \classicalOutputMsmtEqv{11} \classicalOutputMsmtEqv{21} \classicalOutputMsmtEqv{31} &=+1,\\
 \classicalOutputMsmtEqv{12} \classicalOutputMsmtEqv{22} \classicalOutputMsmtEqv{32} &=+1,\\
 \classicalOutputMsmtEqv{13} \classicalOutputMsmtEqv{23} \classicalOutputMsmtEqv{33} &=-1.
\end{split}\end{align}
(To be clear, we are not here assuming that the particular triple of values that are assigned to, say, $\classicalOutputMsmtEqv{11}, \classicalOutputMsmtEqv{12},$ and $\classicalOutputMsmtEqv{13}$, need to be assigned deterministically.  Even if the assignment is indeterministic, so that this triple of values is sampled from some probability distribution, the compatibility relations defining the Peres-Meremin construction dictate that only those triples of values that satisfy $\classicalOutputMsmtEqv{11} \classicalOutputMsmtEqv{12} \classicalOutputMsmtEqv{13} =1$ have a nonzero probability of occurrence.)

Consequently, if the permutation $\pi$ is such that the third column is mapped to itself, then for all triples,
 \begin{align}\label{alltriplesplus}
 \zeta_a  \zeta_b \zeta_c = +1.
 \end{align}

If, on the other hand, the permutation $\pi$ maps the third column to a different triple,
then
 \begin{align}\begin{split}\label{twotriplesminus}
 &\hspace{-\mathindent}\zeta_a  \zeta_b \zeta_c = \begin{dcases}-1 & \!\!\!\text{for this distinguished pair of triples}\\
  +1 & \!\!\!\text{for the other four triples.}\end{dcases}
 \end{split}\end{align}
The same constraints hold for the $\gamma$'s.

It is useful at this stage to distinguish three sorts of permutations of the Peres-Mermin square. Some notation is required to do so. Let the 2-cycle that interchanges the $(ij)$th element of the square with the $(i'j')$th element be denoted $((ij)(i'j))$.  Let the permutation of the first and second row be denoted by $(R_1 R_2)$, and similarly for other cases, that is,  $(R_1 R_2)= ((11)(21))((12)(22))((13)(23))$), etcetera.  Let the transpose of the Peres-Mermin square be denoted $T$, that is,  $T=((13)(31))((12)(21))((23)(32))$. Finally, let the group closure of a set of permutations be denoted ``${\rm clos}$''.

The set of all permutations of the Peres-Mermin square that map the third column to itself is:
\begin{align}\begin{split}
\mathfrak{P}_A := \Big\{ \pi = &\tilde{\pi}_C \pi_R : \\
 &\pi_R \in {\rm clos}\{ (R_1 R_2) , (R_2 R_3)\},\\
 &\tilde{\pi}_C \in {\rm clos}\{ (C_1 C_2)\} \Big\}.
\end{split}\end{align}
Here, $ \pi_R$ runs over all permutations of the rows, and $\tilde{\pi}_C$ is either identity or the swap of the first two columns.

Of those permutations that do not take the third column to itself, it is useful to distinguish those that take it to another column, denoted $\mathfrak{P}_B$, and those that take it to a row, denoted $\mathfrak{P}_C$.  These are defined as follows:
\begin{align}\begin{split}
\mathfrak{P}_B:= \Big\{ \pi = &\tilde{\pi}_C \pi_R (C_2 C_3) : \\
& \pi_R \in {\rm clos}\{ (R_1 R_2) , (R_2 R_3)\},\\
& \tilde{\pi}_C \in {\rm clos}\{ (C_1 C_2)\} \Big\},
\end{split}\end{align}
where again $ \pi_R$ runs overs all permutations of the rows, and $\tilde{\pi}_C$ is either identity or the swap of the first two columns, and
\begin{align}\begin{split}
\mathfrak{P}_C := \Big\{ \pi = &\pi_C \pi_R  : \\
& \pi_R \in {\rm clos}\{ (R_1 R_2) , (R_2 R_3)\},\\
& \pi_C \in {\rm clos}\{ (C_1 C_2) , (C_2 C_3)\} \Big\},
\end{split}\end{align}
where $ \pi_R$ runs overs all permutations of the rows, and $\pi_C$ runs over all permutations of the columns.

We can now specify what sorts of deterministic processings of the outcomes (for measurements and for sources) are allowed for each of these types of permutations.

The fact noted at Eq.~\eqref{alltriplesplus} implies that if $\pi \in \mathfrak{P}_A$, then the $\zeta$'s must satisfy
\begin{align}\begin{split}
\zeta_{11}  \zeta_{12} \zeta_{13} = +1,\\
\zeta_{21}  \zeta_{22} \zeta_{23} = +1,\\
\zeta_{31}  \zeta_{32} \zeta_{33} = +1,\\
\zeta_{11}  \zeta_{21} \zeta_{31} = +1,\\
\zeta_{12}  \zeta_{22} \zeta_{32} = +1,\\
\zeta_{13}  \zeta_{23} \zeta_{33} = +1.
\end{split}\end{align}
The solutions of these equations include the case of $\zeta_{ij}=+1$ for all $ij$ as well as those of the form
\begin{align}
\left(\begin{matrix}
  \zeta_{11}&       \zeta_{12}   &       \zeta_{13} \\
           \zeta_{21}  &       \zeta_{22}   &       \zeta_{23} \\
      \zeta_{31}  &       \zeta_{32}   &       \zeta_{33}
 \end{matrix}\right)
 = \left(\begin{matrix}
 -1 &     -1 &     +1 \\
 -1  &     -1  &   +1 \\
   +1 &   +1   &    +1
 \end{matrix}\right)
\end{align}
and any matrix of $\zeta$'s obtained from this one by permutation of the rows and columns.
The set of such $\zeta$ matrices, denoted $\mathfrak{Z}_A$, can be expressed as
\begin{align}\begin{split}
\mathfrak{Z}_A :=  \{& (\zeta_{11},\dots, \zeta_{33}): \forall i,j: \zeta_{ij}=+1\}\\
\cup  \{& (\zeta_{11},\dots, \zeta_{33}):
 \exists i\ne i',j \ne j',\;\\
 & \zeta_{ij} = \zeta_{i'j}=\zeta_{ij'}=\zeta_{i'j'} = -1,\;\\
&  \zeta_{kl}=+1\; {\rm otherwise}\}.
\end{split}\end{align}

Precisely the same analysis holds for the $\gamma$'s, so the set of $\gamma$ matrices that preserve the compatibility relations if $\pi \in \mathfrak{P}_A$
is also $\mathfrak{Z}_A$.
%are
%\begin{align}
%\mathfrak{G}_A &:= \{ (\gamma_{11},\dots, \gamma_{33}):\nonumber\\
%& \exists i\ne i',j \ne j',\; \gamma_{ij} = \gamma_{i'j}=\gamma_{ij'}=\gamma_{i'j'} = -1,\;\nonumber\\
%&  \gamma_{kl}=+1\; {\rm otherwise}\}.
%\end{align}

It follows that if
\begin{align}
(\pi,\{ \zeta_{ij}\},\{ \gamma_{ij}\}) \in \mathfrak{P}_A \times \mathfrak{Z}_A \times \mathfrak{Z}_A,
\end{align}
then the primed measurements and sources have the compatibility relations of the Peres-Mermin scenario.

Next, consider the case where  $\pi \in \mathfrak{P}_B$.  It is useful to focus on a specific example, namely, $\pi = (C_2 C_3)$.
The fact noted at Eq.~\eqref{twotriplesminus} implies that the $\zeta$'s must satisfy
\begin{align}\begin{split}
\zeta_{11}  \zeta_{12} \zeta_{13} = +1,\\
\zeta_{21}  \zeta_{22} \zeta_{23} = +1,\\
\zeta_{31}  \zeta_{32} \zeta_{33} = +1,\\
\zeta_{11}  \zeta_{21} \zeta_{31} = +1,\\
\zeta_{12}  \zeta_{22} \zeta_{32} = -1,\\
\zeta_{13}  \zeta_{23} \zeta_{33} = -1.
\end{split}\end{align}
Solutions of these equations are of the form
\begin{align}
\left(\begin{matrix}
  \zeta_{11}&       \zeta_{12}   &       \zeta_{13} \\
           \zeta_{21}  &       \zeta_{22}   &       \zeta_{23} \\
      \zeta_{31}  &       \zeta_{32}   &       \zeta_{33}
 \end{matrix}\right)
 = \left(\begin{matrix}
 +1 &     +1 &     +1 \\
 +1  &     +1  &   +1 \\
   +1 &   -1   &    -1
 \end{matrix}\right)
\end{align}
and any matrix of $\zeta$'s obtained from this one by component-wise multiplication with an element of $\mathfrak{Z}_A$.
The set of such $\zeta$ matrices, denoted $\mathfrak{Z}_B$, can therefore be expressed as
\begin{align}
\mathfrak{Z}_B &:= \left\{ Z \circ \left(\begin{matrix}
 +1 &     +1 &     +1 \\
 +1  &     +1  &   +1 \\
   +1 &   -1   &    -1
 \end{matrix}\right): Z \in \mathfrak{Z}_A\right\}
%(\zeta_{11},\dots, \zeta_{33}): \nonumber\\ &\exists Z \in \mathfrak{Z}_A,\; Z \circ \}.
\end{align}

Similarly, the set of $\gamma$ matrices that preserve the compatibility relations if $\pi \in \mathfrak{P}_B$
are those in $\mathfrak{Z}_B$.
%\begin{align}
%\mathfrak{G}_A &:= \{ (\gamma_{11},\dots, \gamma_{33}):\nonumber\\
%& \exists i\ne i',j \ne j',\; \gamma_{ij} = \gamma_{i'j}=\gamma_{ij'}=\gamma_{i'j'} = -1,\;\nonumber\\
%&  \gamma_{kl}=+1\; {\rm otherwise}\}.
%\end{align}

It follows that if
\begin{align}
(\pi,\{ \zeta_{ij}\},\{ \gamma_{ij}\}) \in \mathfrak{P}_B \times \mathfrak{Z}_B \times \mathfrak{Z}_B,
\end{align}
then the primed measurements and sources have the compatibility relations of the Peres-Mermin scenario.

Finally, we consider the third class of permutations, $\mathfrak{P}_C$.  Again, we begin by focusing on a particular permutation in that class, namely, the transpose, $\pi = T$.  The fact noted at Eq.~\eqref{twotriplesminus} implies that in this case the $\zeta$'s must satisfy
\begin{align}\begin{split}
\zeta_{11}  \zeta_{12} \zeta_{13} = +1,\\
\zeta_{21}  \zeta_{22} \zeta_{23} = +1,\\
\zeta_{31}  \zeta_{32} \zeta_{33} = -1,\\
\zeta_{11}  \zeta_{21} \zeta_{31} = +1,\\
\zeta_{12}  \zeta_{22} \zeta_{32} = -1,\\
\zeta_{13}  \zeta_{23} \zeta_{33} = -1.
\end{split}\end{align}
Solutions of these equations are of the form
\begin{align}
\left(\begin{matrix}
  \zeta_{11}&       \zeta_{12}   &       \zeta_{13} \\
           \zeta_{21}  &       \zeta_{22}   &       \zeta_{23} \\
      \zeta_{31}  &       \zeta_{32}   &       \zeta_{33}
 \end{matrix}\right)
 = \left(\begin{matrix}
 +1 &     +1 &     +1 \\
 +1  &     +1  &   +1 \\
   +1 &   +1   &    -1
 \end{matrix}\right)
\end{align}
and any matrix of $\zeta$'s obtained from this one by component-wise multiplication with an element of $\mathfrak{Z}_A$.
The set of such $\zeta$ matrices, denoted $\mathfrak{Z}_C$, is therefore
\begin{align}
\mathfrak{Z}_C &:= \left\{ Z \circ \left(\begin{matrix}
 +1 &     +1 &     +1 \\
 +1  &     +1  &   +1 \\
   +1 &   +1   &    -1
 \end{matrix}\right): Z \in \mathfrak{Z}_A\right\}
\end{align}

The set of $\gamma$ matrices that preserve the compatibility relations if $\pi \in \mathfrak{P}_C$
are also those in $\mathfrak{Z}_C$.
%\begin{align}
%\mathfrak{G}_A &:= \{ (\gamma_{11},\dots, \gamma_{33}):\nonumber\\
%& \exists i\ne i',j \ne j',\; \gamma_{ij} = \gamma_{i'j}=\gamma_{ij'}=\gamma_{i'j'} = -1,\;\nonumber\\
%&  \gamma_{kl}=+1\; {\rm otherwise}\}.
%\end{align}

It follows that if
\begin{align}
(\pi,\{ \zeta_{ij}\},\{ \gamma_{ij}\}) \in \mathfrak{P}_C \times \mathfrak{Z}_C \times \mathfrak{Z}_C,
\end{align}
then the primed measurements and sources have the compatibility relations of the Peres-Mermin scenario.

We have now exhaustively enumerated all of the choices of symmetry operations $(\pi,\{ \zeta_{ij}\},\{ \gamma_{ij}\})$ that preserve the compatibilty relations of the Peres-Mermin scenario.  We turn to the question of how these symmetry operations transform the noncontextuality inequalities that we derive.

Recall that our noncontextuality inequalities will ultimately only refer to the following nine products of a measurement outcome variable with its corresponding source outcome variable:
\begin{align}
\left(\begin{matrix}
  \mathcal{s}_{11} \mathcal{m}_{11} &       \mathcal{s}_{12}  \mathcal{m}_{12} &       \mathcal{s}_{13}\mathcal{m}_{13} \\
           \mathcal{s}_{21}  \mathcal{m}_{21}  &       \mathcal{s}_{22} \mathcal{m}_{22}   &       \mathcal{s}_{23} \mathcal{m}_{23} \\
      \mathcal{s}_{31} \mathcal{m}_{31}   &       \mathcal{s}_{32}  \mathcal{m}_{32}  &       \mathcal{s}_{33} \mathcal{m}_{33}
 \end{matrix}\right)
 \end{align}
If we perform a deterministic processing $(\pi,\{ \zeta_{ij}\},\{ \gamma_{ij}\})$
then the relevant products of outcome variables for the primed experiment are given by\begin{align}\label{relabelpairs}
&\left(\begin{matrix}
  \mathcal{s}'_{11} \mathcal{m}'_{11} &       \mathcal{s}'_{12}  \mathcal{m}'_{12} &       \mathcal{s}'_{13}\mathcal{m}'_{13} \\
           \mathcal{s}'_{21}  \mathcal{m}'_{21}  &       \mathcal{s}'_{22} \mathcal{m}'_{22}   &       \mathcal{s}'_{23} \mathcal{m}'_{23} \\
      \mathcal{s}'_{31} \mathcal{m}'_{31}   &       \mathcal{s}'_{32}  \mathcal{m}'_{32}  &       \mathcal{s}'_{33} \mathcal{m}'_{33}
 \end{matrix}\right) \nonumber\\\begin{split}
 &\hspace{-\mathindent}=
  \left(\begin{matrix}
  \eta_{11}  \zeta_{11}&       \eta_{12}  \zeta_{12} &       \eta_{13} \zeta_{13}\\
           \eta_{21}   \zeta_{21}  &       \eta_{22} \zeta_{22}   &       \eta_{23} \zeta_{23} \\
      \eta_{31} \zeta_{31}  &       \eta_{32} \zeta_{32}   &       \eta_{33} \zeta_{33}
 \end{matrix}\right)\\
&\circ
\left(\begin{matrix}
  \mathcal{s}_{\pi(11)} \mathcal{m}_{\pi(11)}&       \mathcal{s}_{\pi(12)} \mathcal{m}_{\pi(12)}   &      \mathcal{s}_{\pi(13)}   \mathcal{m}_{\pi(13)} \\
          \mathcal{s}_{\pi(21)}   \mathcal{m}_{\pi(21)}  &      \mathcal{s}_{\pi(22)}   \mathcal{m}_{\pi(22)}   &       \mathcal{s}_{\pi(23)} \mathcal{m}_{\pi(23)} \\
      \mathcal{s}_{\pi(31)}   \mathcal{m}_{\pi(31)}  &       \mathcal{s}_{\pi(32)}   \mathcal{m}_{\pi(32)}   &      \mathcal{s}_{\pi(33)}  \mathcal{m}_{\pi(33)}
 \end{matrix}\right)\hspace{-10pt}
 \end{split}\end{align}

Note that only the component-wise {\em product} of the $\zeta$ matrix and the $\gamma$ matrix is relevant for the products of outcomes of interest.
The sets of such products, in each of the three cases considered above, are
\begin{align}\begin{split}
\mathfrak{D}_A &:= \left\{ Z \circ Z': Z,Z' \in \mathfrak{Z}_A\right\},\\
\mathfrak{D}_B &:= \left\{ Z \circ Z': Z,Z' \in \mathfrak{Z}_B\right\},\\
\mathfrak{D}_C &:= \left\{ Z \circ Z': Z,Z' \in \mathfrak{Z}_C\right\}.
\end{split}\end{align}
However, it is straightforward to verify that these three sets are equivalent,
\begin{align}
\mathfrak{D}_A =\mathfrak{D}_B =\mathfrak{D}_C,
% =: \mathfrak{D}.
\end{align}
and easy to determine.
%This set $\mathfrak{D}_A$ is easy to determine.
They are simply the union of $\mathfrak{Z}_A$ and all matrices of the form
\begin{align}
\left(\begin{matrix}
 -1 &     -1 &     +1 \\
 -1  &     +1  &   -1 \\
   +1 &   -1   &    -1
 \end{matrix}\right),
\end{align}
or obtainable from this one by permutations of the rows and columns.

\section{The trivial facet inequalities of the noncontextual correlation polytope}\label{TrivialIneqs}

Here we justify the claim that the inequalities in the symmetry class of Eq.~\eqref{trivialineq} are trivial in the sense that they hold for any correlations arising in the operational Peres-Mermin scenario, independently of whether they admit of a noncontextual model or not.
%We pause here to explain why these inequalities are trivial.
Consider the binary-outcome variables defined as products of outcome variables for source-measurement pairs in the first row of the Peres-Mermin square, $\mathcal{s}_{11}\mathcal{m}_{11}$, $\mathcal{s}_{12}\mathcal{m}_{12}$ and $\mathcal{s}_{13}\mathcal{m}_{13}$.    Recalling Eqs.~(\ref{cc1}-\ref{cc3}), for every run of the experiment we have $\mathcal{m}_{11}\mathcal{m}_{12}\mathcal{m}_{13}=1$, and recalling Eqs.~(\ref{ss1}-\ref{ss3}),  for each run we have $\mathcal{s}_{11}\mathcal{s}_{12}\mathcal{s}_{13}=1$. This implies that for each run we have $(\mathcal{s}_{11}\mathcal{m}_{11})(\mathcal{s}_{12}\mathcal{m}_{12})(\mathcal{s}_{13}\mathcal{m}_{13})=1$. Given this constraint, it follows that the expectation values of the three variables, $\mathcal{s}_{11}\mathcal{m}_{11}$, $\mathcal{s}_{12}\mathcal{m}_{12}$ and $\mathcal{s}_{13}\mathcal{m}_{13}$, are determined by a joint probability distribution over two of them.  But then, positivity  of the probabilities in this distribution implies, by an argument articulated around \cref{jjj1,jjj2}, that
\begin{align}\begin{split}
 &\hspace{-1ex}\text{for all }a,b \in \{-1,+1\}\,:
 \\&\hspace{-1ex}- a \langle \mathcal{s}_{11}\mathcal{m}_{11} \rangle - b \langle \mathcal{s}_{12}\mathcal{m}_{12} \rangle - ab \langle \mathcal{s}_{13}\mathcal{m}_{13} \rangle \le   1.
\end{split}\end{align}
Eq.~\eqref{trivialineq} is of this form.

Note that for the third column of the Peres-Mermin square, Eqs.~(\ref{comp1}-\ref{comp3}) imply that for every run of the experiment we have $\mathcal{m}_{13}\mathcal{m}_{23}\mathcal{m}_{33}=-1$, and Eqs.~(\ref{sss1}-\ref{sss3}) imply that for each run we have $\mathcal{s}_{13}\mathcal{s}_{23}\mathcal{s}_{33}=-1$, but these together yield $(\mathcal{s}_{13}\mathcal{m}_{13})(\mathcal{s}_{23}\mathcal{m}_{23})(\mathcal{s}_{33}\mathcal{m}_{33})=1$, exactly as we have for the other columns and the rows.

\section{Criticism of a previous proposal for a noncontextuality inequality based on the Peres-Mermin square}\label{subsec:prevProposals}

%\color{red} [Say more about the distinction between NC inequalities and KS inequalities?  Tell reader: for more details, see Appendix C of Ref.~\cite{KunjwalSpekkens}.]\color{black}

In this section, a prior proposal for an experimental test of
noncontextuality based on the Peres-Mermin proof of the Kochen-Specker theorem, that of Cabello \cite{cabello08experimentally},  will be reviewed and criticized. Our criticisms here parallel those provided   in Appendix C of Ref.~\cite{KunjwalSpekkens} for a similar proposal that was based on a different proof of the Kochen-Specker theorem.
%and discuss their shortcomings.

%The first proposal related to the Peres-Mermin construction was one that started with a state-dependent proof of the Kochen-Specker theorem  involving a subset of the observables from the Peres-Mermin construction~\cite{CFRH}.
%Cabello \cite{cabello08experimentally} subsequentlyproposed an experimental test of noncontextuality based on the state-independent version of Peres-Mermin.  The latter, therefore, is the appropriate proposal to compare with our work.
%As  and we shall begin with a discussion of this proposal.

%The first such proposal appears in Cabello, Filipp, Rauch and Hasegawa \cite{CFRH}.  Cabello \cite{cabello08experimentally} subsequently proposed an experimental test of noncontextuality for the state-independent version of Peres-Mermin and we shall begin with a discussion of this proposal.

We describe the proposal of Ref.~\cite{cabello08experimentally} using the notation introduced in this article.  There are nine operational quantities appearing in the inequality derived therein, corresponding to the expectation value, relative to an arbitrary preparation $P$, of the product of the outcomes of each of the six triples of compatible measurements in the Peres-Mermin scenario: $\avg{ \mathcal{m}_{11} \mathcal{m}_{12} \mathcal{m}_{13} }_P, \avg{ \mathcal{m}_{21} \mathcal{m}_{22} \mathcal{m}_{23} }_P, \ldots, \avg{ \mathcal{m}_{13} \mathcal{m}_{23} \mathcal{m}_{33} }_P$.
%In this model, the expectation value assigned to a random variable by an ontic state depends only on the equivalence class of the random variable.
Only the average of these expectation values is considered in Ref.~\cite{cabello08experimentally}; we denote it
%In Ref~\cite{cabello08experimentally}, it is proposed that the operational quantity of interest is
\begin{dmath} \label{RofP}
  R(P) \equiv \avg{ \mathcal{m}_{11} \mathcal{m}_{12} \mathcal{m}_{13} }_P
  +\avg{ \mathcal{m}_{21} \mathcal{m}_{22} \mathcal{m}_{23} }_P
  +\avg{ \mathcal{m}_{31} \mathcal{m}_{32} \mathcal{m}_{33} }_P
  +\avg{ \mathcal{m}_{11} \mathcal{m}_{21} \mathcal{m}_{31} }_P
  +\avg{ \mathcal{m}_{12} \mathcal{m}_{22} \mathcal{m}_{32} }_P
  -\avg{ \mathcal{m}_{13} \mathcal{m}_{23} \mathcal{m}_{33} }_P,
\end{dmath}
%for some fixed preparation $P$ and that it should satisfy the following inequality in a noncontextual model
The inequality derived in Ref.~\cite{cabello08experimentally} is
\begin{align}
R(P) \le 4.
\label{operationalboundonRPM}
\end{align}
%It is claimed  that this bound delimits the boundary between contextual and noncontextual theories.
We dispute the notion that this inequality delimits the boundary between contextual and
noncontextual theories, and therefore also the notion that experimental violations thereof, such as those achieved in Refs.~\cite{amselem2009, liu2009,dambrosio2013}, have any bearing on the possibility of noncontextual models, for reasons that we presently outline.

%We dispute the claim that this bound delimits the boundary between contextual and noncontextual theories.
We begin by reviewing the argument presented in favour of Eq.~\eqref{operationalboundonRPM}.
%To begin, we repeat the argument presented in Ref.~\cite{cabello08experimentally}  present the  derivation of this inequality in the notation introduced here.
Ref.~\cite{cabello08experimentally}  presumes that outcomes are assigned deterministically by the ontic state.  Suppose that $R(\lambda)$ denotes a particular noncontextual deterministic assignment to the sum of the triple-products.  Recalling our notational convention that $\lfloor \cdot \rfloor_{\lambda}$ denotes a deterministic assignment (See below Eq.~\eqref{gptPM6}), we have
\begin{dmath} \label{Roflambda0}
  R(\lambda) \equiv
  \lfloor \mathcal{m}_{11} \mathcal{m}_{12} \mathcal{m}_{13} \rfloor_{\lambda}
  +\lfloor \mathcal{m}_{21}\mathcal{m}_{22} \mathcal{m}_{23} \rfloor_{\lambda}
  +\lfloor \mathcal{m}_{31} \mathcal{m}_{32} \mathcal{m}_{33} \rfloor_{\lambda}
  +\lfloor \mathcal{m}_{11} \mathcal{m}_{21} \mathcal{m}_{31} \rfloor_{\lambda}
  +\lfloor \mathcal{m}_{12} \mathcal{m}_{22} \mathcal{m}_{32} \rfloor_{\lambda}
  -\lfloor \mathcal{m}_{13} \mathcal{m}_{23}\mathcal{m}_{33} \rfloor_{\lambda}.
\end{dmath}
Any preparation procedure $P$ corresponds to a convex sum of such assignments,
\begin{align}
R(P) = \sum_{\lambda} R(\lambda) \mu(\lambda|P).
\label{defnRofP}
\end{align}
The inequality of Eq.~\eqref{operationalboundonRPM} is inferred from the inequality
%is meant to follow from the fact that all noncontextual deterministic assignments satisfy
%derived through an argument wherein it is claimed that in any noncontextual model,
\begin{align}
R(\lambda)\le 4,
\label{ontologicalboundRPM}
\end{align}
%and then using Eq.~\eqref{defnRofP}, one deduces that
%\begin{align}
%  R(P) \leq 4
%  \label{eq:CFRH}
%\end{align}
and the argument presented in favour of the latter
%Eq.~\eqref{ontologicalboundRPM} in Ref.~\cite{cabello08experimentally}
is simply this:
recalling that each of the nine $\mathcal{m}_{ij}$ is assigned a value in $\{-1,+1\}$ in an outcome-deterministic assignment, there are $2^9$ possible joint assignments of values to
$(\mathcal{m}_{11},\dots, \mathcal{m}_{33})$,
%by an ontic state $\lambda$  are chosen from all $2^9$
%possible combinations of assignments of an element of $\{-1,+1\}$ to each of the nine $\mathcal{m}_{i,j}$. For every such combination,
and for each of these, one can verify that $R(\lambda)\le 4$.

%The problem is that for each of these $2^9$ assignments, one can identify a compatible triple which are assigned values that are {\em inconsistent} with the compatibility relation that they are meant to satisfy.

We now explain the problem with this argument. The measurement compatibility relations that serve to {\em define} the operational Peres-Mermin scenario, described around Eqs.~(\ref{cc1}-\ref{cc3}), imply that for each of the six compatible triples of measurements, the associated triple of outcomes that one can obtain in any run of the experiment has a fixed product.  Specifically, this product is $-1$ for the triple corresponding to the third column, and +1 for the other five triples, as noted in Eq.~\eqref{tripleproducts}.

However, for each of the $2^9$ joint assignments to
$(\mathcal{m}_{11},\dots, \mathcal{m}_{33})$, one can identify a compatible triple which are assigned values that are {\em inconsistent} with the compatibility relation that they are meant to satisfy.  For instance, the assignment of the +1 outcome to each of the nine measurements fails to give the correct product of outcomes for the third column.
In other words, if a set of nine measurements satisfy the compatibility relations of the operational Peres-Mermin construction, then {\em none} of the $2^9$ joint assignments of outcomes to these measurements are logically possible assignments.

More formally, the fact that the assignments are assumed to be deterministic implies that the assignment to the product of the outcome is
%Indeed, another way to see this fact is to note that the $2^9$ joint assignments being considered are {\em deterministic} assignments.   The determinism implies that the assignment to the product of outcomes is
the product of the assignments to each outcome, $\lfloor \mathcal{m}_{11} \mathcal{m}_{12} \mathcal{m}_{13} \rfloor_{\lambda} = \lfloor \mathcal{m}_{11}\rfloor_{\lambda}  \lfloor \mathcal{m}_{12}\rfloor_{\lambda}  \lfloor \mathcal{m}_{13} \rfloor_{\lambda} $,
%, and therefore that
%\begin{dmath} \label{Roflambda}
%  R(\lambda) \equiv \lfloor \mathcal{m}_{11}\rfloor_{\lambda}  \lfloor \mathcal{m}_{12} \rfloor_{\lambda}  \lfloor\mathcal{m}_{13} \rfloor_{\lambda}
%  +\lfloor \mathcal{m}_{21} \rfloor_{\lambda}  \lfloor\mathcal{m}_{22}\rfloor_{\lambda}  \lfloor \mathcal{m}_{23} \rfloor_{\lambda}
 % +\lfloor \mathcal{m}_{31} \rfloor_{\lambda}  \lfloor\mathcal{m}_{32}\rfloor_{\lambda}  \lfloor \mathcal{m}_{33} \rfloor_{\lambda}
%  +\lfloor \mathcal{m}_{11} \rfloor_{\lambda}  \lfloor\mathcal{m}_{21}\rfloor_{\lambda}  \lfloor \mathcal{m}_{31} \rfloor_{\lambda}
%  +\lfloor \mathcal{m}_{12} \rfloor_{\lambda}  \lfloor\mathcal{m}_{22}\rfloor_{\lambda}  \lfloor \mathcal{m}_{32} \rfloor_{\lambda}
%  -\lfloor \mathcal{m}_{13}\rfloor_{\lambda}  \lfloor \mathcal{m}_{23} \rfloor_{\lambda}  \lfloor\mathcal{m}_{33} \rfloor_{\lambda}.
%\end{dmath}
and
%Furthermore, for deterministic assignments,
 the compatibility relations that are assumed to hold among the measurements imply that the $(\lfloor \mathcal{m}_{12} \rfloor_{\lambda} , \ldots, \lfloor \mathcal{m}_{33} \rfloor_{\lambda})$ must satisfy the constraints of Eqs.~(\ref{gptPM1}-\ref{gptPM6}), which have no solution.

%Here is another take on this criticism.
Furthermore, the compatibility relations imply that  for {\em any} preparation $P$, whatever triple of compatible measurements one implements, the outcomes always satisfy Eq.~\eqref{tripleproducts}, and therefore
the expectation values satisfy these relations as well:
\begin{align}\begin{split}\label{tripleproductsexp}
 \langle \classicalOutputMsmtEqv{11} \classicalOutputMsmtEqv{12} \classicalOutputMsmtEqv{13}\rangle_P  &=+1,\\
 \langle \classicalOutputMsmtEqv{21} \classicalOutputMsmtEqv{22} \classicalOutputMsmtEqv{23} \rangle_P&=+1,\\
 \langle \classicalOutputMsmtEqv{31} \classicalOutputMsmtEqv{32} \classicalOutputMsmtEqv{33}\rangle_P &=+1,\\
 \langle \classicalOutputMsmtEqv{11} \classicalOutputMsmtEqv{21} \classicalOutputMsmtEqv{31}\rangle_P &=+1,\\
 \langle \classicalOutputMsmtEqv{12} \classicalOutputMsmtEqv{22} \classicalOutputMsmtEqv{32}\rangle_P &=+1,\\
  \langle \classicalOutputMsmtEqv{13} \classicalOutputMsmtEqv{23} \classicalOutputMsmtEqv{33} \rangle_P&=-1.
\end{split}\end{align}
In other words,
just as the observables $\1 \otimes \1$ and $-\1 \otimes \1$ in quantum theory are trivial insofar as they take the same value for all states, the triple-product of outcomes $ \classicalOutputMsmtEqv{11} \classicalOutputMsmtEqv{12} \classicalOutputMsmtEqv{13}, \ldots, \classicalOutputMsmtEqv{13} \classicalOutputMsmtEqv{23} \classicalOutputMsmtEqv{33}$ are trivial operational quantities insofar as they take the same value for all preparations.
Substituting the identities in Eq.~\eqref{tripleproductsexp} into Eq.~\eqref{RofP}, one obtains
\begin{align}\label{RP6}
R(P)=6.
\end{align}
Therefore, for {\em any} set of measurements satisfying the compatibility relations of the operational Peres-Mermin scenario, one will necessarily find this equality to hold.
% $R(P)=6$.

%This is true regardless of whether the experiment admits of a noncontextual ontological model.  The operational quantity $R(P)$, therefore, contains no information about whether or not a noncontextual model can describe the experiment.

%One more criticism is in order.
In particular, we expect Eq.~\eqref{RP6} to hold {\em no matter how noisy the measurements are}.
%regardless of how noisy the measurements are.
To see this, it suffices to note that the equalities of Eq.~\eqref{tripleproducts} hold for any set of noisy measurements satisfying the compatibility relations, and this implies that Eq.~\eqref{RP6} holds for such measurements as well.
% {\em no matter how noisy the measurements are}.
  For instance, Eqs.~\eqref{tripleproducts} and \eqref{RP6} hold for the noisy quantum realization of the operational Peres-Mermin scenario, described in Sec.~\ref{noisyQrealization}, for any amount of depolarization noise.  Consequently, if measuring $R(P)$ to have a value greater than 4 could constitute evidence for the failure of noncontextuality, then this evidence could be obtained even in the presence of arbitrarily large amounts of noise.  This is an indictment of the proposal of Ref.~\cite{cabello08experimentally} because a minimal constraint on any reasonable notion of noncontextuality (first articulated in Ref.~\cite{KunjwalSpekkens}) is that it should not be possible to demonstrate its failure in a completely incoherent experiment.

%Summarizing,
To summarize then, our criticism is as follows.  The inequality $R(P) \le 4$ should {\em not} be expected to hold for any experiment satisfying the compatibility structure of the operational Peres-Mermin scenario, while the equality $R(P)=6$ (and hence the violation of $R(P) \le 4$) is expected to hold {\em trivially} in all such experiments.  And this is the case {\em regardless} of whether the experiment admits of a noncontextual ontological model.  As such, the operational quantity $R(P)$ contains no information about whether or not a noncontextual model can describe the experiment.

The most significant point of contrast between the proposal of this article and that of Ref.~\cite{cabello08experimentally} is that we assume the notion of universal noncontextuality proposed in Ref.~\cite{Spekkens05}, rather than the notion of KS-noncontextuality.
Because universal noncontextuality, unlike KS-noncontextuality, does not assume outcome determinism, we are led to consider {\em indeterministic} noncontextual  assignments to the measurements.  As we saw above, the fact that there are strictly no deterministic noncontextual  assignments respecting the compatibility relations of the operational Peres-Mermin scenario is what makes it futile to attempt to derive constraints on operational statistics from the assumption of such assignments.  The assumption is logically ruled out, so an operational test is neither necessary nor conceivable.
%and the notion that experimental violations thereof~\cite{} warrant conclusions about noncontextuality,
On the other hand, there {\em are} many indeterministic noncontextual assignments that respect the compatibility relations, such as the example provided in Eq.~\eqref{examplevertex}.  These, therefore, {\em do} impose nontrivial constraints on operational statistics, constraints that are encoded in the noncontextuality inequalities that we have derived.

\end{document}